\documentclass[article]{aa}
\usepackage{array}
\usepackage{graphicx}
\usepackage[graphicx]{realboxes}
\usepackage{placeins}
\usepackage[varg]{txfonts}
\usepackage{subcaption}
\usepackage{multirow}
\usepackage{natbib}
\usepackage[percent]{overpic}
\usepackage{xcolor}

\usepackage{hyperref}
\hypersetup{
    colorlinks=true,
    linkcolor=blue,
    citecolor=blue,
    filecolor=magenta,      
    urlcolor=blue,
    bookmarks=true}

\begin{document} 

        \title{X TrA through the eyes of MATISSE: More evidence of clumpy molecular layers around C-type asymptotic giant branch stars\thanks{Based on observations collected at the European Organisation for Astronomical Research in the Southern Hemisphere under ESO Program ID 108.22E9.}}
\titlerunning{X TrA through the eyes of MATISSE: More evidence of clumpy molecular layers around C-type AGB stars}
   
        \author{V.~Răstău\inst{\ref{inst1}} 
                \and C.~Paladini\inst{\ref{inst2}} 
                \and J.~Drevon\inst{\ref{inst2}}
                \and J.~Hron\inst{\ref{inst1}} 
                \and F.~Kerschbaum\inst{\ref{inst1}}
                \and M.~Wittkowski\inst{\ref{inst3}}
                \and J.P.~Fonfria\inst{\ref{inst4}}
                \and M.~Montargès\inst{\ref{inst5}}
                \and T.~Khouri\inst{\ref{inst6}}
                \and W.~Vlemmings\inst{\ref{inst6}}
                \and H.~Olofsson\inst{\ref{inst6}}
                \and K.~Ohnaka\inst{\ref{inst7}}
                \and J.~Alonso-Hernandez\inst{\ref{inst8}}
                \and C.~Sánchez Contreras\inst{\ref{inst8}}
                \and L.~Velilla-Prieto\inst{\ref{inst9}}
                \and W.C.~Danchi\inst{\ref{inst10}}
                \and G.~Rau\inst{\ref{inst11}, \ref{inst12}}
                \and F.~Lykou\inst{\ref{inst13}}
                \and J.~Sanchez-Bermudez\inst{\ref{inst14}}
                \and B.~Lopez\inst{\ref{inst15}}
                \and S.~Höfner\inst{\ref{inst16}}
                \and B.~Aringer\inst{\ref{inst1}}
                \and L.~Planquart\inst{\ref{inst6}}
                \and P.~Cruzalèbes\inst{\ref{inst15}}
                \and G.~Weigelt\inst{\ref{inst17}}
                }
                
        \institute{
        Department of Astrophysics, University of Vienna, Türkenschanzstrasse 17, 1180 Vienna, Austria \label{inst1}
        \and
            European Southern Observatory, Alonso de Córdova, 3107 Vitacura, Santiago, Chile \label{inst2}
        \and 
        European Southern Observatory, Karl-Schwarzschild-Str. 2, 85748 Garching bei München, Germany \label{inst3}
        \and
        Institute of Fundamental Physics (CSIC), Department of Molecular Astrophysics, Serrano 123, 28006 Madrid, Spain \label{inst4}
        \and
        LIRA, Observatoire de Paris, Université PSL, Sorbonne Université, Université Paris Cité, CY Cergy Paris Université, CNRS, 92190 Meudon, France \label{inst5}
        \and
        Department of Space, Earth and Environment, Chalmers University of Technology, Gothenburg, Sweden \label{inst6}
        \and
        Instituto de Astrofísica, Facultad de Ciencias Exactas, Universidad Andrés Bello, Chile \label{inst7}
        \and
        Centro de Astrobiología (CAB), CSIC-INTA. ESAC-campus, Camino Bajo del Castillo s/n, E-28692, Villanueva de la Ca\~nada, Madrid, Spain \label{inst8}
        \and
        Institute of Fundamental Physics (CSIC), Department of Molecular Astrophysics. Serrano 123, 28006 Madrid, Spain \label{inst9}
        \and
        NASA Goddard Space Flight Center, Greenbelt, MD, USA \label{inst10}
        \and
        Schmidt Sciences, New York, NY \label{inst11}
        \and
        National Science Foundation, Alexandria, VA \label{inst12}
        \and
        HUN-REN Research Centre for Astronomy and Earth Sciences, Konkoly Observatory, Budapest, Hungary \label{inst13}
        \and
        Universidad Nacional Autónoma de México. Instituto de Astronomía. A.P. 70-264, 04510, Ciudad de México, 04510, México \label{inst14}
        \and
        Université Côte d’Azur, Observatoire de la Côte d’Azur, CNRS, Laboratoire Lagrange, France \label{inst15}
        \and
        Department of Physics and Astronomy, Division of Astronomy and Space Physics, Theoretical Astrophysics, Uppsala University, Box 516, 75120 Uppsala, Sweden \label{inst16}
        \and
        Max-Planck-Institut fur Astrophysik, Karl-Schwarzschild-Straße 1, 85741 Garching, Germany \label{inst17}
        }
        \date{}
 
        \abstract
        {}
        {The goal of this study is to further the understanding of the wind formation mechanism in asymptotic giant branch (AGB) stars through the analysis of the close environment (within a few stellar radii) of the carbon star \object{X TrA}.}
        {\object{X TrA} was observed for the first time with the Mid-Infrared SpectroScopic Experiment instrument (MATISSE) in the L and N bands in low spectral resolution mode (R=30), and its close surroundings were mapped in specific wavelength ranges corresponding to specific molecules ($\mathrm{C_2H_2}$ and $HCN$, at 3.1 and 3.8 $\mu$m) and dust (amorphous carbon and, for example, SiC at 11.3 $\mu$m), via image reconstruction techniques.}
        {The angular diameter of the star ranges from 10 mas in the L band pseudo-continuum (3.5 $\mu$m) to 20 mas at 3.1 and 11.3 $\mu$m. The reconstructed images show some mild elongated features (along the east-west direction) and asymmetric protrusions, which are most evident around 3.1 $\mu$m. Imaging results highlight the clumpy nature of the circumstellar environment, starting from the photospheric region up to more distant layers.}
        {The angular diameters found for X TrA in the image data are in agreement with previous photospheric diameter estimates (following VLTI/MIDI 8--13$\mu$m observations), and their wavelength dependence is similar to values found for other carbon stars observed with MATISSE (\object{R Scl} and \object{V Hya}). The 3.1 $\mu$m images presented here show highly asymmetric features, another case of a C-rich star with irregular morphologies close to the stellar disk; this supports the notion that the $\mathrm{C_2H_2 + HCN}$ abundance distribution usually originates from a clumpy layer around carbon stars.}


        \maketitle

\section{Introduction} \label{s1}

\indent\indent For low- to intermediate-mass stars (0.8 < M < 8 $M_{\sun}$), the asymptotic giant branch (AGB) phase is one of the final evolutionary stages, before they end up as white dwarfs \citep{herwig2005}. They experience slow but pronounced dusty winds \citep{holof2018} and strong convective dredge-up. The high mass-loss-rate winds remove the outer layers of the stars, creating envelopes that are expected to be roughly spherically symmetric on large scales. 

\par On smaller scales, inhomogeneous gas and/or dust structures (i.e. geometrically thin shells, also called detached shells, filaments, and clumps) can form due to variations in the mass-loss rate \citep{kerschbaum2017}, episodic and anisotropic ejections \citep{velilla2023}, or binary interactions \citep{decin2020}. Many of these departures from spherical symmetry are attributed to (sub)stellar companions interacting with AGB winds, but the brightness and variability of AGB stars make companion identification challenging \citep[with only a few such objects confirmed so far;][and \textcolor{blue}{Drevon et al., in prep.}]{ramstedt2014, kervella2016, doan2020, planquart, montarges2025}. Close to the photosphere, convection plays an important role in shaping asymmetries \citep{freytag2023}, as evidenced by clumpy circumstellar envelopes seen in stars such as \object{W Hya} \citep{vlemmings2017}, \object{$\pi^1$ Gru} \citep{paladini2018}, \object{IRC+10$\degr$216} \citep{velilla2023}, \object{R Car} \citep{guzman2023}, and \object{R Dor} \citep{vlemmings2024}. 

\par Analysing both small- and large-scale departures from symmetry is therefore crucial for understanding the evolution of AGB stars and their winds. To address the small-scale deviations close to the stellar photosphere, the BINary AGB ESO Large Programme (BIN-AGB LP; PI Paladini, programme ID: 108.22E9) was designed to image the close environment (the first 10 stellar radii) of a collection of ten AGB stars that span different chemical and variability types, including both stars with known companions and those thought to be single.

\par \object{X Trianguli Australis} (X TrA) is a carbon-rich (C-type) AGB star with an effective temperature of 2\,700 K \citep{kipper2004}. It is located approximately 354 pc away \citep[based on a \textit{Gaia} parallax of $\mathrm{2.86 \pm 0.15\,mas;}$][]{abia2022}, although \citet{miora_deathstar} placed the star at a distance of 292 pc using an updated period-luminosity relation calibrated with very long-baseline interferometry parallaxes. The star has been reported as a single object \citep{eggtok2008}, and even though it is considered to be an irregular variable \citep[][type C5,5]{walkn1998}, \citet{percy2009} estimate its variability period to be 500 days. It has an estimated mass-loss rate ($\dot{M}$) of $\mathrm{1.3\,to\, 1.8 \times 10^{-7}}$ $M_{\sun}/yr$ \citep{scholof2001, bergchev2005}. 

\par Asymptotic giant branch stars are major contributors to the enrichment of the interstellar medium \citep{schneider2014}, and the winds colliding with the interstellar medium at large scales create interaction regions \citep[see e.g.][]{cox2012}. First reported by \citet{izu1995}, a detached shell expected to be created by an increased mass-loss rate during a thermal pulse was detected around X TrA with the Infrared Astronomical Satellite (IRAS) at a radial distance of 2.3$\arcmin$ (corresponding to approximately 49\,000 au), in the form of infrared excess at 60 and 100 $\mathrm{\mu m}$. Later, \citet{cox2012} found a faint ring-like structure in the far-infrared around the object (seen with \textit{Herschel}/PACS at 70 and 160 $\mathrm{\mu m}$), with a radius of about 2.5$\arcmin$ (or about 53\,000 au). These two different estimates for the size could indicate the presence of two different shells, and assuming a typical expansion velocity of $\mathrm{10\,km\,s^{-1}}$ for such outflows would give dynamical ages of the shells of approximately 23\,000 and 25\,000 years, respectively. 

\par X TrA was also observed with the Very Large Telescope Interferometer (VLTI) MID-infrared interferometric Instrument (MIDI; 8--13 $\mu$m) between April and August 2011 by \citet{paladini2017}, who were able to confirm previous reports of SiC dust features \citep[IRAS spectra;][]{sloan1998} and located them at a distance of four stellar radii. Obtained via the V-K relation from \citet{vb99}, they estimated the photospheric diameter of the star to be 11 mas. More recently, \citet{miora2021} observed the star with the Atacama Large Millimeter/submillimeter Array and determined the CO(2--1) and CO(3--2) envelope size, finding the CO(3--2) envelope to be slightly elongated (following Gaussian fitting of the CO lines and resulting in a minor to major axis ratio of $0.88\pm0.42$). Its irregular pulsation classification, along with the so far uncontested single status, makes X TrA a good case to be compared against the suspected and/or confirmed binary C-type stars of the BIN-AGB LP sample.

\section{Observations and data description} \label{s2}

\indent\indent \object{X TrA} observations were obtained as part of the large programme 108.22E9 with the Multi Aperture Mid-Infrared SpectroScopic Experiment instrument \citep[MATISSE;][]{matisse}. MATISSE combines the beams of four telescopes of the Very Large Telescope (VLT) at Cerro Paranal -- either the Unit Telescopes (8m aperture) or the Auxiliary Telescopes (1.8 m aperture) -- and allows for a wide coverage of the $(u, v)$ plane with baselines as long as 200 m. It works simultaneously in three wavelength bands: the L band (3.0 to 4.0 $\mu$m), the M band (4.5 to 5.0 $\mu$m), and the N band (8.0 to 13.0 $\mu$m). This allows the gas and dust features to be probed at infrared wavelengths with a spatial resolution ranging from 3 to 10 mas, making the instrument well suited to the study of AGB winds and dust formation regions. In this case, the goal of the programme is to determine the overall geometry of the targets. As such, the observations were done with the Auxiliary Telescopes in the L and N bands in low spectral resolution mode ($\mathrm{R=30}$). X TrA was observed between 2 April 2022 and 5 July 2022, with science observations bracketed between calibrators. The calibrators were used for both visibility and flux calibration, with Tables~\ref{calibdetails} and \ref{obslog} showing the full observation log and all relevant calibrator data. It is worth noting that the N band data did not provide sufficient angular resolution for our purposes. This limitation arises from the telescope configurations, which reached baseline lengths of only about 130 m (yielding estimated angular resolutions of 5--6 mas in the L band and $\sim$16 mas in the N band) and from the use of relatively faint calibrators in the N band.

\par
\begin{figure}
\centering
        \includegraphics[width = \linewidth]{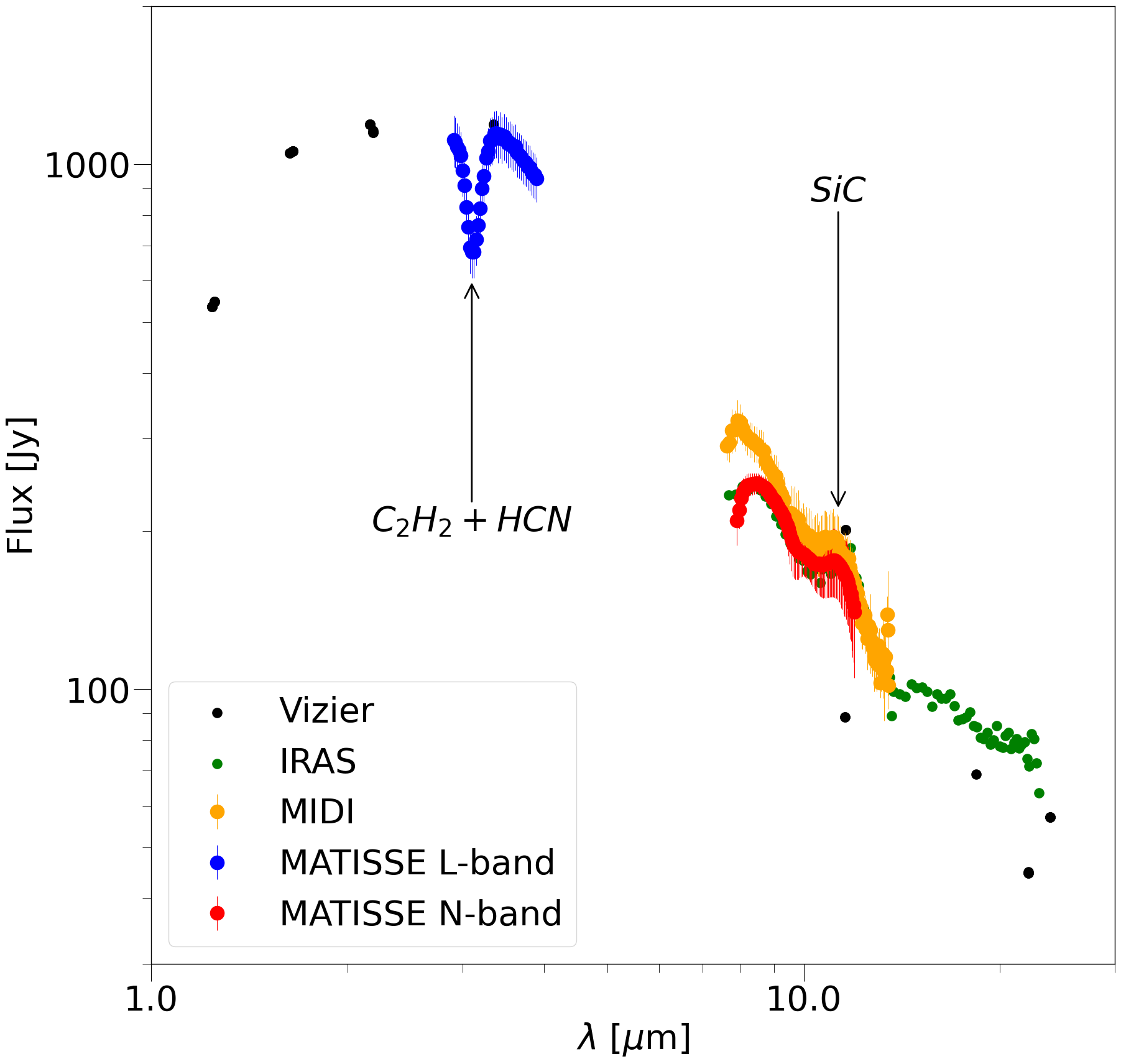}
        \caption{Spectral energy distribution of \object{X TrA}. The black points represent the photometric data listed in Table~\ref{phot}, and the green points correspond to IRAS data. Red and blue data points come from the current MATISSE observations, and the yellow dataset represents the MIDI observations of \citet{paladini2017}. The black arrows mark some of the spectral features we are interested in.} \label{sed}
\end{figure}

 The processing of the data (reduction and calibration) was performed using the Python package \texttt{mattools}\footnote{Available at \href{https://gitlab.oca.eu/MATISSE/tools}{https://gitlab.oca.eu/MATISSE/tools}} developed by the MATISSE consortium (a python wrapper for the official ESO pipeline), following the method of \citet{drevon2024}. The calibrated data were combined into one dataset; the visibility and closure phase data can be seen in Appendix~\ref{appB}. The L band visibility (Fig.~\ref{l_v2}) shows a clear dip at 3.1 $\mu$m, indicating the presence of spectral features, while the closure phases are non-zero (see Fig.~\ref{l_cp}), a typical indicator of asymmetries. The N band visibility and near-zero closure phase data (Figs.~\ref{n_v} and \ref{n_cp}) indicate the source is mostly symmetric in this wavelength range within the limits of our spatial resolution. Based on this, some wavelength windows were selected for the image reconstruction process. Appendix~\ref{appC} shows the complex squared visibility curves and closure phase data that were used. The selected wavelengths cover the spectral signatures of a few molecular species ($\mathrm{C_2H_2}$ and HCN) and dust compounds (SiC) that are typical of C-rich environments \citep{jorgensen2000, pitman2008}. $\mathrm{C_2H_2}$ is thought to play a key role in the formation and growth of amorphous carbon dust \citep{gailsed2013}. In addition to being excellent gas tracers, $\mathrm{C_2H_2\,+\,HCN}$ produces a strong absorption band at 3.1 $\mu$m. Probing this feature can provide valuable information about the distribution and physical conditions of the extended molecular layers above the photosphere of carbon stars. The wavelength intervals that were singled out are as follows: 3.05 -- 3.13 $\mu$m in order to probe the $\mathrm{C_2H_2\,+\,HCN}$ molecular band, 3.45  -- 3.55 $\mu$m for the L band pseudo-continuum, 3.75 -- 3.85 for the $\mathrm{C_2H_2}$ molecular band, 8.4 -- 8.6 $\mu$m as our N band pseudo-continuum (dominated by thermal emission of warm dust) and 11.25 -- 11.55 $\mu$m to probe any potential dust emission. The more pronounced features are marked in Fig.~\ref{sed}. 

\par A correction factor was applied to the MATISSE N band spectrum presented in Fig.~\ref{sed}. This factor is derived from the ratio between the model flux used for each calibrator during the flux calibration process done by the \texttt{mattools} package and their corresponding catalogue fluxes. For this work, the 8.6 $\mu$m measurement from the the AKARI satellite (formerly known as the InfraRed Imaging Surveyor) was used to determine that ratio. The scatter in the Vizier data past 10 $\mu$m, as well as their offset relative to the IRAS flux, is likely caused by similar correction procedures applied in those datasets.

\begin{figure}
\centering
        \includegraphics[width = \linewidth]{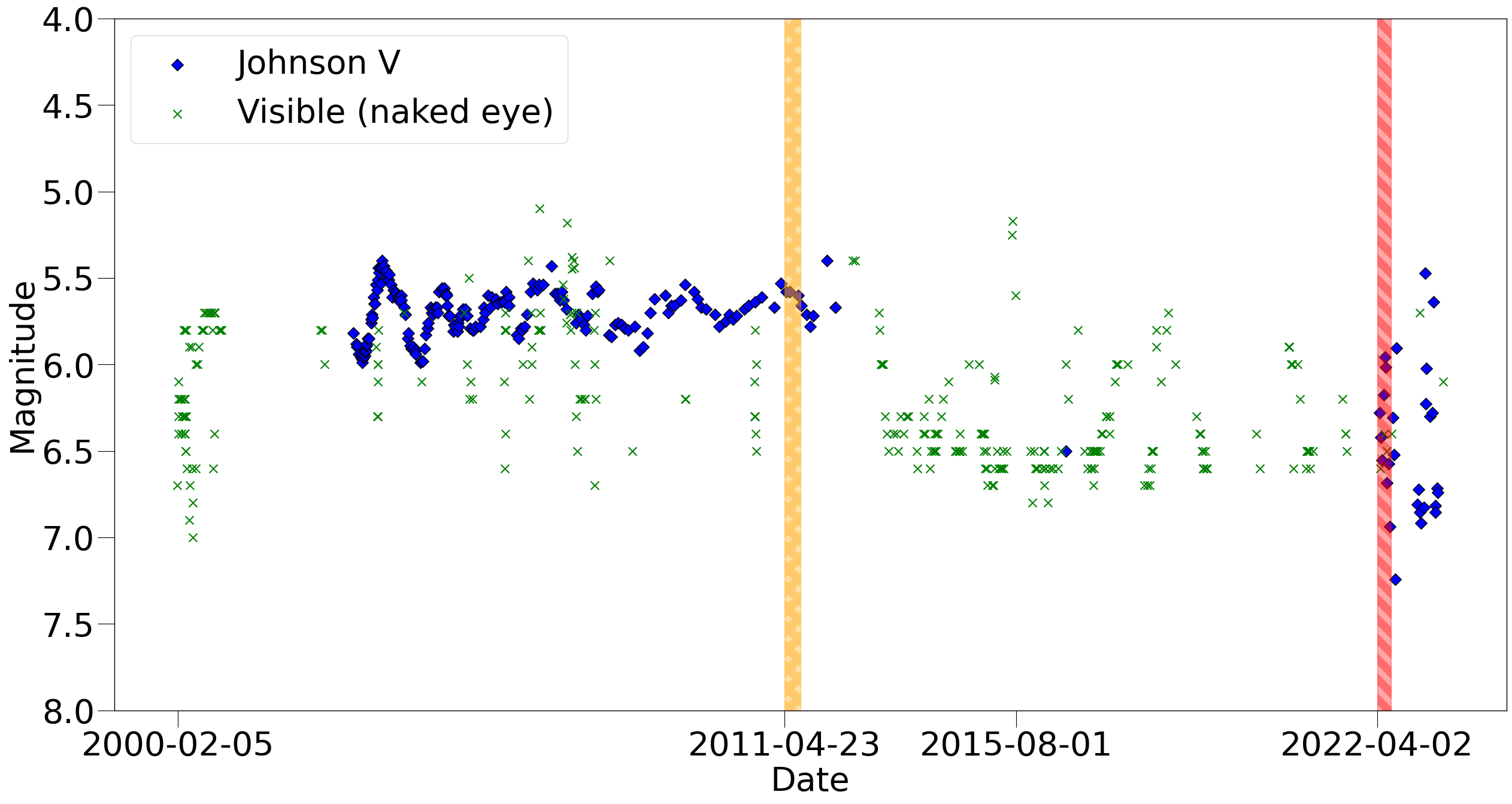}
        \caption{AAVSO \citep{aavso} light curve of \object{X TrA} (blue and green symbols), from 5  February 2000 to 13 February 2025. The MIDI and MATISSE epochs of observation are highlighted in orange and red, respectively.} \label{light_curve}
\end{figure}

\section{Image reconstruction} \label{s3}

\indent\indent The datasets obtained from the observations were merged before the image reconstruction process for two reasons. For one, the observations were performed over a period shorter than the star's variability period (95 days, corresponding to $\sim$20\% of the total period of 500 days). Since available photometric data do not show very strong variations (around 0.5 amplitude variability in Johnson V magnitude; see Fig.~\ref{light_curve}) and since the MATISSE observations are in the near- to mid-infrared, the expected variability is even lower \citep[see e.g. ][]{lebertre1993}. The Fourier transform of the sky brightness distribution is sampled in the $(u,v)$-plane, where each point corresponds to a telescope baseline and its measured spatial frequency. Combining different observations increases the $(u,v)$-coverage by adding more spatial frequency measurements, improving the filling of gaps done by image reconstruction techniques and reducing artefacts. The reconstruction process was carried out with the Multi-aperture Image Reconstruction Algorithm \citep[MiRA;][]{mira}, an adaptive gradient-based method, across the different wavelength windows listed in Sect.~\ref{s2} via PYRA (Python for MiRA)\footnote{\url{https://github.com/jdrevon/PYRA/tree/main}} and MYTHRA (Mean Astrophysical Images with PYRA \footnote{\url{https://github.com/jdrevon/MYTHRA/tree/main}}, two newly developed Python-based tools that interface with MiRA (\textcolor{blue}{Drevon et al., in prep.}).

\par MiRA requires a starting image and/or model and offers a few different types of regularisation functions. For the purpose of this study, the reconstruction setup consists of a Dirac distribution as the starting image, 500 iterations and two regularisation functions, quadratic compactness and edge-preserving smoothness. The reconstruction algorithm \texttt{SQUEEZE} \citep{squeeze} was also used as an additional check of the results; additional image reconstruction efforts are discussed in Appendices~\ref{squeezults} and \ref{rot_tests}.

\begin{figure*}[htb!]
    \centering
        \begin{minipage}[c]{0.98\textwidth}
           \centering
            \begin{subfigure}{.32\linewidth}
                \begin{overpic}[width=\linewidth]{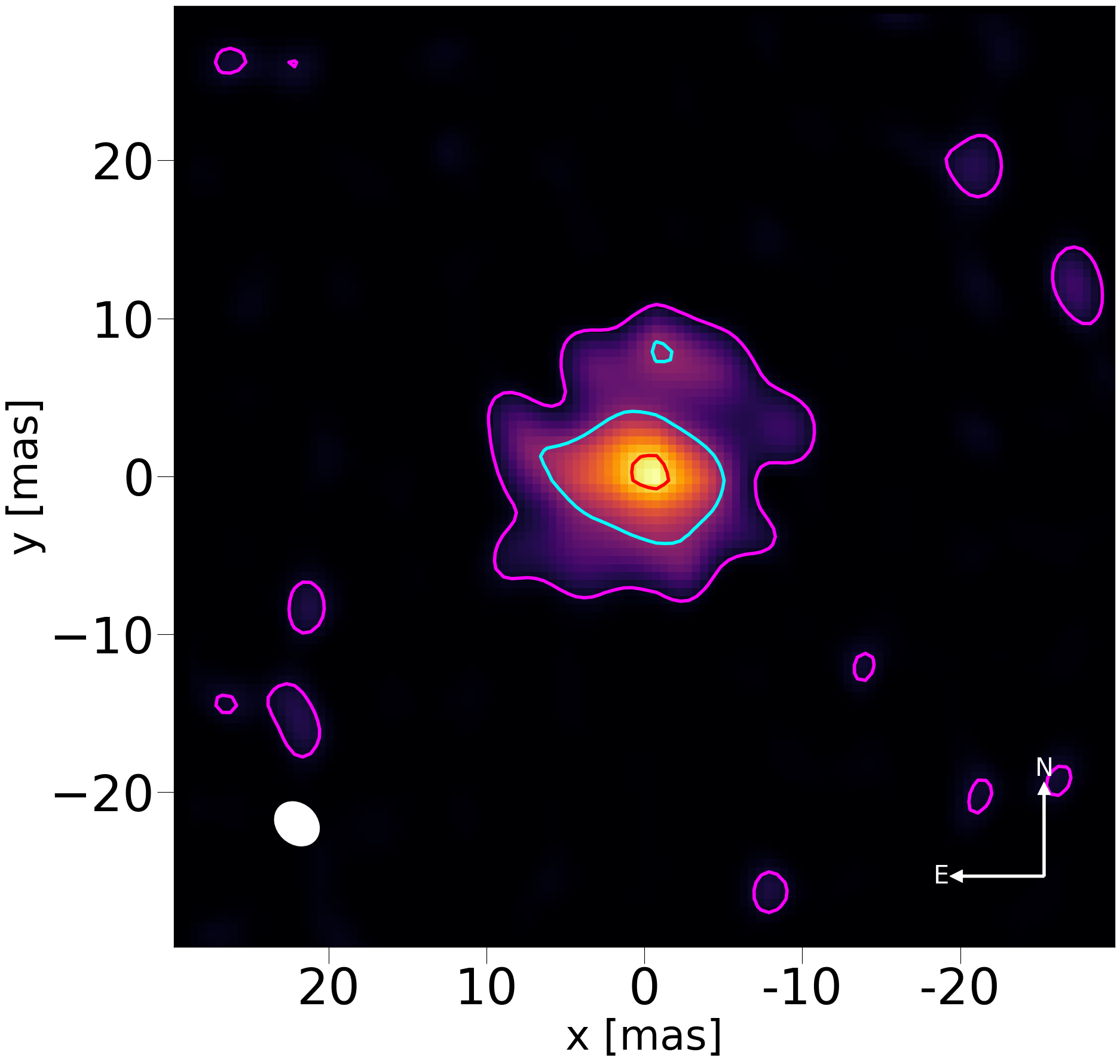}
                    \put(20,85){\color{white}\bfseries 3.1 $\mu$m}
                    \put(65,85){\color{white}\bfseries $\mathrm{C_2H_2+HCN}$}
                \end{overpic}
            \end{subfigure}
            \begin{subfigure}{.32\linewidth}
                \begin{overpic}[width=\linewidth]{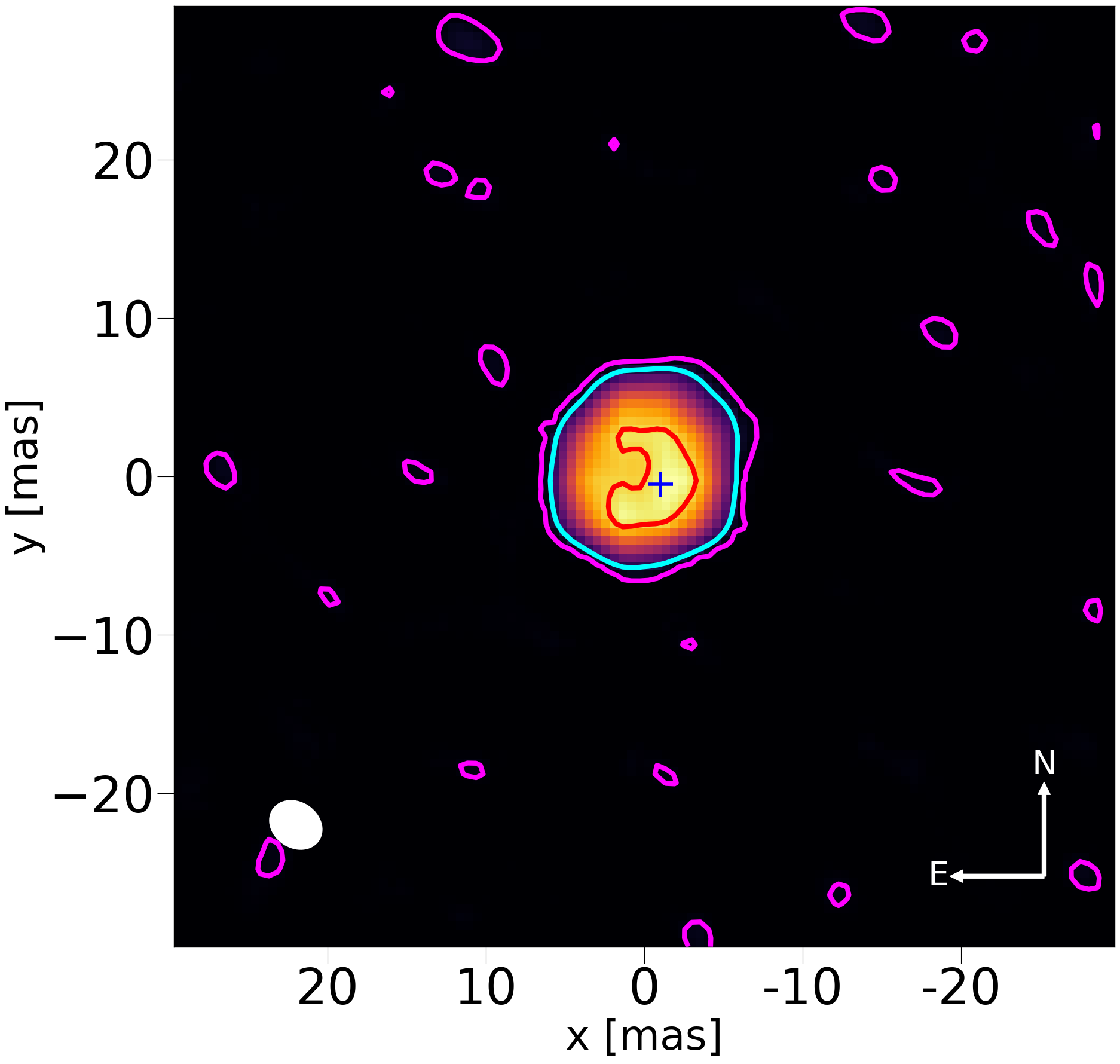}
                    \put(20,85){\color{white}\bfseries 3.5 $\mu$m}
                \end{overpic}
            \end{subfigure}
            \begin{subfigure}{.32\linewidth}
                \begin{overpic}[width=\linewidth]{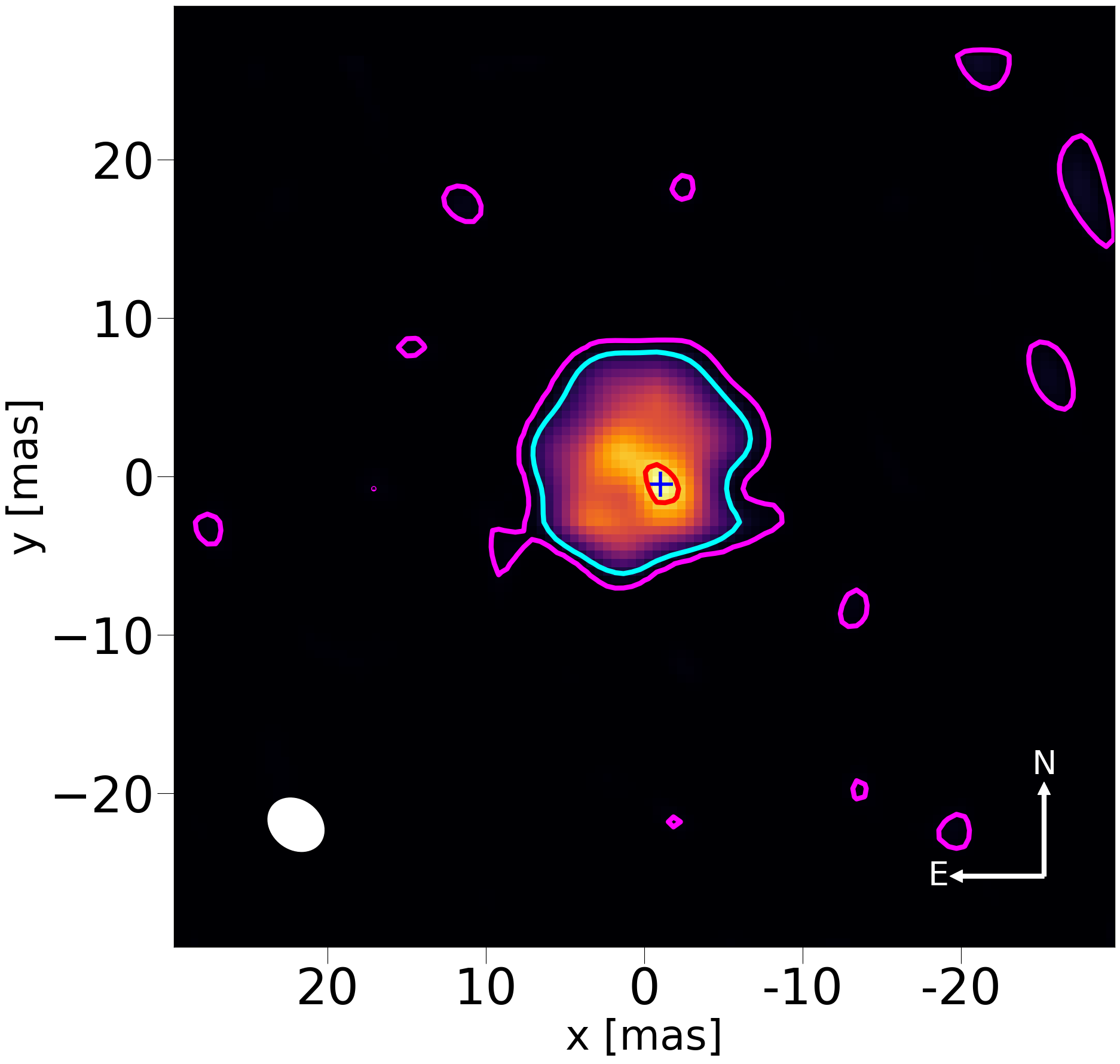}
                    \put(20,85){\color{white}\bfseries 3.8 $\mu$m}
                    \put(85,85){\color{white}\bfseries $\mathrm{C_2H_2}$}
                \end{overpic}
            \end{subfigure}           
            \par
            \begin{subfigure}{.32\linewidth}
                \begin{overpic}[width=\linewidth]{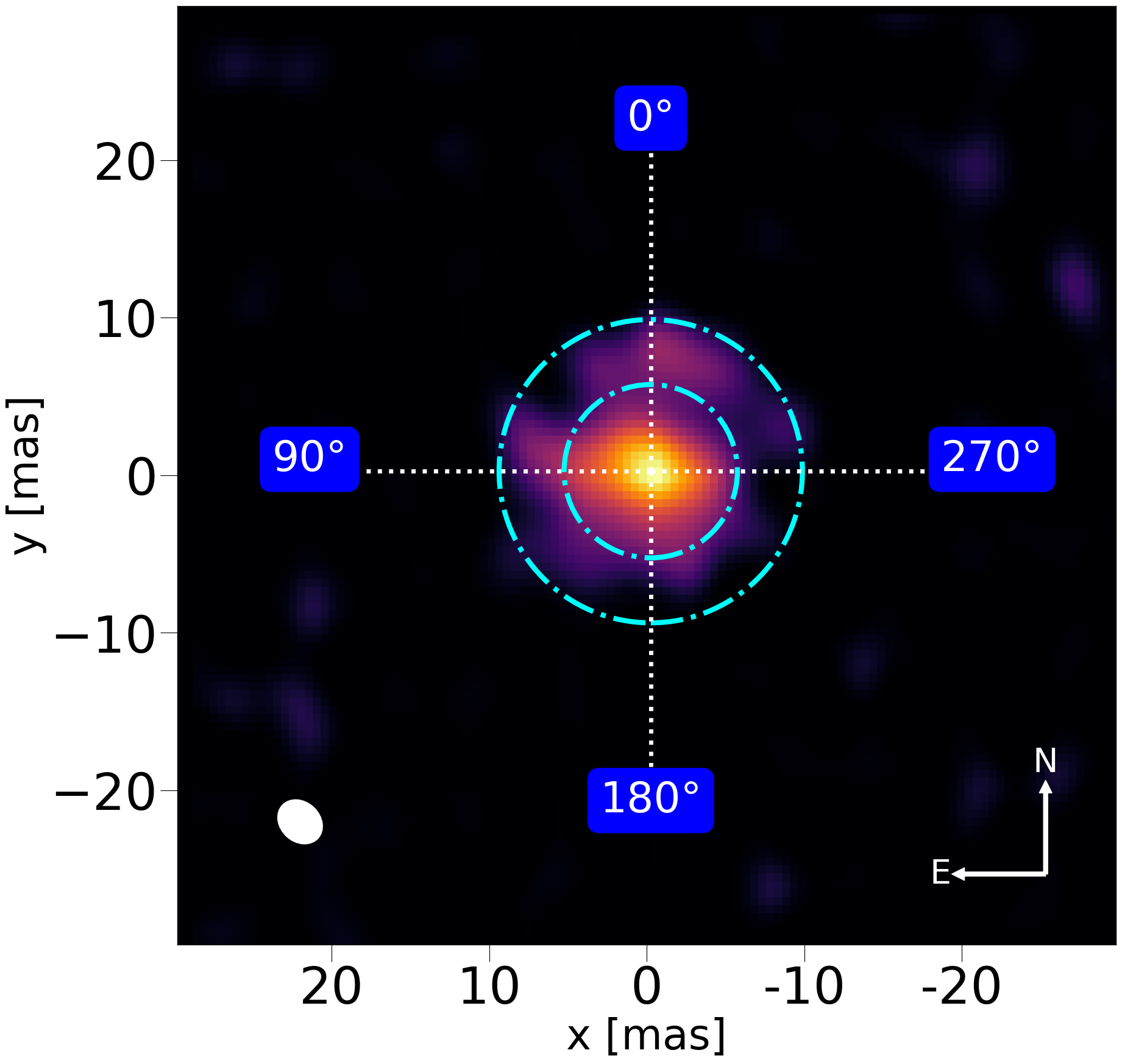}
                    \put(20,85){\color{white}\bfseries 3.1 $\mu$m}
                    \put(65,85){\color{white}\bfseries $\mathrm{C_2H_2+HCN}$}
                \end{overpic}
            \end{subfigure}
            \begin{subfigure}{.32\linewidth}
                \begin{overpic}[width=\linewidth]{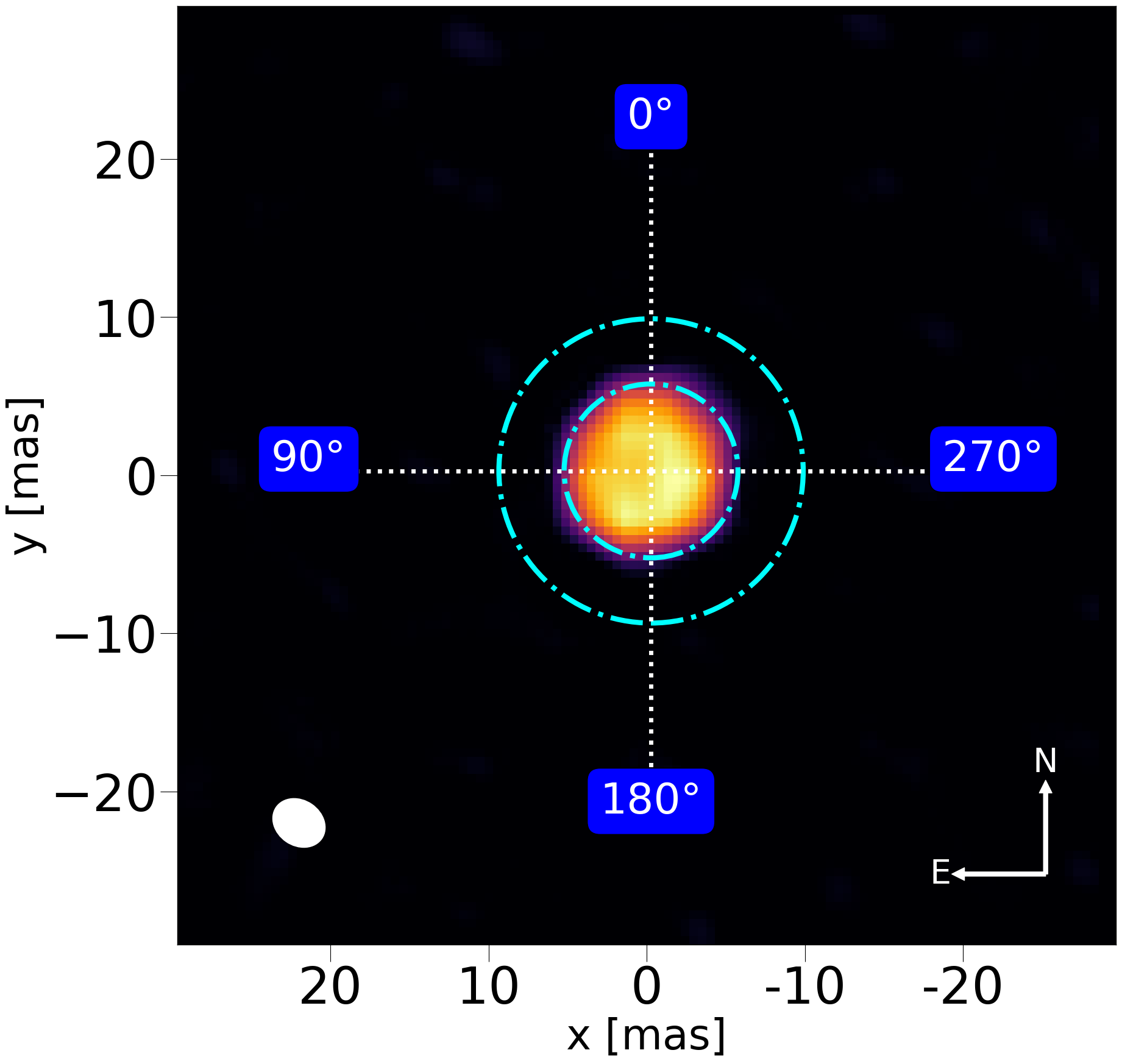}
                    \put(20,85){\color{white}\bfseries 3.5 $\mu$m}
                \end{overpic}
            \end{subfigure}        
            \begin{subfigure}{.32\linewidth}
                \begin{overpic}[width=\linewidth]{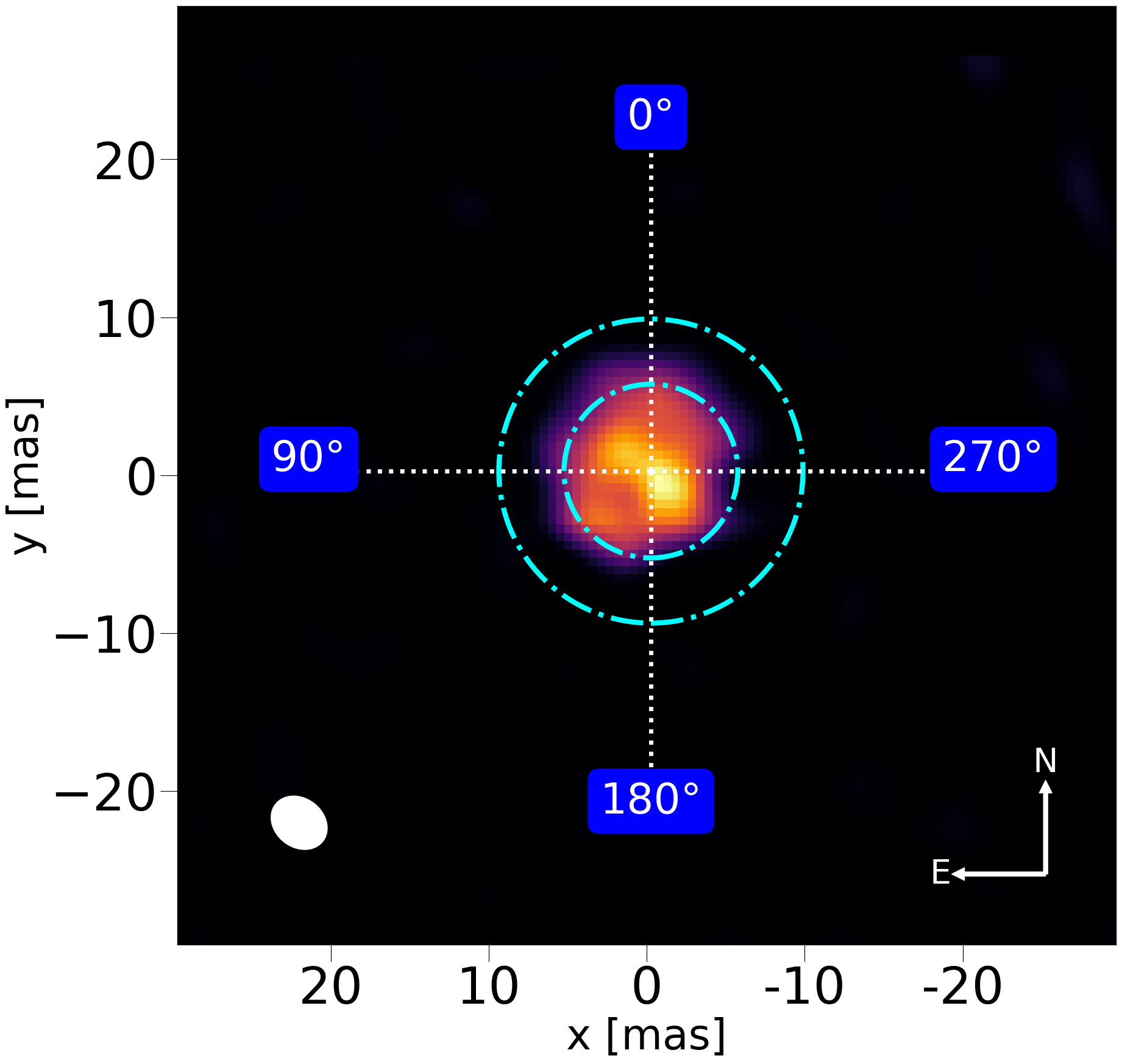}
                    \put(20,85){\color{white}\bfseries 3.8 $\mu$m}
                    \put(85,85){\color{white}\bfseries $\mathrm{C_2H_2}$}
                \end{overpic}
            \end{subfigure}        
            \par     
            \begin{subfigure}{.99\linewidth}
            \centering
            \hspace{0.85cm}
            \includegraphics[width=0.5\linewidth]{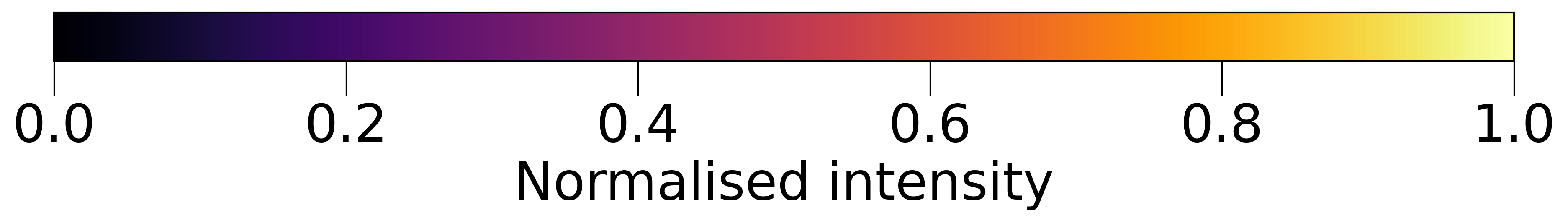}
            \end{subfigure}
            \par  
            \hspace{0.55cm}
              \begin{subfigure}{.3\linewidth}
                      \includegraphics[width = \linewidth]{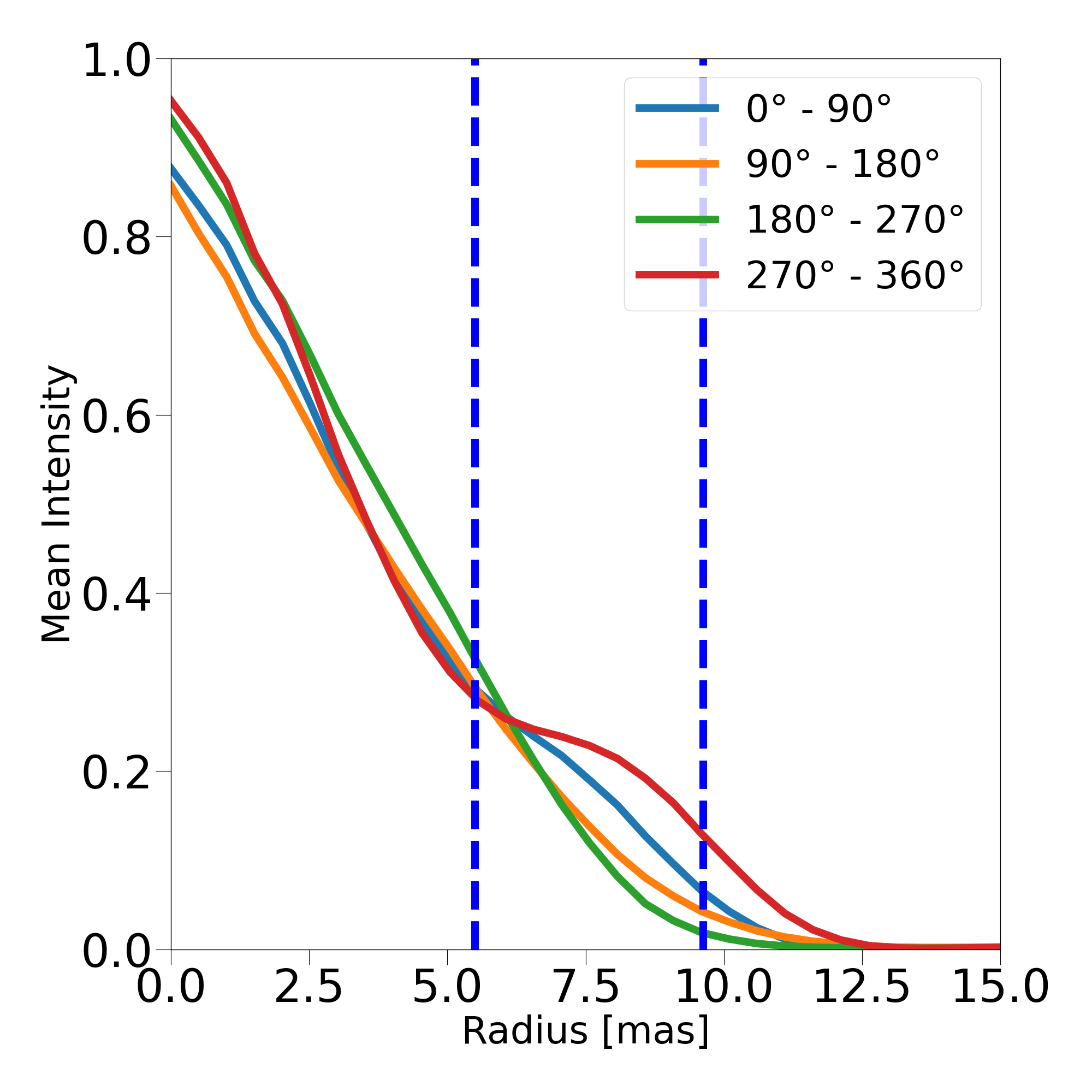}
                \caption{3.1 $\mu$m ($\mathrm{C_2H_2 + HCN}$)} \label{l_b1_4p_rp}
            \end{subfigure}     \hspace{0.25cm}       
        \begin{subfigure}{.3\linewidth}
                      \includegraphics[width = \linewidth]{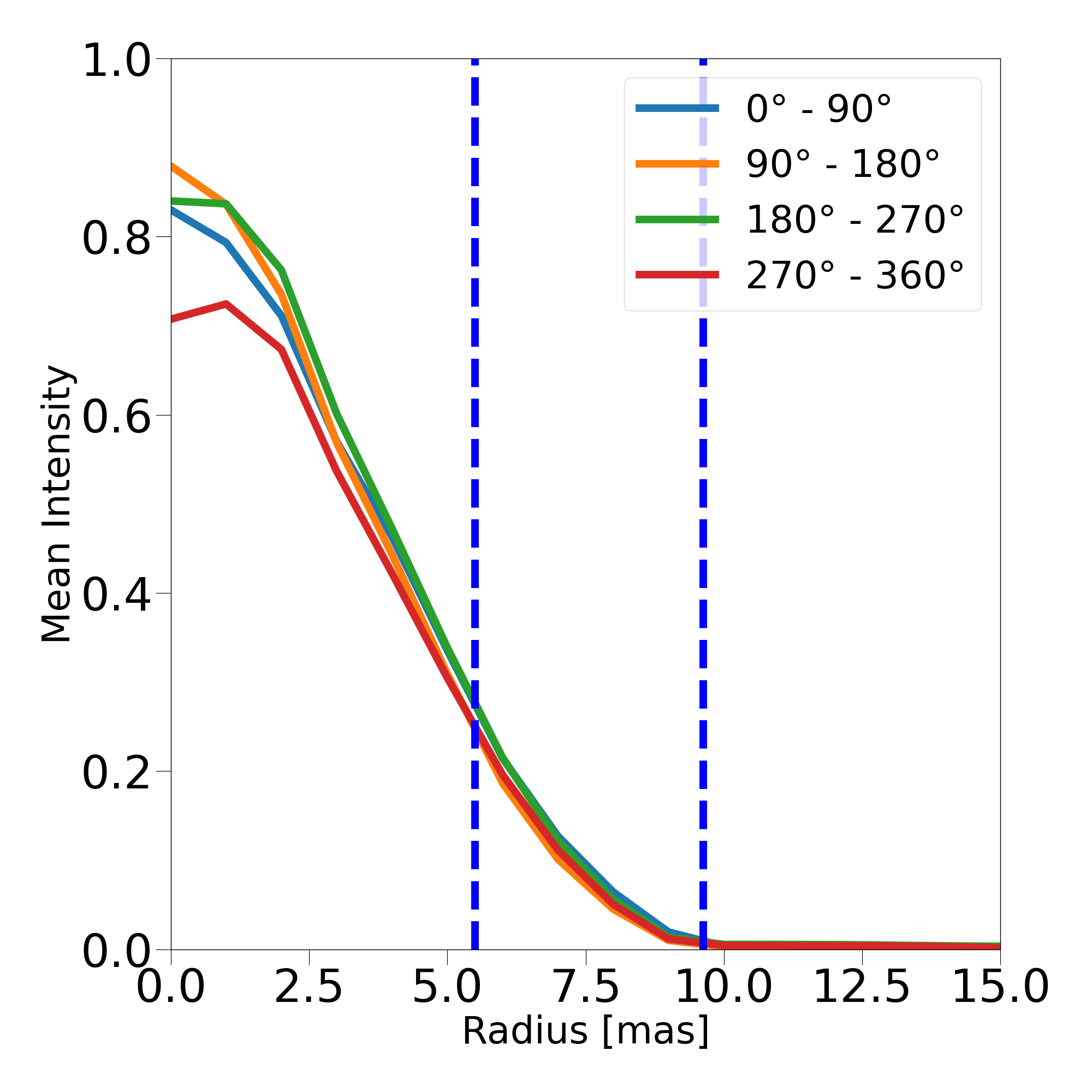}
                      \caption{3.5 $\mu$m (pseudo-continuum)} \label{l_b2_rp}
        \end{subfigure} \hspace{0.15cm}
            \begin{subfigure}{.3\linewidth}
                \includegraphics[width = \linewidth]{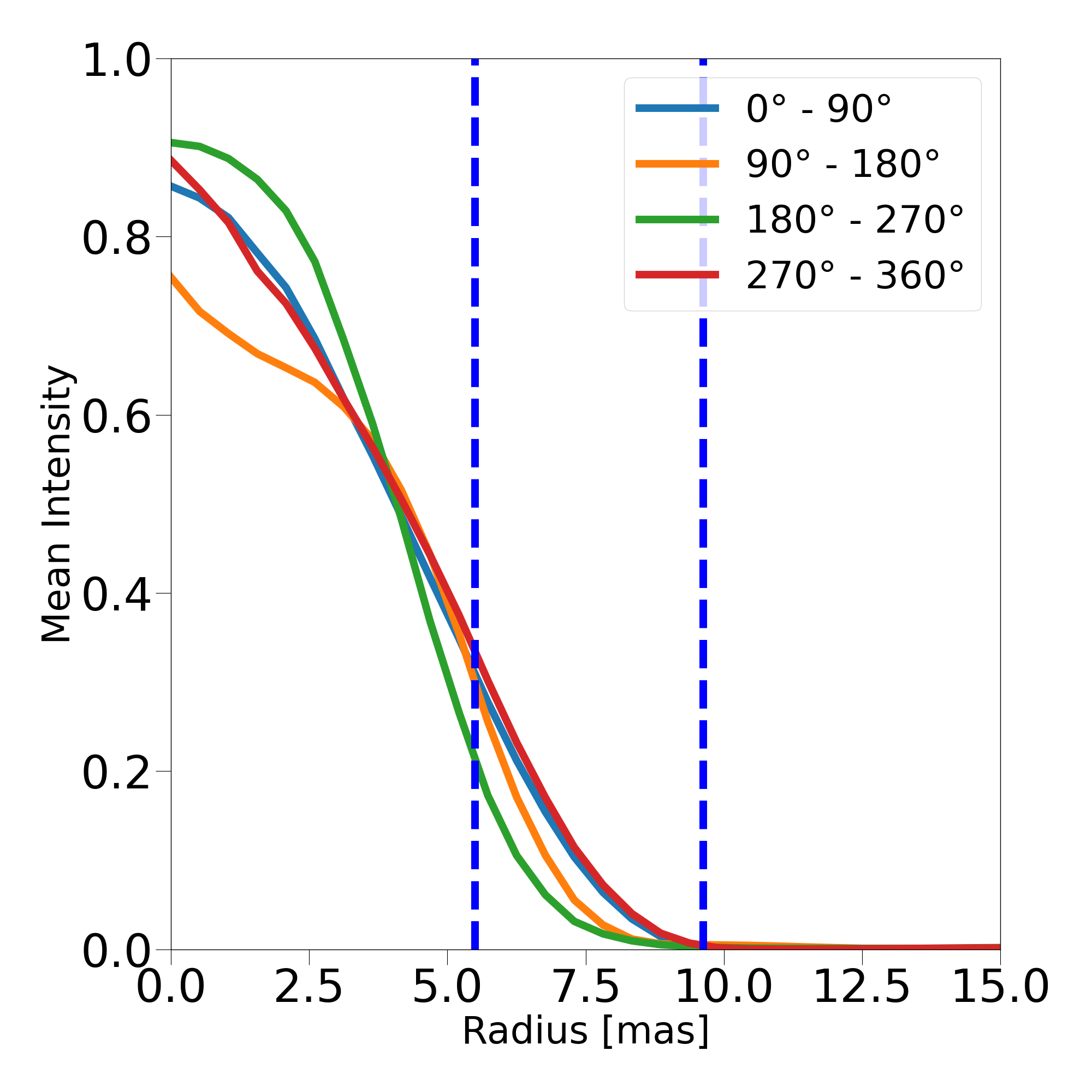}
                \caption{3.8 $\mu$m ($\mathrm{C_2H_2}$)} \label{l_b4_rp}
            \end{subfigure} \label{radprofs_plot_mira_l}
        \end{minipage}
        \caption{MiRA image reconstructions for the MATISSE L band for the 3.1 $\mu$m wavelength region (leftmost column), the  3.5 $\mu$m interval (middle), and the reconstructed data for the 3.8 $\mu$m region (right). In the top row, magenta (cyan) contours indicate where the signal is 5 times (30 times) above the estimated background noise, and the inner red contour encapsulates the regions where the S/N is in the top 10\% of all values across the image. The blue cross indicates the position of the emission peak seen at 3.8 $\mu$m. In the middle row, the innermost dashed cyan circle represents the photospheric diameter of \citet[see our Sect.~\ref{s2}]{paladini2017}, while the outer circle corresponds to the photospheric diameter scaled to 12 $\mu$m, based on the size ratio of the 2 and 12 $\mu$m mass-losing dynamic models in \citet{paladini2009}. Both are marked with a dashed blue line in the radial profile plots. The sectors marked by the dotted white lines show the regions used in the plotting of the radial profiles presented in the bottom row. The images were normalised with respect to the highest intensity pixel and re-centred by computing the local average intensity using a square kernel corresponding to 3 mas in radius. The white ellipse shows the estimated beam size and its inclination at each wavelength interval.} \label{mira_images_lband} 
\end{figure*}

\begin{figure*}[htb!]
    \centering
    \begin{minipage}[c]{0.98\textwidth}
        \centering 
        \begin{subfigure}{.32\linewidth}
            \begin{overpic}[width=\linewidth]{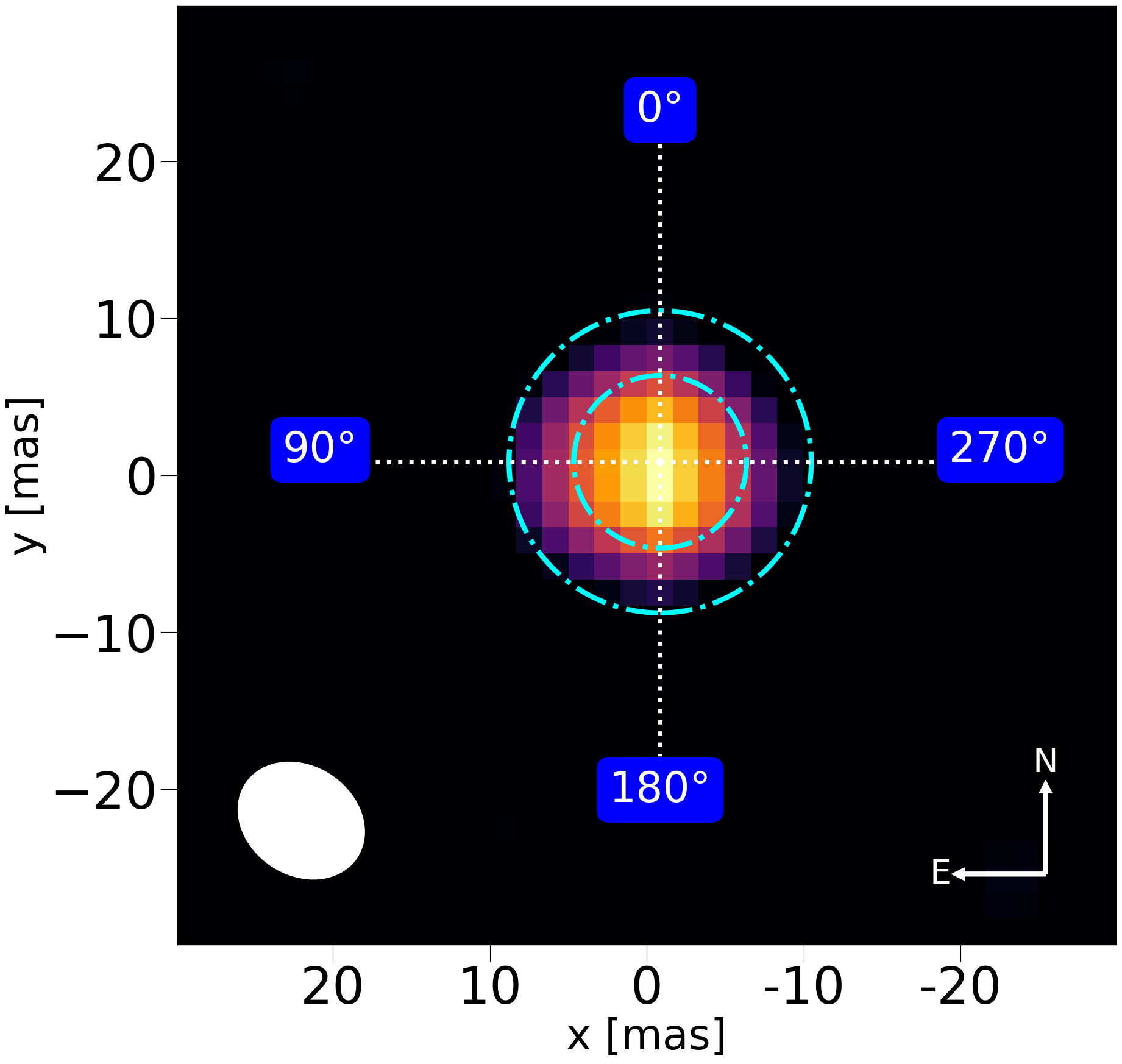}
                \put(20,85){\color{white}\bfseries 8.5 $\mu$m}
            \end{overpic}
        \end{subfigure}
        \begin{subfigure}{.32\linewidth}
            \begin{overpic}[width=\linewidth]{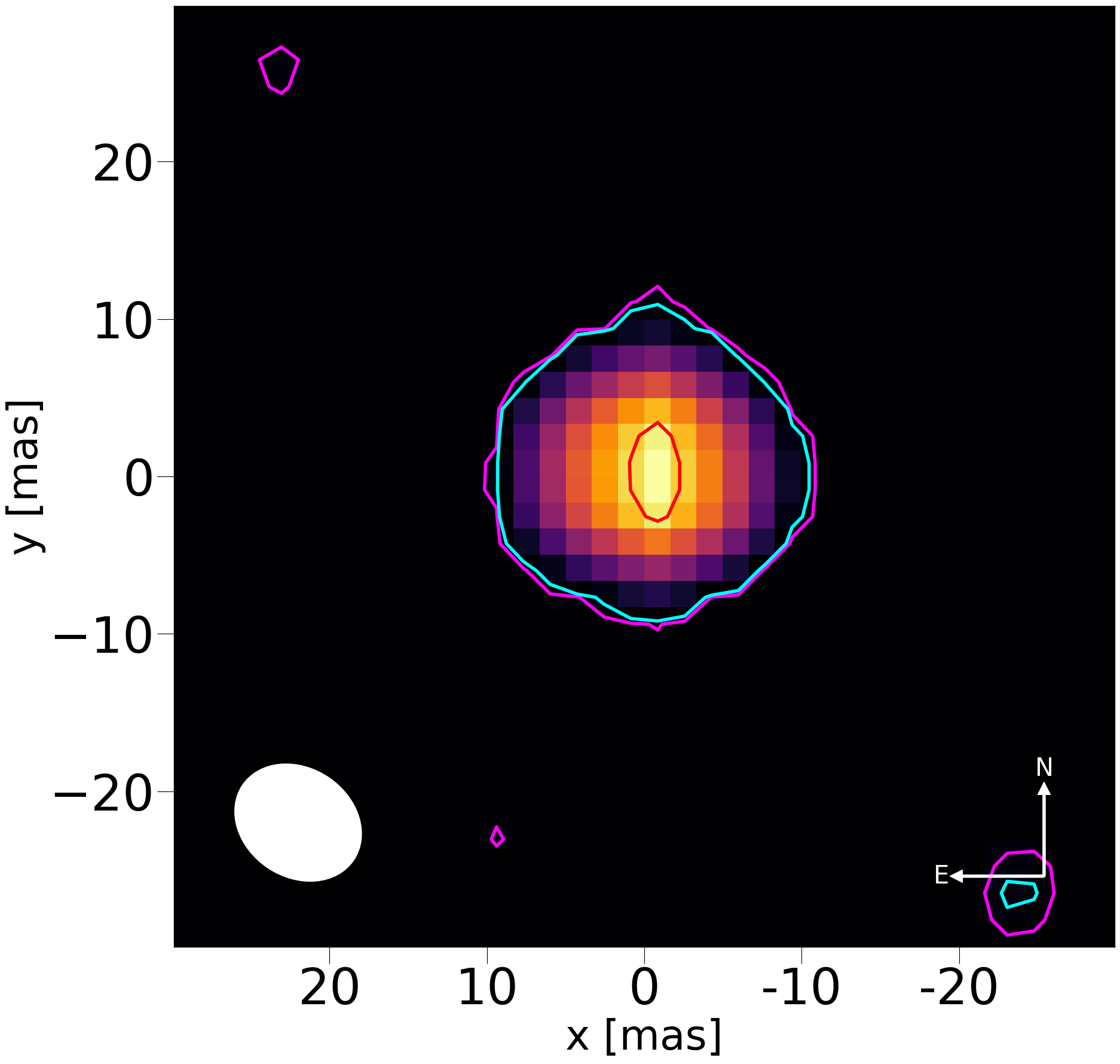}
                \put(20,85){\color{white}\bfseries 8.5 $\mu$m}
                \put(47.5,85){\color{white}\bfseries pseudo-continuum}
            \end{overpic}
        \end{subfigure}
        \begin{subfigure}{.32\linewidth}
            \begin{overpic}[width = \linewidth]{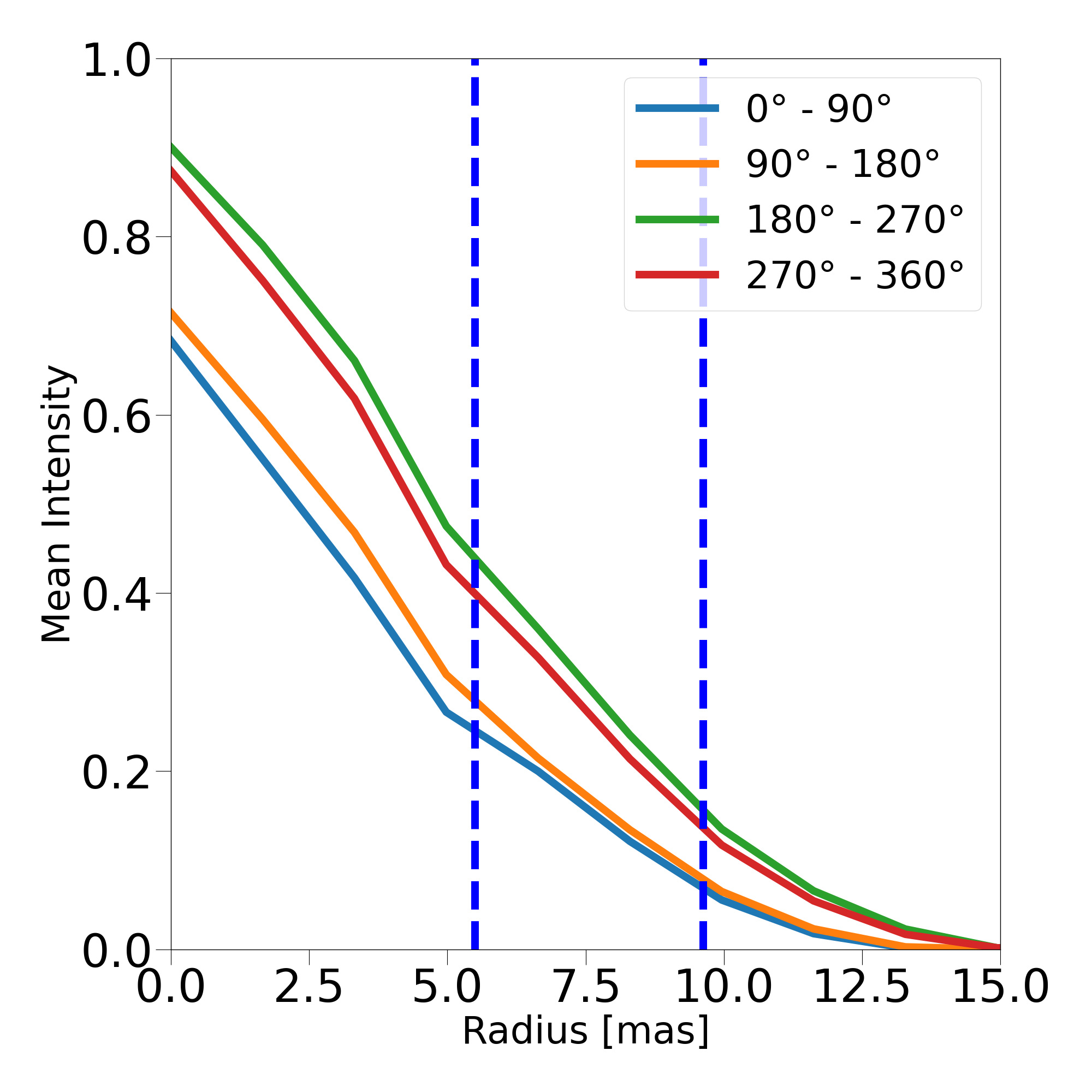}
                \put(20,87.5){\color{black}\bfseries 8.5 $\mu$m}
            \end{overpic}
        \end{subfigure}
        \par
        \begin{subfigure}{.32\linewidth}
            \begin{overpic}[width=\linewidth]
            {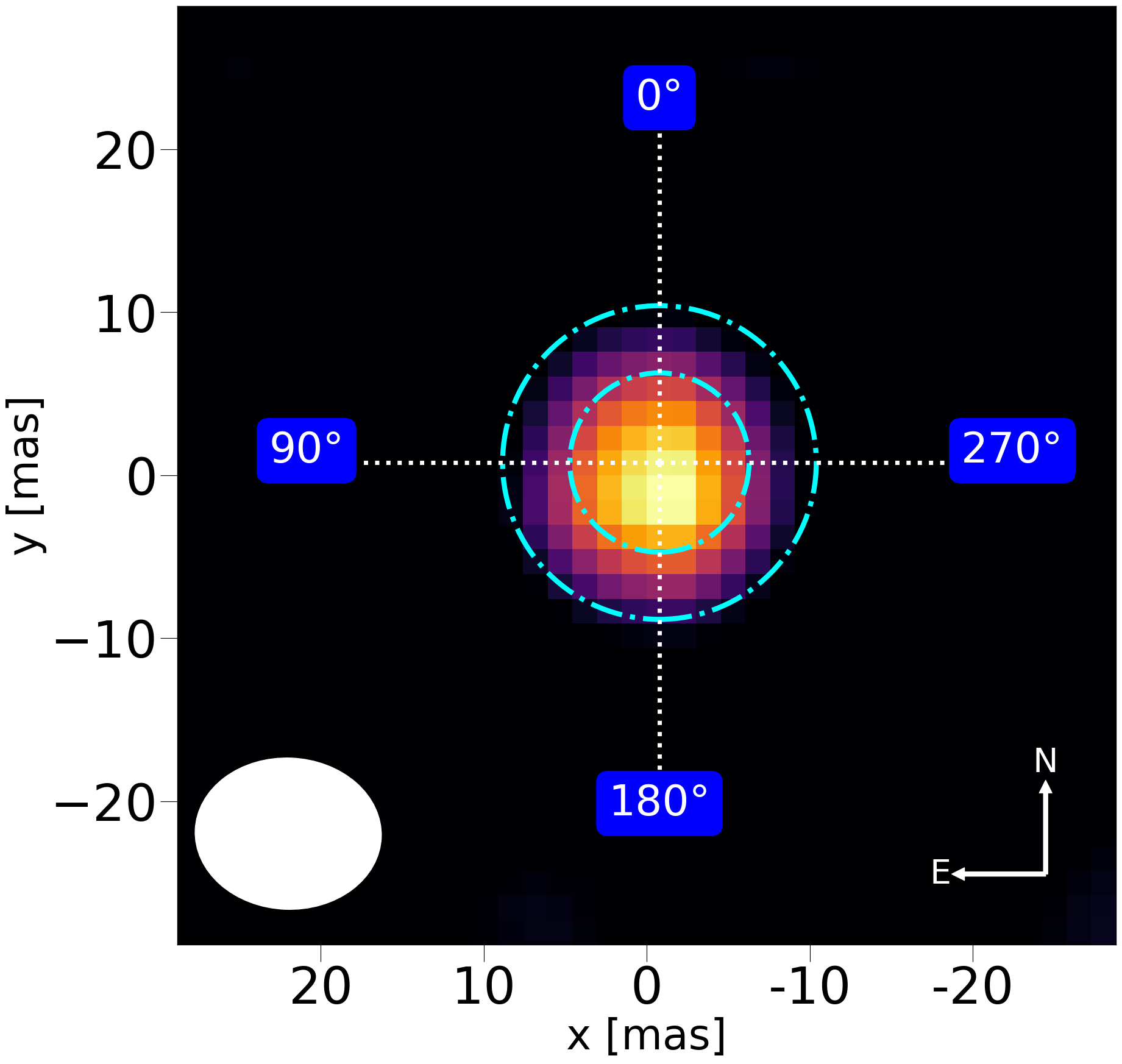}
                \put(20,85){\color{white}\bfseries 11.3 $\mu$m}
            \end{overpic}
        \end{subfigure}
        \begin{subfigure}{.32\linewidth}
            \begin{overpic}[width=\linewidth]{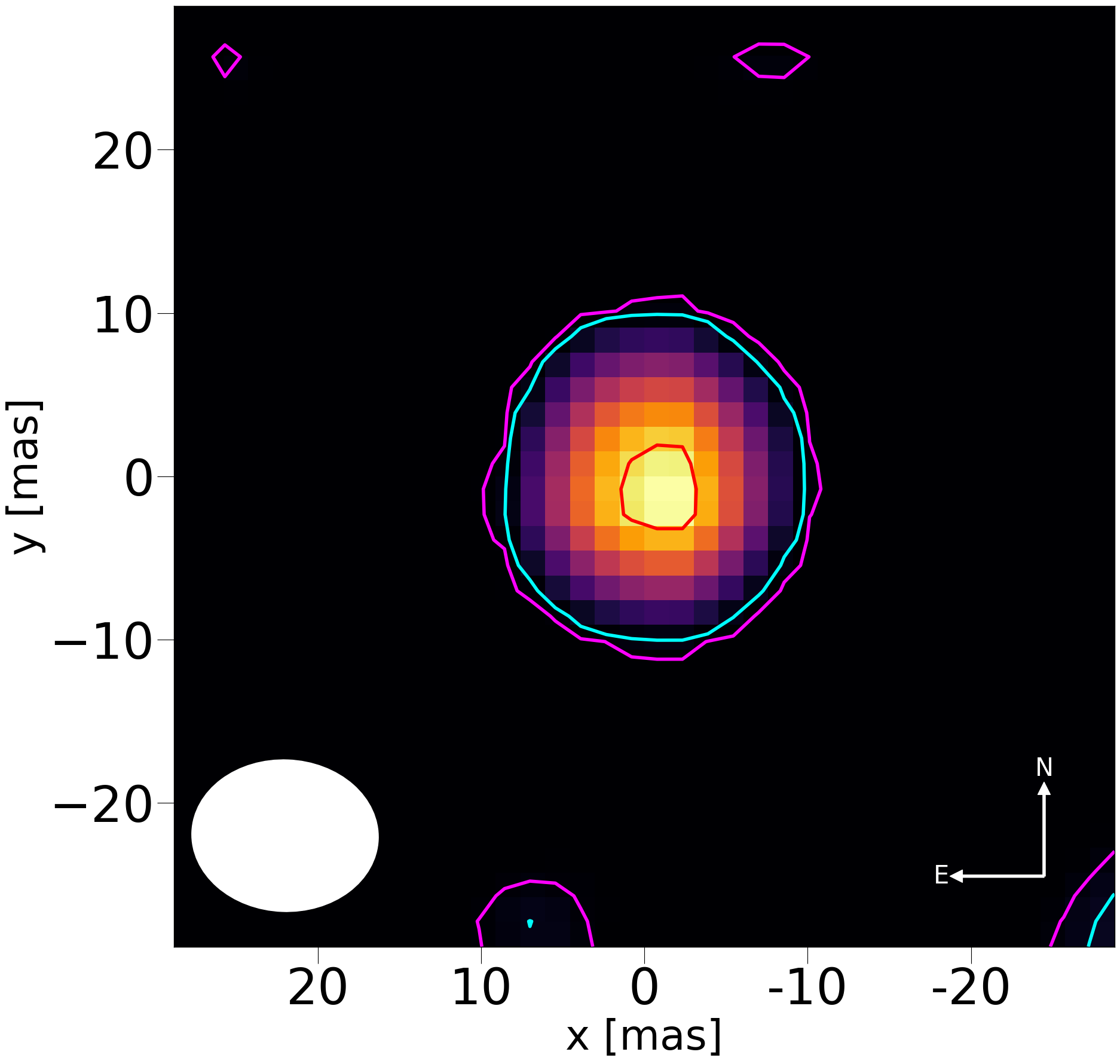}
                \put(20,85){\color{white}\bfseries 11.3 $\mu$m}
                \put(85,85){\color{white}\bfseries SiC}
            \end{overpic}
        \end{subfigure}
        \begin{subfigure}{.32\linewidth}
            \begin{overpic}[width = \linewidth]{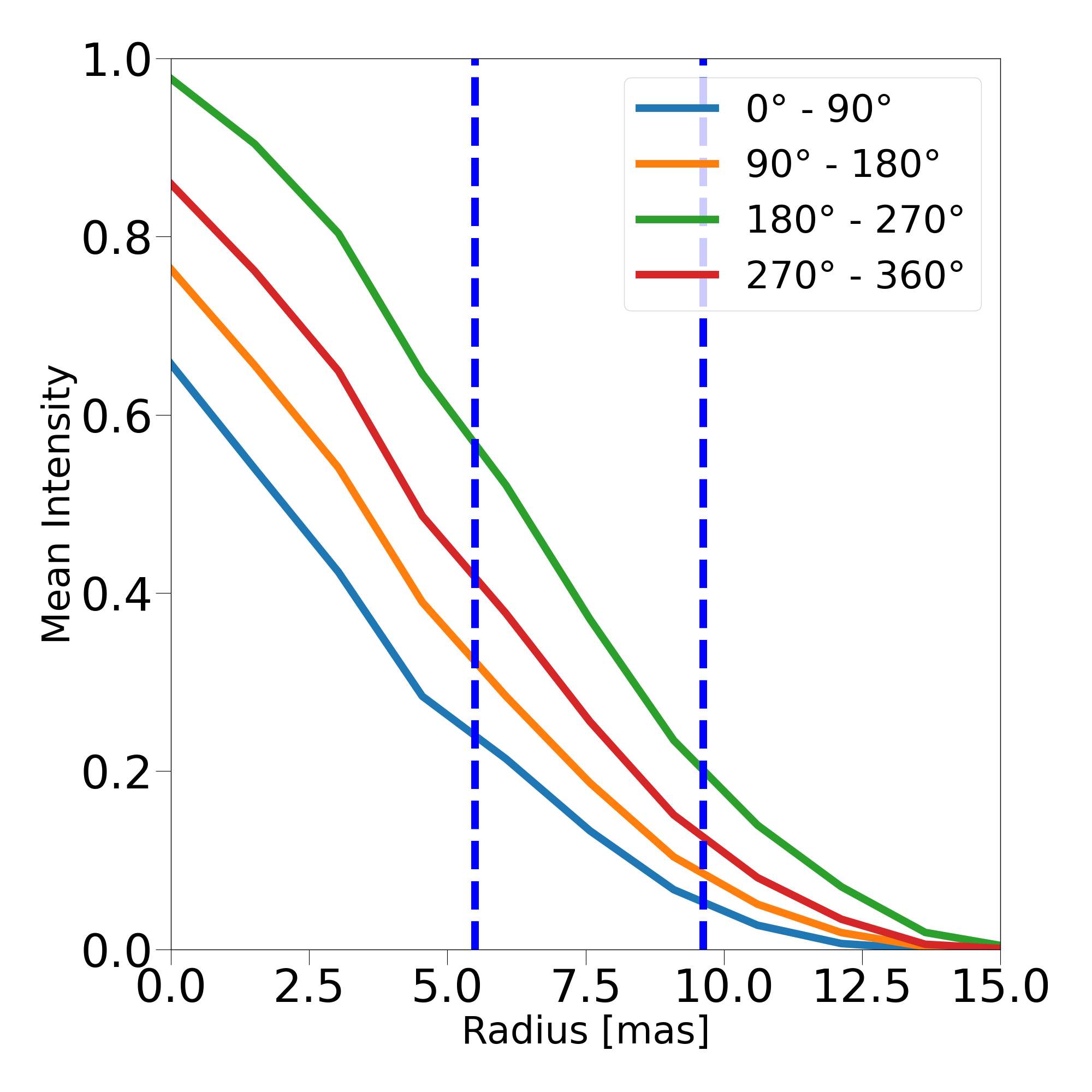}
                \put(20,87.5){\color{black}\bfseries 11.3 $\mu$m}
            \end{overpic}
        \end{subfigure} 
        \par
        \begin{subfigure}{.99\linewidth}
            \hspace{1.75cm}
            \includegraphics[width=0.5\linewidth]{colorbar.png}
           \end{subfigure}
    \end{minipage}
    \caption{MiRA image reconstructions for the MATISSE N band for the 8.5 $\mu$m (top row) and 11.3 $\mu$m  (bottom row) wavelength region. Magenta (cyan) contours indicate where the signal is 5 times (30 times) the estimated background noise, and the inner red contour encapsulates the regions where the S/N is in the top 10\% of all values across the image. The innermost dashed cyan circle represents the photospheric diameter from \citet[see our Sect.~\ref{s2}]{paladini2017}, while the outer circle corresponds to the photospheric diameter scaled to 12 $\mu$m, based on the size ratio of the 2 and 12 $\mu$m mass-losing dynamic models in \citet{paladini2009}. Both are marked with a dashed blue line in the radial profile plots. The sectors marked by the dotted white lines are the regions used in the plotting of the radial profiles presented in the bottom row. The images were normalised with respect to the highest intensity pixel and re-centred on the maximum local average intensity, computed using a square kernel with a radius of 3 mas. The white ellipse shows the estimated beam size and its inclination at each wavelength interval.} \label{mira_images_nband} 
\end{figure*}

\section{Results} \label{res}

\indent\indent The images obtained following the reconstruction process are shown in Figs.~\ref{mira_images_lband} and \ref{mira_images_nband}. The continuum contribution was not subtracted from any of the images, as there is no near-continuum bandpass available in our data. Radial profiles were also calculated for four 90$\degr$-wide sectors, as a means of quantifying any potential asymmetries in the intensity maps. The background noise in each image was taken as the standard deviation of the pixels located within an annulus spanning 70--90\% of the image's radius, measured from the centre to the outer edge. Signal-to-noise ratio (S/N) contours are overplotted in the images, starting with a 5$\sigma$ level (in magenta).

\par From the L band MiRA results several photospheric and circumstellar layers of X TrA can be seen. Starting with the pseudo-continuum image at 3.5 $\mu$m (middle column in Fig.~\ref{mira_images_lband}), a region with minimal contributions from $\mathrm{C_2H_2}$ and HCN, the star is at its smallest, with a radius of $\sim$2 au. The photospheric diameter computed by \citet{paladini2017} is marked by the inner dashed circle in Figs.~\ref{mira_images_lband} and \ref{mira_images_nband}, so clearly the angular extent of X TrA in the pseudo-continuum region coincides well with this. The other thing visible in this wavelength window is a horseshoe-like intensity peak in the central region, marked by a red contour in the top row of Fig.~\ref{mira_images_lband}.

\par Moving outwards, the $\mathrm{C_2H_2}$ layer at 3.8 $\mu$m (right column in Fig.~\ref{mira_images_lband}) starts to show a more complex structure, being patchier. Of note is a south-westerly intensity peak (marked by the blue cross in the images from the top row of Fig.~\ref{mira_images_lband}) that seems to be part of a SW-NE elongated feature and almost a complete lack of emission outside the extent of the photosphere. The peak here is cospatial with parts of the horseshoe shape seen at 3.5 $\mu$m, indicated by a higher intensity radial profile in the 180$\degr$--270$\degr$ sector in both cases, suggesting that the molecular layer probed here is inhomogeneous/clumpy. Farther away from the star, the $\mathrm{C_2H_2}$ layer is quite tenuous (has a low opacity), showing a very small amount of emission outside the photosphere in the 270$\degr$--360$\degr$ and 0$\degr$--90$\degr$ sectors (see the bottom-right radial profiles in Fig.~\ref{l_b4_rp}). One or two protrusions (towards the E--SE and one towards the west) can be seen in the S/N contours, but these cannot necessarily be considered real as they are smaller than the beam size.

\par The $\mathrm{C_2H_2\,+\,HCN}$ molecular band is probed at 3.1 $\mu$m and the circumstellar envelope reaches its greatest angular size here (left column in Fig.~\ref{mira_images_lband}), with a diameter of $\sim$20 mas. The intensity map shows a horizontally elongated feature inside the photospheric region (along the E-W direction). Numerous features/protrusions are visible in most directions beyond the photosphere, up to radii of 10 mas (e.g. radial profile in the 270$\degr$--360$\degr$ sector), and are characterised by a smooth, diffuse morphology. In combination with the high intensity of the central region, it is likely that this layer is not very optically thick due to a low column density. 

\par The MIDI and MATISSE observations sample similar epochs, as seen in Fig.~\ref{light_curve}. A comparison between the imaging results presented here and the MIDI studies of \citet{rauxtra} and \citet{paladini2017} finds the angular size estimates to be in good agreement. The MATISSE images at 3.5 $\mu$m give a diameter of 12 mas, while \citet{paladini2017} estimate it to be 11 mas \citep[based on the V-K relation of][]{vb99} and \citet{rauxtra} calculate it to be 9.8 mas \citep[based on the V-K relation of][]{vb2013}. The reconstruction at 8.5 $\mu$m gives a diameter of about 19 mas, while \citet{rauxtra} compute a diameter of $21.9\pm2.5$ mas at 8 $\mu$m. The latter study mentions not finding any well-fitting models for the SiC signature seen in the visibility data, while \citet{paladini2017} state that the visibility data indicate SiC detection around 4 stellar radii. None of the images in the L and N bands show any significant features indicative of companion presence such as disks, spirals, arcs, and tails \citep[with typical sizes of the order of tens to a few hundred au and even up to a few thousand au;][]{mayer2015, kervella2016, decin2020}.

\section{Discussion} \label{dis}

\subsection{Morphology} \label{dismorph}

\indent\indent The reconstructed L band images reveal an asymmetric structure in the molecular layers, with elongations and protrusions especially visible in the $\mathrm{C_2H_2\,+\,HCN}$ layer. This result is not unexpected, as \citet{hulberg2025} and \citet{drevon2022} attribute the nature of the spectrum and morphology of the images to the molecular influence of $\mathrm{C_2H_2}$ and $\mathrm{C_2H_2\,+\,HCN}$, the latter study exhibiting similar morphological features in the image of \object{R Scl} at 3.1 $\mu$m (i.e. elongated features and asymmetric protrusions).

\par It has previously been found that the intensity of the spectral features of molecules such as $\mathrm{C_2H_2}$ in the L and N bands are correlated with stellar pulsation phase \citep{loidl2000, ohnaka2007}. This would translate to changes in size when looking at the corresponding specific wavelengths. Other more recent studies of carbon stars \citep[][all done with MATISSE]{drevon2022, planquart, hulberg2025} find similar angular sizes and a similar wavelength dependence regarding the scale of the emission as presented here, namely that the extent of the source is at its largest in the 3.1 $\mu$m region and in the N band and at its smallest in the L band pseudo-continuum at 3.5 $\mu$m. Table~\ref{sizerat} shows a comparison between stellar parameters for three different carbon stars that have been imaged with MATISSE and the ratio of their size in the 3.1 micron $\mathrm{C_2H_2 + HCN}$ feature and the L band pseudo-continuum at 3.5 micron ($\mathrm{\theta_{3.1/3.5}}$). When observed at different pulsation phases, the ratios follow the expected behaviour of being larger close to maximum light \citep{woodruff2008}. Since X TrA's irregular aspect makes it difficult to determine the pulsation phase, this comparison between the $\mathrm{\theta_{3.1/3.5}}$ values suggests X TrA could have been close to maximum light at the time of MATISSE observations.

\subsection{Clumpiness}

\indent\indent Closer inspection of the 3.1 and 3.8 $\mu$m layers indicates that the molecular distribution is inhomogeneous, pointing to clumpy structures within the circumstellar environment. The 3D models of dust-driven winds in AGB stars by \citet{freytag2023} support the idea that an inhomogeneous gas distribution due to convection and pulsation results in anisotropic and clumpy dust formation. Although these models are based on an oxygen-rich chemistry, the mechanisms of structure formation in the circumstellar environment are independent of the C/O ratio and also apply to carbon stars. Similarly, \citet{velilla2023} detect HCN emission in the close vicinity of IRC+10$\degr$216, with a morphology reminiscent of larger-scale structures previously identified in its extended circumstellar environment, reaching the same conclusion that localised conditions can drive clumpy outflows. All this suggests that the $\mathrm{C_2H_2 + HCN}$ feature might typically appear as a clumpy layer around the target object, and the resulting dust structures may exhibit similar morphologies.

\begin{table*}[htbp]
\centering
\caption{Carbon stars recently imaged with MATISSE.} \label{sizerat}
\begin{tabular}{c*{6}c}                 \hline \rule[+0.3ex]{0pt}{2.5ex}
Object  & $\mathrm{T_{eff}}$ & L & $\mathrm{\dot{M}}$ & C/O & $\phi$ & $\mathrm{\theta_{3.1/3.5}}$        \\ 
& [K] & $\mathrm{[L_{\sun}]}$ & $\mathrm{[M_{\sun}yr^{-1}]}$ & & & \\ \hline \rule[+0.1ex]{0pt}{2.5ex}
R Scl   & 2700 \tablefootmark{\,a}      & 8000 \tablefootmark{\,a}      & $1.6 \times 10^{-6}$ \tablefootmark{\,b}                        
& 1.4 \tablefootmark{\,c} & 0.95 \tablefootmark{\,a} & 2 \tablefootmark{\,a}    \\
X TrA   & 2700 \tablefootmark{\,d}      & 8600 \tablefootmark{\,e}      & 1.3 -- 1.8 $\times 10^{-7}$ \tablefootmark{\,f}   & 1.35 \tablefootmark{\,g}    & -- & 1.8        \\
V Hya   & 2650 \tablefootmark{\,h}      & 18\,000 \tablefootmark{\,h}   & $\sim10^{-5}$ \tablefootmark{\,i} & 1.4 \tablefootmark{\,h}     & 0.10 -- 0.27 \tablefootmark{\,h} & 1.4 \tablefootmark{\,h} \\ \hline
\end{tabular}
\tablefoot{Some stellar parameters of C-rich stars imaged with MATISSE: effective temperature, luminosity, mass-loss rate, carbon-to-oxygen ratio and pulsation phase of MATISSE observation epoch. The final column lists the overall angular size ratio of the 3.1 and 3.5 $\mu$m reconstructed images of each star ($\mathrm{C_2H_2 + HCN}$ feature versus pseudo-continuum).\\
\textbf{References.} \tablefoottext{a}{\citet{drevon2022}}
\tablefoottext{b}{\citet{debeck2010}}
\tablefoottext{c}{\citet{hron1998}}
\tablefoottext{d}{\citet{kipper2004}}
\tablefoottext{e}{\citet{rauxtra}}
\tablefoottext{f}{\citet{scholof2001, bergchev2005}}
\tablefoottext{g}{\citet{rauxtra}}
\tablefoottext{h}{\citet{planquart}}
\tablefoottext{i}{\citet{knapp1997}}
}
\end{table*}

\par The stark difference between the $\mathrm{C_2H_2\,+\,HCN}$ and $\mathrm{C_2H_2}$ emission beyond the extent of the photosphere is also of interest. The difference could arise because the $\mathrm{C_2H_2}$ band at 3.1 $\mu$m is significantly stronger than that at 3.8 $\mu$m, but HCN could also contribute. The latter molecule is believed to form in the inner wind of AGB stars \citep[within a few stellar radii,][]{cherchneff2012}, and the observed emission difference outside the photosphere may provide some initial constraints on the HCN column density in the close environment of X TrA. Detailed chemical modelling of the wind would therefore be required to quantify the significance of the HCN contribution.

\subsection{Dust layer}

\indent\indent Due to the spatial resolution of the instrument in the N band (8-13 mas), data only probe the first lobe of the visibility curve without entirely resolving the central source (see Fig.~\ref{mira_inf_vs_input_nband}). Nevertheless, even here the intensity map takes a slightly elliptical shape with an elongated central peak (Fig.~\ref{mira_images_nband}), although in the north-south direction, and the SiC band at 11.3 $\mu$m (bottom row in Fig.~\ref{mira_images_nband}) shows a higher intensity towards the south-west (more clearly shown by the 180$\degr$--270$\degr$ sector in the rightmost plot on the bottom row of Fig.~\ref{mira_images_nband}). The reconstructed source has a diameter of 19 mas (3.5 au) in the N band, similar to the $\mathrm{C_2H_2\,+\,HCN}$ layer, increasing ever so slightly at 11.3 $\mu$m and corresponding to the inferred diameter at 12 $\mu$m from the relation of \citet[represented by the outer dashed cyan line in the images]{paladini2009}. No significant emission is seen outside the circumstellar dust disk.

\par The MATISSE images at 11.3 $\mu$m do not show any significant features at this range, just as the ones at 8.5 $\mu$m. The dust thermal emission at 8.5 $\mu$m is mostly caused by amorphous carbon and, just as SiC, is expected to form close to the star. It is therefore likely that the N band images show no significant difference due to a similar spatial distribution of the two species.

\section{Conclusions} \label{conc}

\indent\indent In this paper we have presented the first MATISSE images of the irregular variable X TrA. Image reconstruction with MiRA reveals an angular diameter range of 10 to 20 mas (corresponding to a radius of 2 to 3.5 au), with the smallest size in the 3.5 $\mu$m pseudo-continuum band and the largest in the $\mathrm{C_2H_2\,+\,HCN}$ molecular and dust emission bands, at 3.1 and 11.3 $\mu$m, respectively. The reconstructed images confirm previous photospheric diameter estimates of \citet{paladini2017} and \citet{rauxtra} and show some interesting features, such as elongations and some asymmetric protrusions. They are likely caused by convection and pulsation effects, which can result in inhomogeneous gas distributions and subsequently anisotropic and/or clumpy dust formation \citep{freytag2023}. No significant signs of a companion's effects are detected in the image data, down to the 5$\sigma$ level. The angular sizes estimated here, and their wavelength dependence, follow similar values and trends as presented in other recent studies of carbon stars \citep{drevon2022, planquart, hulberg2025}. The highly asymmetric appearance of the $\mathrm{C_2H_2\,+\,HCN}$ layer is not the first case of a carbon-rich star showing a complex morphology in the close stellar environment \citep{drevon2022, velilla2023}, indicating that the $\mathrm{C_2H_2\,+\,HCN}$ molecular spatial distribution is clumpy in nature. The angular resolution of the N band data was unsatisfactory, likely due to the calibrators being too faint in this wavelength range and the baselines too short. Therefore, the images do not show any distinct features within the circumstellar dust disk, suggesting that any SiC present is likely co-spatial with amorphous carbon and, therefore, making the two difficult to distinguish. Additional image data, combined with chemical modelling efforts, could help disentangle the role of HCN in the apparent difference between the $\mathrm{C_2H_2\,+\,HCN}$ and $\mathrm{C_2H_2}$ layers. In the N band, using the 202m baseline offered by the extended Auxiliary Telescope configuration would theoretically increase the angular resolution by $\sim$35\%, potentially enabling a better constraint to be placed on the size of any circumstellar features.

\begin{acknowledgements}
The research leading to these results has received funding from the European Union’s Horizon 2020 research and innovation programme under Grant Agreement 101004719 (ORP). V.R. acknowledges the support from the ESO Early-Career Scientific Visitor Programme. K.O. acknowledges the support of the Agencia Nacional de Investigación Científica y Desarrollo (ANID) through the FONDECYT Regular grant 1240301. J.S.-B. acknowledges the support received by the UNAM DGAPA-PAPIIT project AG101025. The work of J.A.H. and C.S.C. is part of the I+D+i project PID2023-146056NB-C22 funded by Spanish MCIN/AEI/10.13039/501100011033. L.V-P. acknowledges Spanish Ministerio de Ciencia, Innovación y Universidades for funding support through project PID2020-117034RJ-I00, PID2023-147545NB-I00 and grant RYC2023-045648-I. S.H. acknowledges funding from the European Research Council (ERC) under the European Union's Horizon 2020 research and innovation programme (Grant agreement No. 883867, project EXWINGS) and the Swedish Research Council (Vetenskapsrådet, grant number 2019-04059). J.P.F. acknowledges funding support from Spanish Ministerio de Ciencia, Innovación, y Universidades through grants PID2023-147545NB-I00, PID2023-146056NB-C21, and PID2023-146056NB-C22.
\end{acknowledgements}

\bibliographystyle{aa_url}
\bibliography{refs}

\begin{appendix}

\section{Observational details} \label{appA}
\begin{table}[ht] 
\caption{Calibrator details.} \label{calibdetails}
\centering
\resizebox{\linewidth}{!}{
\begin{tabular}{l*{6}l}                 \hline\hline
Name 
& Spectral type         
& $\mathrm{\theta_{UD, L}}$
& $\mathrm{\theta_{UD, N}}$
& $\mathrm{F_L}$
& $\mathrm{F_N}$ \\ 
& & [mas] & [mas] & [Jy] & [Jy] \\ \hline\hline
$\beta$ TrA (HD 141891) & F1V & 1.446 & 1.453 & 90  & 5  \\
$\epsilon$ TrA (HD 138538) & K1/2III & 2.472 & 2.481 & 77  & 9  \\
$\eta$ Ara (HD 151249) & K5III & 5.623 & 5.669 & 353 & 77 \\
$\zeta$ Ara (HD 152786) & K3III & 7.146 & 7.205 & 522 & 40 \\ \hline
\end{tabular}
}
\tablefoot{$\mathrm{\theta_{UD, L}}$ and $\mathrm{\theta_{UD, N}}$ are the uniform disk diameters in the L and N band, respectively, adopted from \citet{bourges}. $\mathrm{F_L}$ and $\mathrm{F_N}$ are the L and  N band fluxes of the stars, taken from \citet{cruzalebes}}
\end{table}

\begin{table}[ht]
\caption{Observation log for target \object{X TrA.}} \label{obslog}
\centering
\resizebox{\linewidth}{!}{
\begin{tabular}{l*{7}l}                 \hline\hline
Date & Configuration & Time & Object & $\tau_{0}$ & Seeing & FT \\ 
     &               &      &        & [ms]       & ["]    &    \\ \hline\hline
2 Apr 2022  & A0-G1-J2-J3 & 4:20:05 & $\epsilon$ TrA & 3.88  & 0.97 & 1 \\
            &             & 4:34:15 & X TrA          & 3.89  & 0.86 & 0.9908 \\
            &             & 4:56:25 & $\eta$ Ara     & 3.96  & 0.86 & 1 \\
3 Apr 2022  & A0-G1-J2-J3 & 7:39:42 & $\epsilon$ TrA & 4.78  & 0.71 & 1 \\
            &             & 7:53:23 & X TrA          & 4.41  & 0.74 & 0.8899 \\
            &             & 8:19:07 & $\eta$ Ara     & 5.19  & 0.62 & 0.9909 \\
5 Apr 2022  & A0-G1-J2-J3 & 6:14:54 & $\beta$ TrA    & 4.77  & 0.66 & 1 \\
            &             & 6:31:07 & X TrA          & 4.98  & 0.63 & 0.9083 \\
            &             & 6:53:11 & $\zeta$ Ara    & 5.35  & 0.52 & 0.9847 \\
7 Apr 2022  & A0-G1-J2-J3 & 4:49:29 & $\epsilon$ TrA & 4.53  & 0.87 & 1 \\
            &             & 5:00:47 & X TrA          & 4.22  & 0.93 & 0.8909 \\
            &             & 5:24:48 & $\eta$ Ara     & 5.41  & 0.65 & 1 \\
            & A0-G1-J2-J3 & 8:34:25 & $\beta$ TrA    & 3.85  & 0.82 & 1 \\
            &             & 8:46:51 & X TrA          & 4.41  & 0.67 & 0.2752 \\
            &             & 9:10:37 & $\zeta$ Ara    & 4.16  & 0.71 & 1 \\
8 Apr 2022  & A0-G1-J2-J3 & 8:16:31 & $\beta$ TrA    & 5.85  & 0.52 & 1 \\
            &             & 8:30:49 & X TrA          & 5.51  & 0.53 & 0.9174 \\
            &             & 8:57:36 & $\zeta$ Ara    & 4.79  & 0.67 & 1 \\
11 Apr 2022 & A0-G1-J2-K0 & 7:12:22 & $\beta$ TrA    & 1.54  & 1.03 & 1 \\
            &             & 7:26:23 & X TrA          & 2.91  & 0.88 & 0.9083 \\
            &             & 7:52:45 & $\zeta$ Ara    & 3.22  & 1.01 & 0.9083 \\
29 Apr 2022 $^N$ & A0-B2-D0-C1 & 4:06:29 & X TrA     & 10.35 & 0.28 & 1 \\
            &             & 4:30:20 & $\zeta$ Ara    & 10.35 & 0.28 & 0.9909 \\
            &             & 6:08:57 & X TrA          & 8.87  & 0.41 & 1 \\
            &             & 6:32:41 & $\zeta$ Ara    & 9.30  & 0.51 & 1 \\
1 May 2022 $^L$ & A0-B2-D0-C1 & 8:48:52 & X TrA      & 3.73  & 0.69 & 1 \\
            &             & 9:12:28 & $\zeta$ Ara    & 3.37  & 0.72 & 1  \\
30 May 2022 $^N$ & K0-G2-D0-J3 & 2:51:17 & $\beta$ TrA & 4.00  & 0.70 & 1 \\
            &             & 3:02:13 & X TrA          & 4.40  & 0.62 & 0.9909 \\
            &             & 3:23:52 & $\zeta$ Ara    & 3.18  & 0.63 & 1 \\
            &             & 4:34:25 & $\beta$ TrA    & 4.37  & 0.59 & 1 \\
            &             & 4:57:49 & X TrA          & 3.90  & 0.59 & 0.9091 \\
            &             & 5:23:54 & $\zeta$ Ara    & 3.84  & 0.49 & 1 \\
5 Jul 2022  & A0-G2-J2-J3 & 3:07:57 & $\beta$ TrA    & 5.49  & 0.41 & 1 \\
            &             & 3:26:39 & X TrA          & 5.14  & 0.50 & 0.9182 \\
            &             & 3:48:57 & $\zeta$ Ara    & 4.96  & 0.46 & 1 \\ 
\hline
\end{tabular}
}
\tablefoot{$\tau_{0}$ is the coherence time, FT the fringe tracking ratio. The superscripts in the date column mark the datasets  that were of poor quality and were not used for the analysis presented in this study, with $L$ and $N$ identifying the MATISSE band the unused data come from.}
\end{table}

\begin{figure}
\centering
        \includegraphics[width = \linewidth]{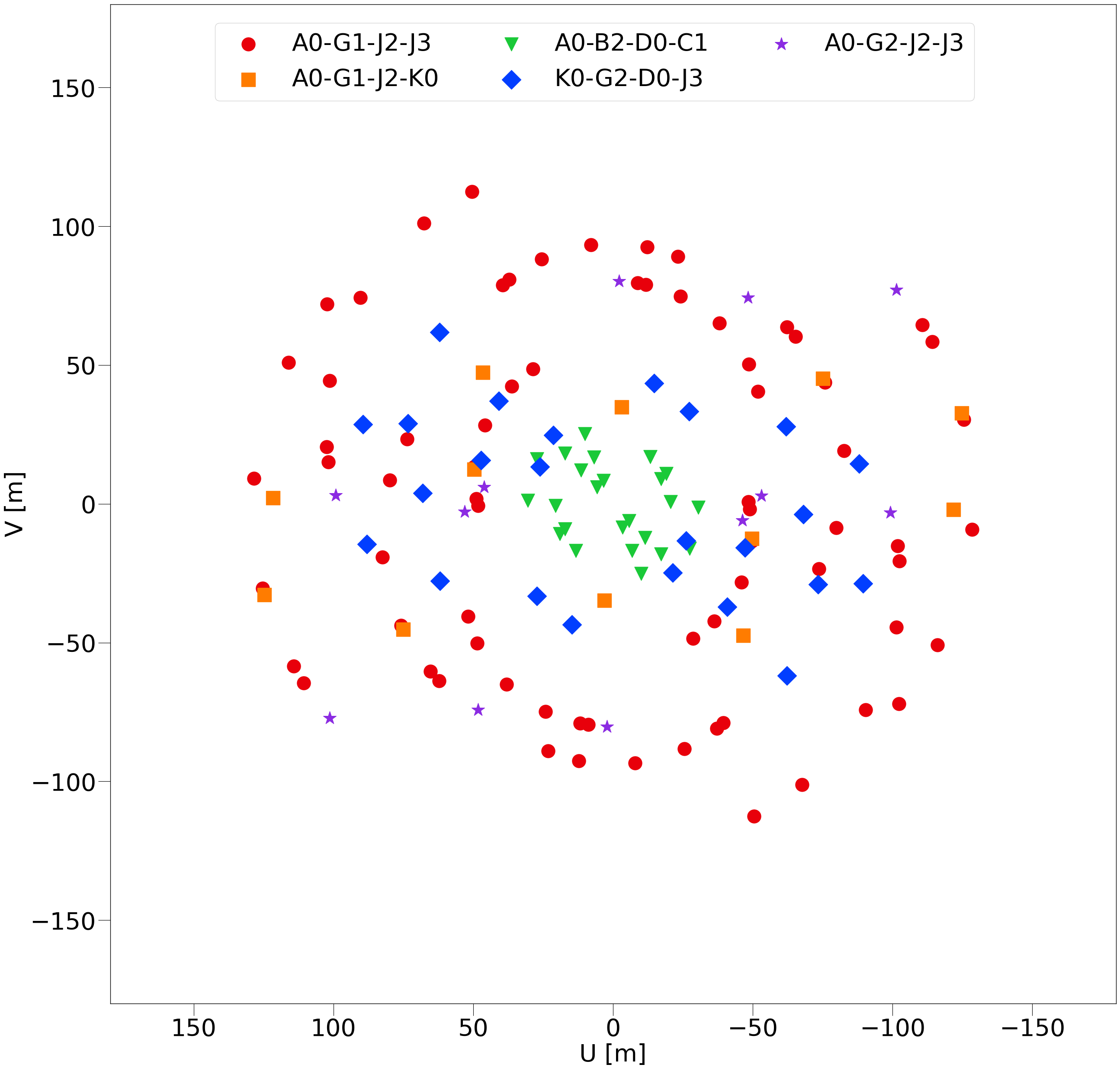}
        \caption{VLTI/MATISSE uv coverage of \object{X TrA}, colour-coded according to the telescope configuration used.} \label{uvcoverage}
\end{figure}

\begin{table}[ht] 
\caption{Photometric data used in the plotting of the spectral energy distribution shown in Fig.~\ref{sed}.} \label{phot}
\centering
\begin{tabular}{l*{2}l} \hline\hline
Filter & Flux \\ 
 & [Jy] \\ \hline\hline
2MASS:J                 & 535 \\
2MASS:H                 & 1060 \\
2MASS:Ks                & 1190 \\
Johnson:J               & 547 \\
Johnson:H               & 1050 \\
Johnson:K               & 1150 \\
WISE:W1                 & 1190 \\
WISE:W3                 & 88.5 \\
WISE:W4                 & 44.6 \\
IRAS:12                 & 201 \\
IRAS:25                 & 57.1 \\
AKARI:L18W          & 68.9 \\ \hline
\end{tabular}
\end{table}

\FloatBarrier
\section{Input MATISSE data visualised}\label{appB}

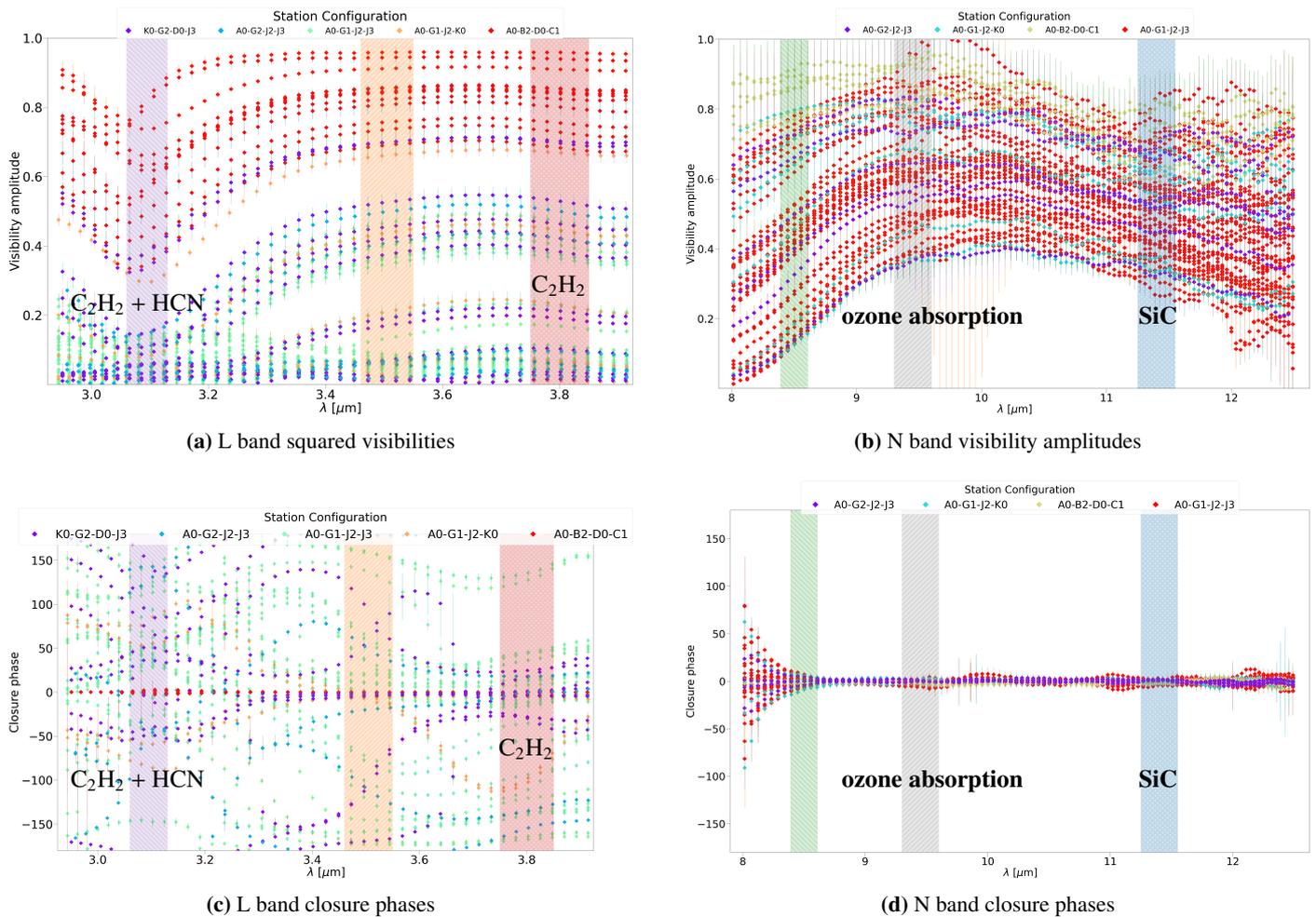
\begin{figure*}[htbp]
\centering
\begin{subfigure}{0.48\linewidth}
    \centering  
     \begin{overpic}[width = \linewidth]{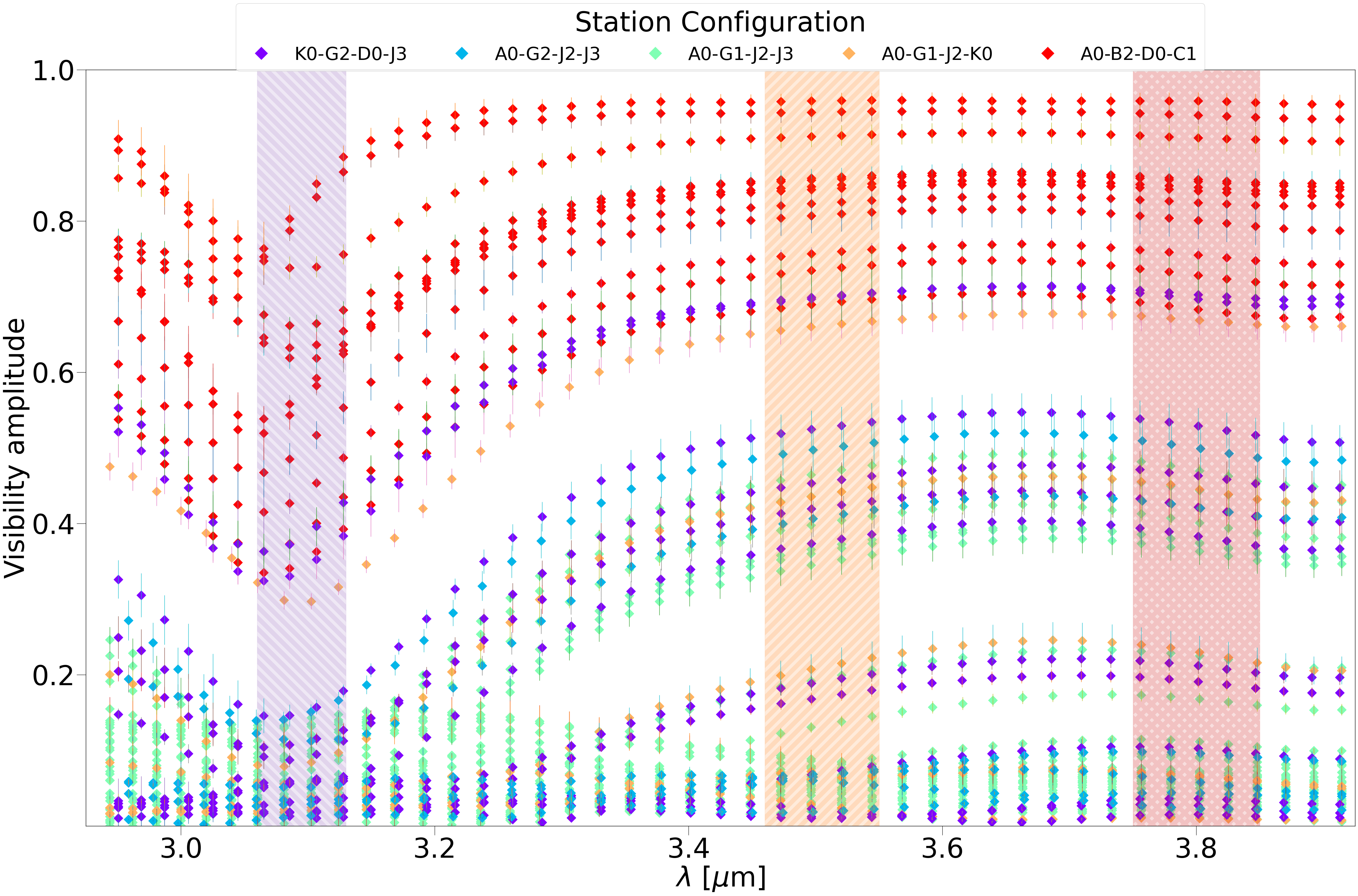}
         \put(10,17){\color{black}\bfseries $\mathrm{C_2H_2+HCN}$}
          \put(83.75,20){\color{black}\bfseries $\mathrm{C_2H_2}$}
    \end{overpic}
    \caption{L band squared visibilities} \label{l_v2}
\end{subfigure}
\hfill
\begin{subfigure}{0.48\linewidth}
    \centering
    \begin{overpic}[width = \linewidth]{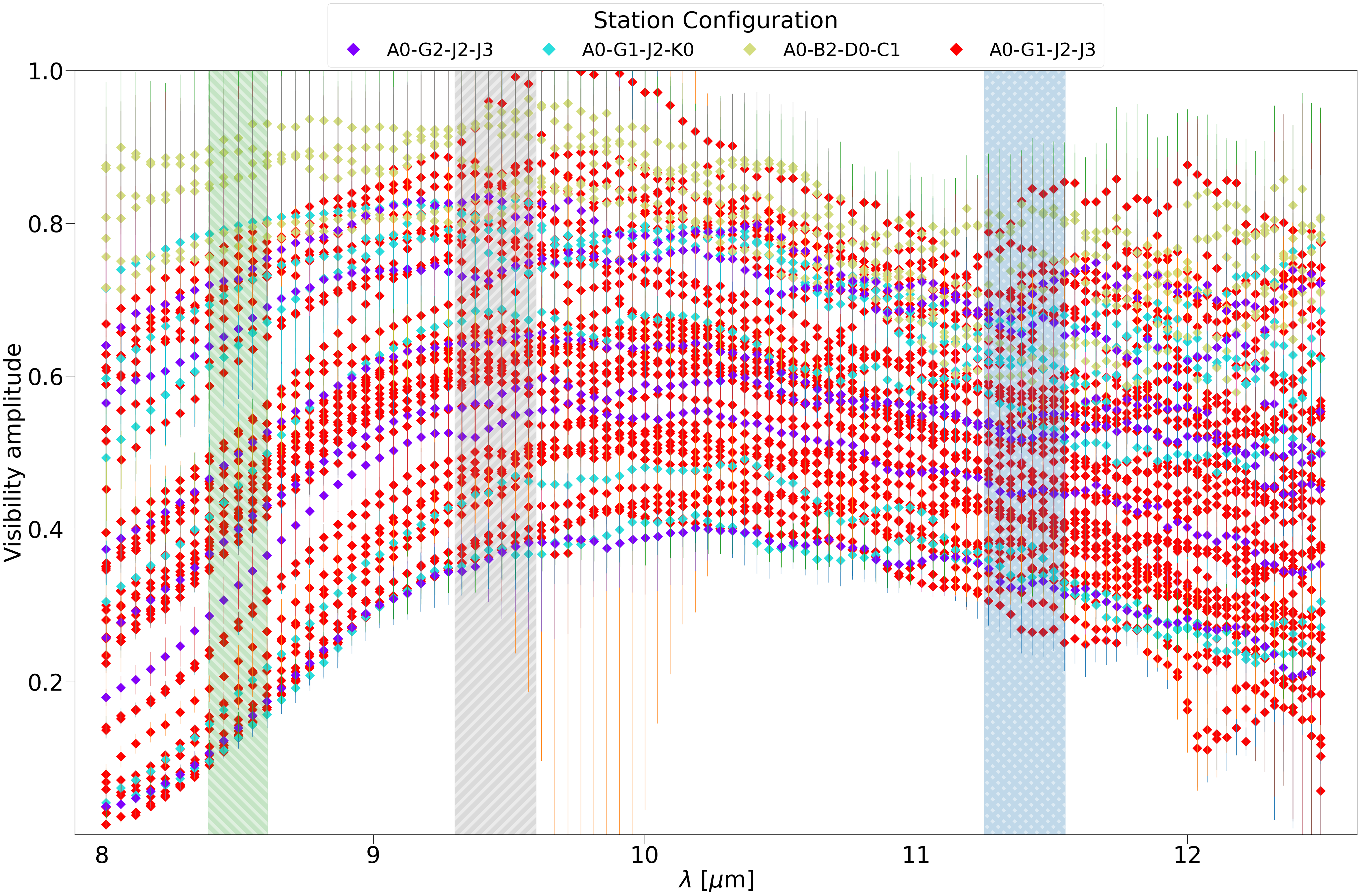}
                \put(25,15){\color{black}\bfseries ozone absorption}
                \put(72.5,15){\color{black}\bfseries SiC}
   \end{overpic}
    \caption{N band visibility amplitudes} \label{n_v}
\end{subfigure}

\vspace{1em}

\begin{subfigure}{0.48\linewidth}
    \centering  
    \begin{overpic}[width = \linewidth]{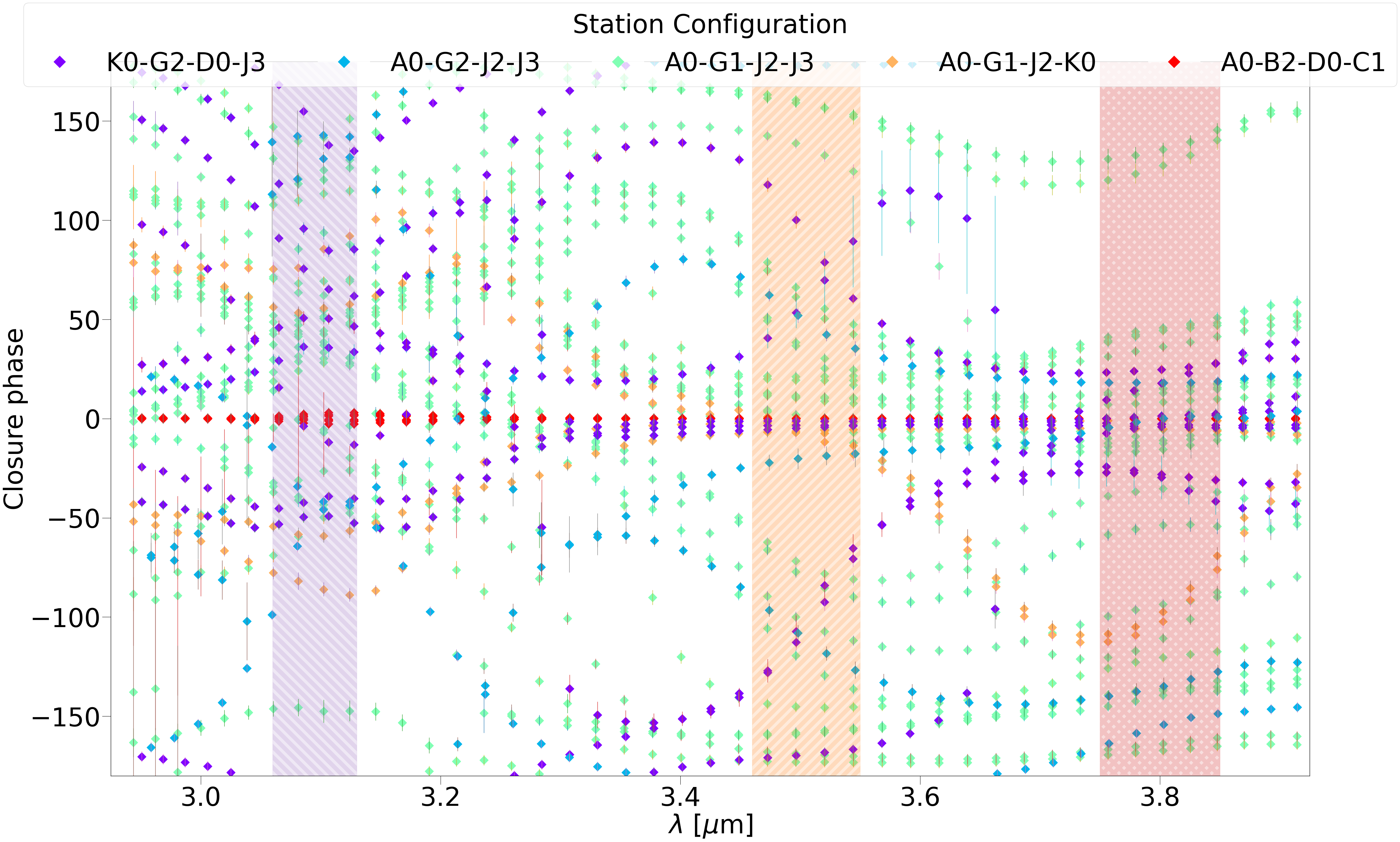}
                 \put(10,15){\color{black}\bfseries $\mathrm{C_2H_2+HCN}$}
          \put(78.5,20){\color{black}\bfseries $\mathrm{C_2H_2}$}
        \end{overpic}
    \caption{L band closure phases} \label{l_cp}
\end{subfigure}
\hfill
\begin{subfigure}{0.48\linewidth}
    \centering  
    \begin{overpic}[width = \linewidth]{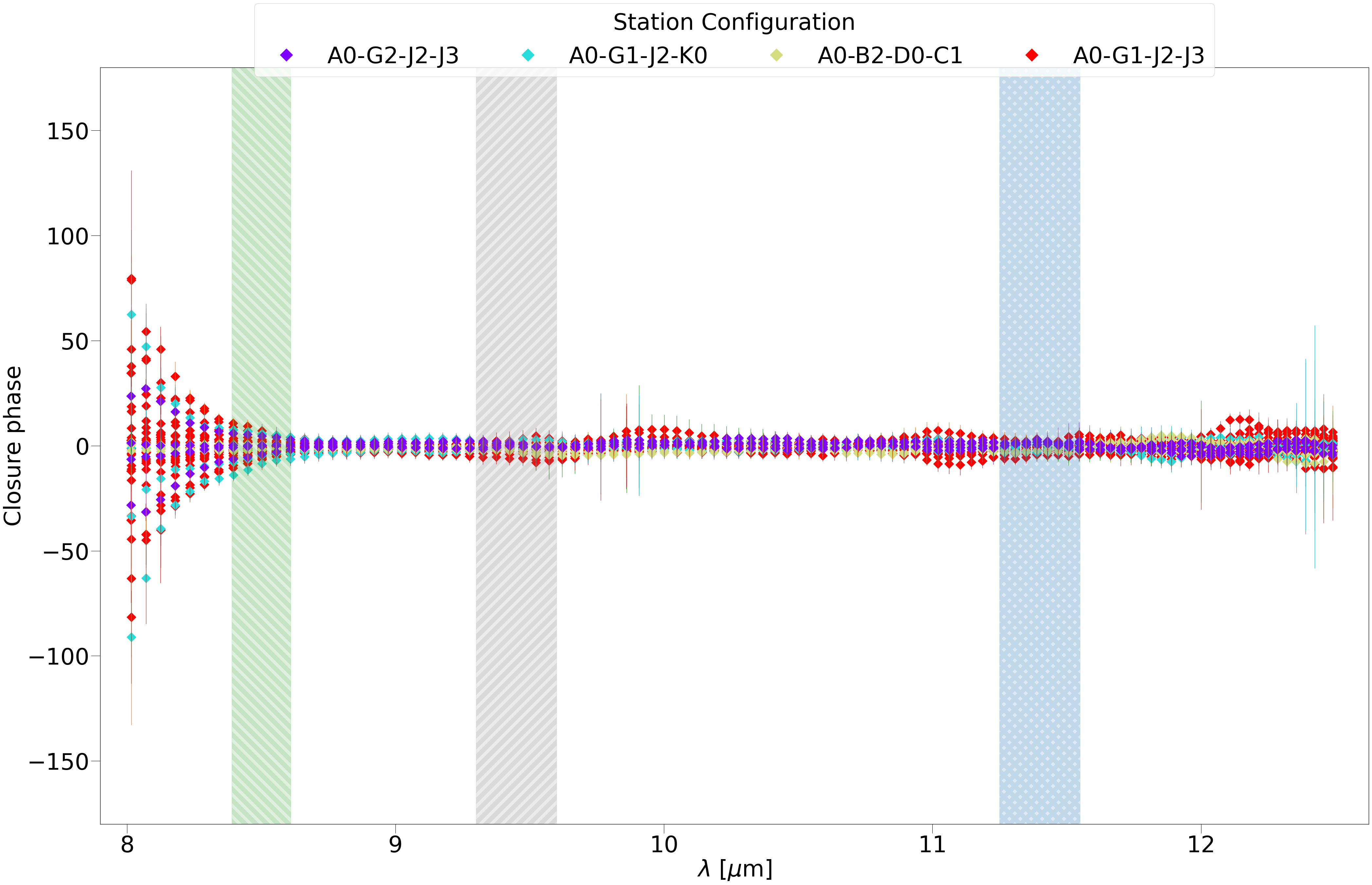}
                \put(25,15){\color{black}\bfseries ozone absorption}
          \put(72.5,15){\color{black}\bfseries SiC}
   \end{overpic}
    \caption{N band closure phases} \label{n_cp}
\end{subfigure}

\caption{Squared visibilities, visibility amplitudes, and closure phases of the input MATISSE data used in the image reconstruction process, colour-coded according to telescope configuration. The shaded areas represent the wavelength regions selected for image reconstruction (described in Sect.~\ref{s2}), except for the grey area shown in the N band plots. This marks a region of strong ozone absorption that falls in the wavelength range covered by the MATISSE instrument. The shaded areas without a label correspond to the selected pseudo-continuum bands.} 
\label{v2_cp_input}
\end{figure*}

\indent\indent Here we present the interferometric data of X TrA that were used to obtain the images presented in this study. The figures shown here present L band squared-visibility curves (Fig.~\ref{l_v2}), N band visibility amplitudes (Fig.~\ref{n_v}) and their corresponding closure phase data (Fig.~\ref{l_cp} and \ref{n_cp}).

\FloatBarrier
\section{Additional MiRA reconstruction results } \label{appC}

\indent\indent The images shown in the article are obtained via the quadratic compactness regularisation function of MiRA, as the edge-preserving smoothness function produced unrealistic results in many cases (e.g. gaps in the centre, spots with nothing in-between). The figures presented in this section show the observed data versus the MiRA reconstructed observables, corresponding to each of the wavelength channels used in the article.

\begin{figure*}[htbp]
    \centering
    \begin{subfigure}{0.48\linewidth}
        \centering
        \makebox[\linewidth][c]{%
            \begin{overpic}[height=7.35cm]{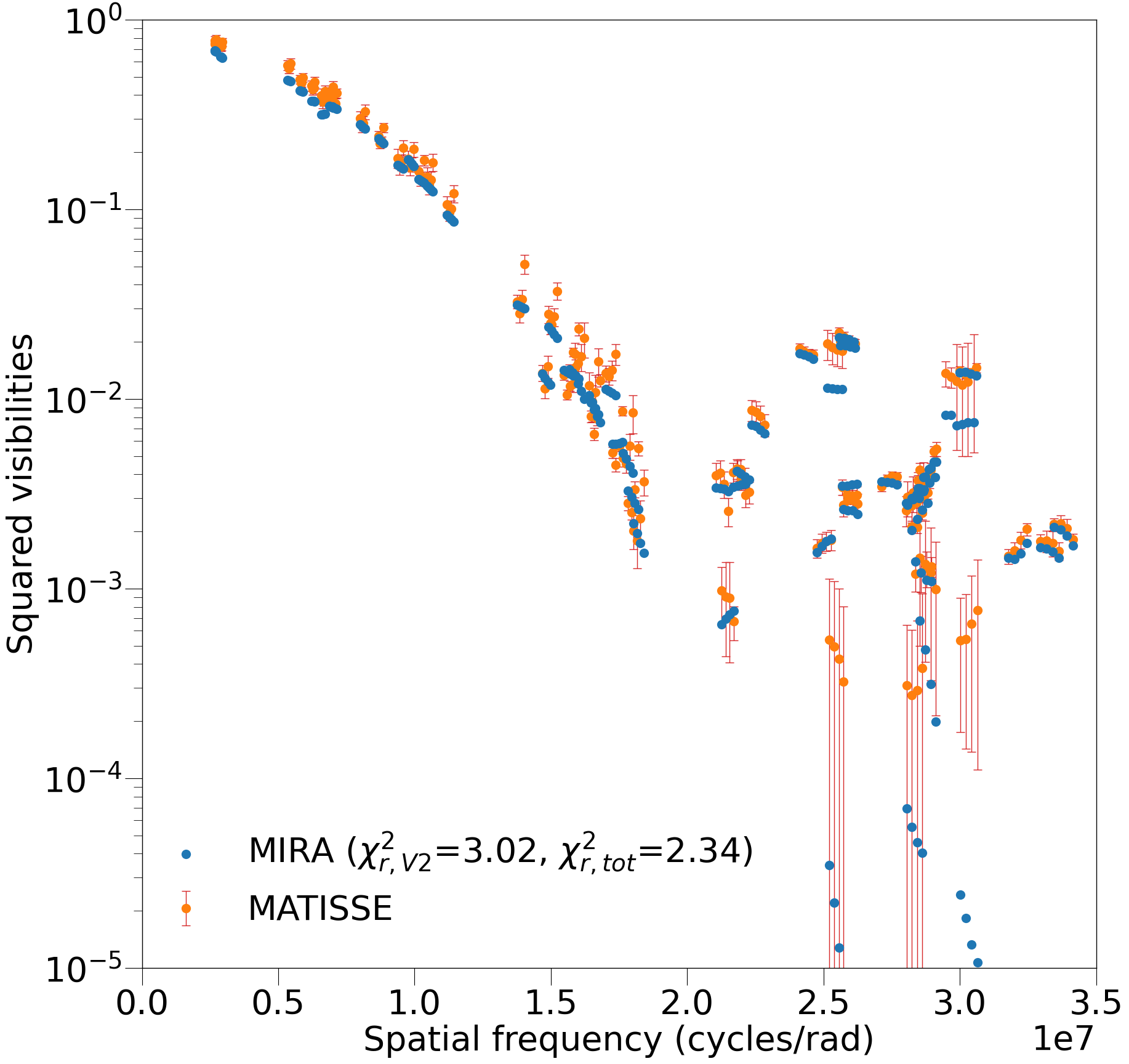}
            \put(17,23){\color{black}\bfseries 3.1 $\mu$m ($\mathrm{C_2H_2+HCN}$)}
            \end{overpic}   \hfill
            \begin{overpic}[height=7.35cm]{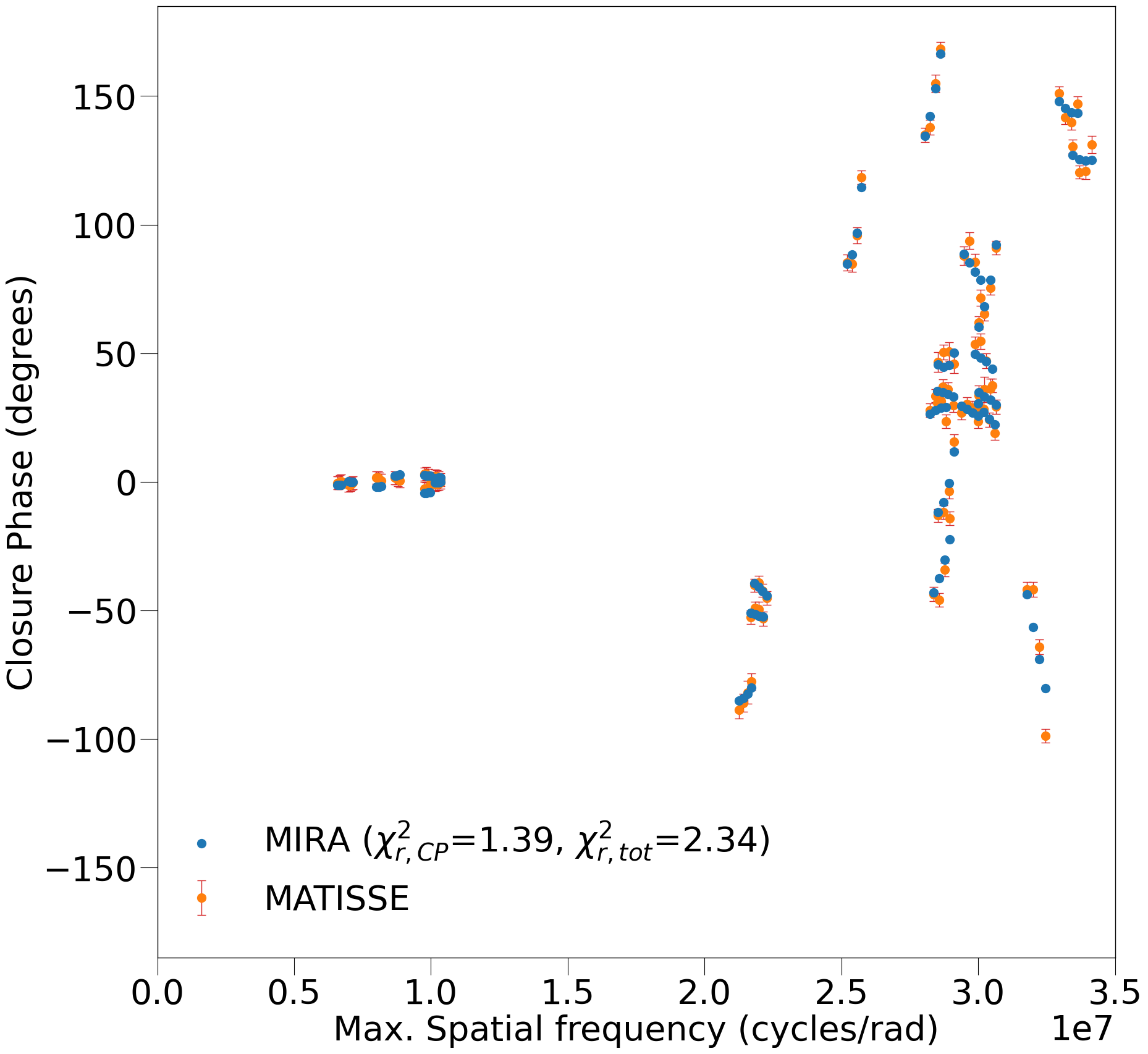}
            \put(17,23){\color{black}\bfseries 3.1 $\mu$m ($\mathrm{C_2H_2+HCN}$)}
            \end{overpic}%
        } \label{}
    \end{subfigure}\\
    \begin{subfigure}{0.48\linewidth}
        \centering
        \makebox[\linewidth][c]{%
            \begin{overpic}[height=7.35cm]{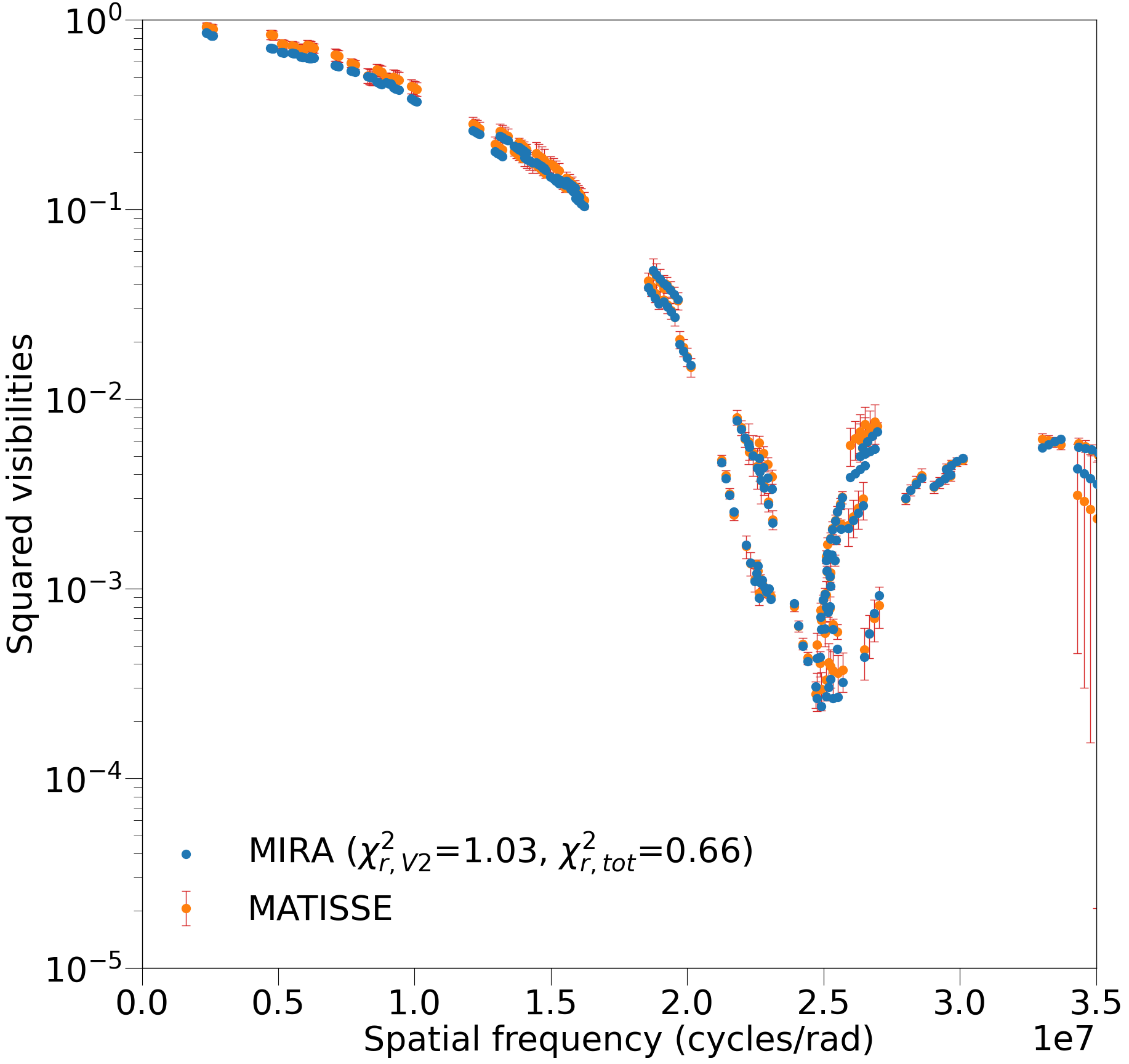}
            \put(17,23){\color{black}\bfseries 3.5 $\mu$m}
            \end{overpic}   \hfill
            \begin{overpic}[height=7.35cm]{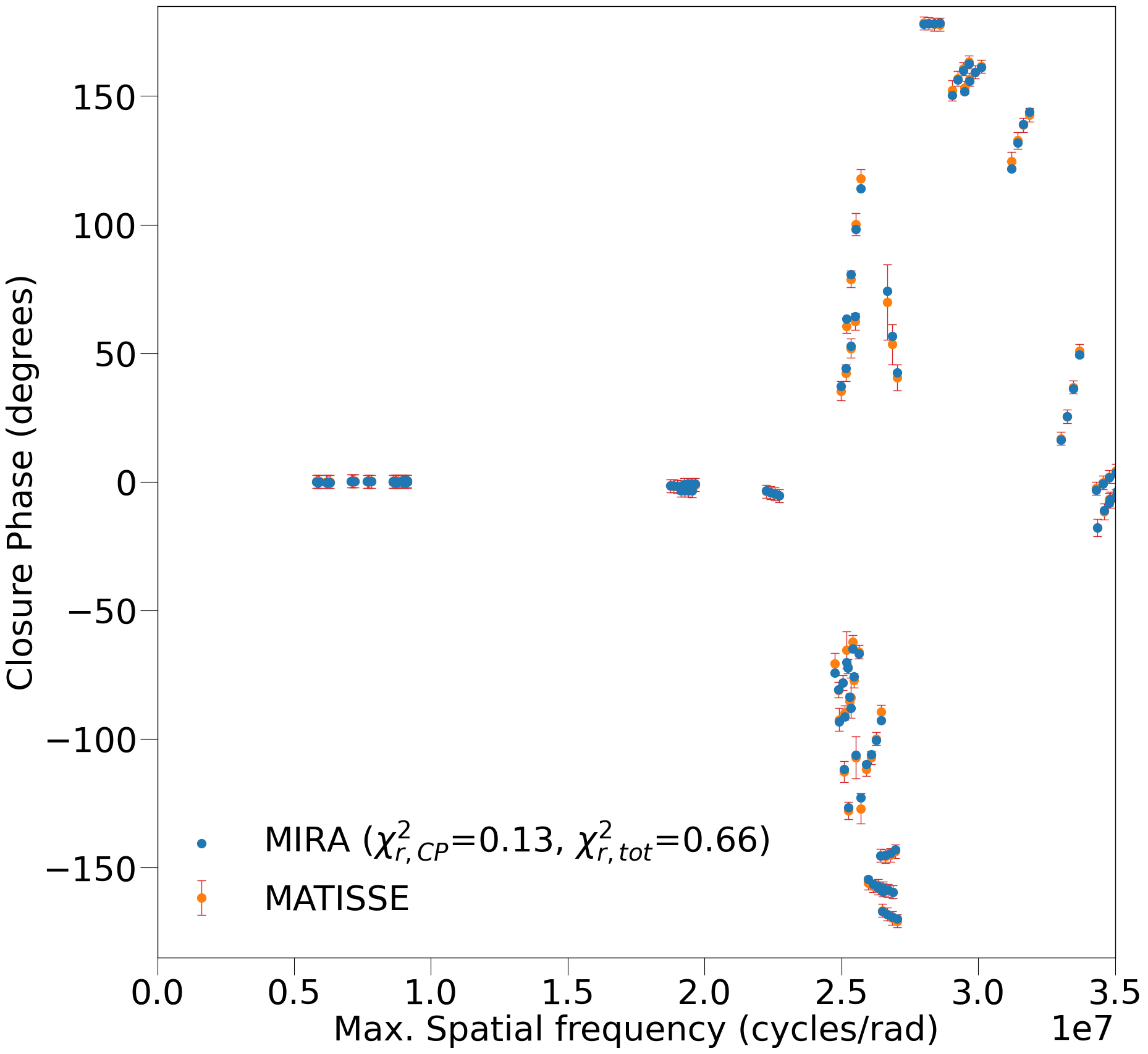}
            \put(17,23){\color{black}\bfseries 3.5 $\mu$m}
            \end{overpic}%
        } \label{}
    \end{subfigure}\\
    \begin{subfigure}{0.48\linewidth}
        \centering
        \makebox[\linewidth][c]{%
            \begin{overpic}[height=7.35cm]{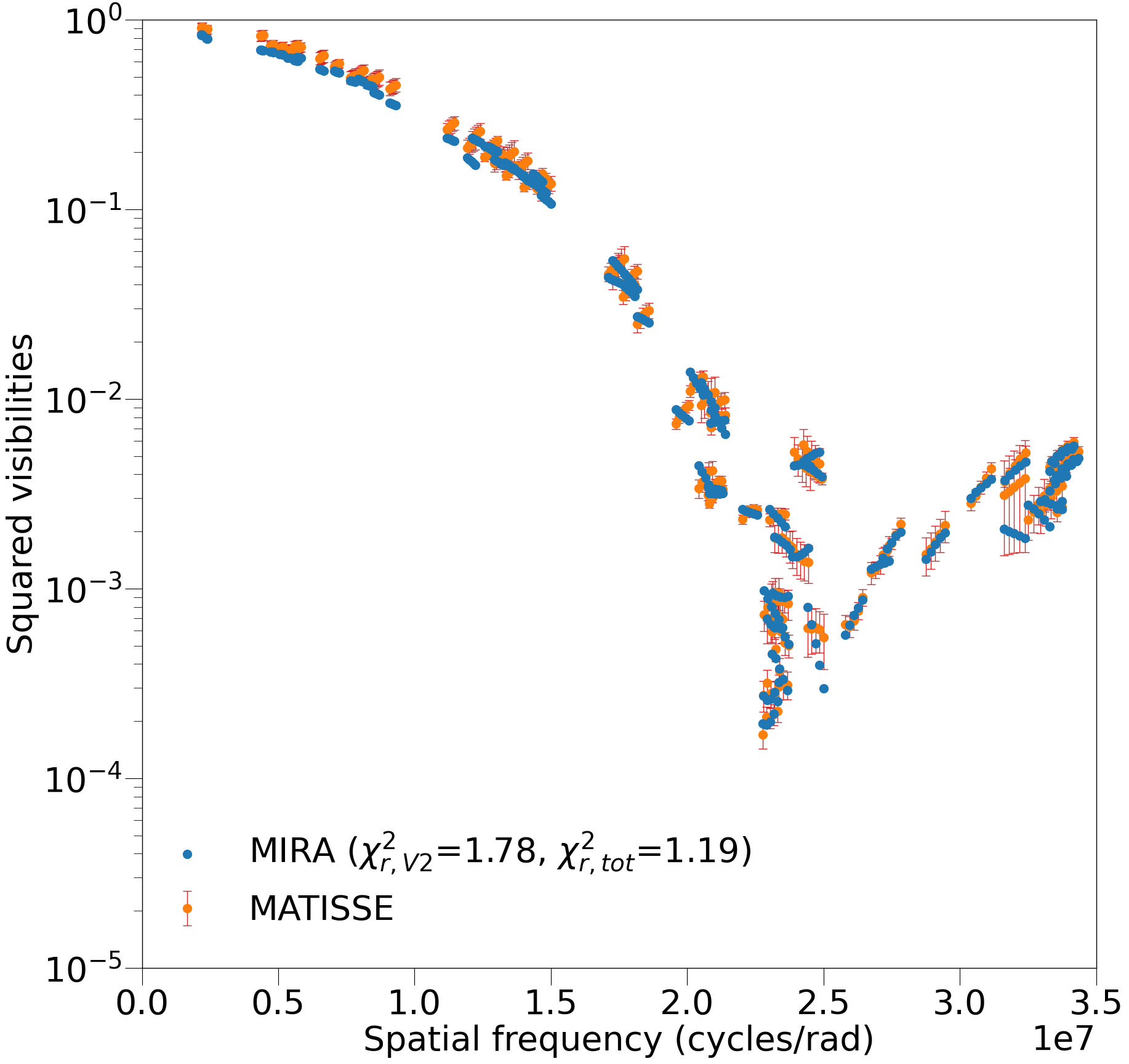}
            \put(17,23){\color{black}\bfseries 3.8 $\mu$m ($\mathrm{C_2H_2}$)}
            \end{overpic}   \hfill
            \begin{overpic}[height=7.35cm]{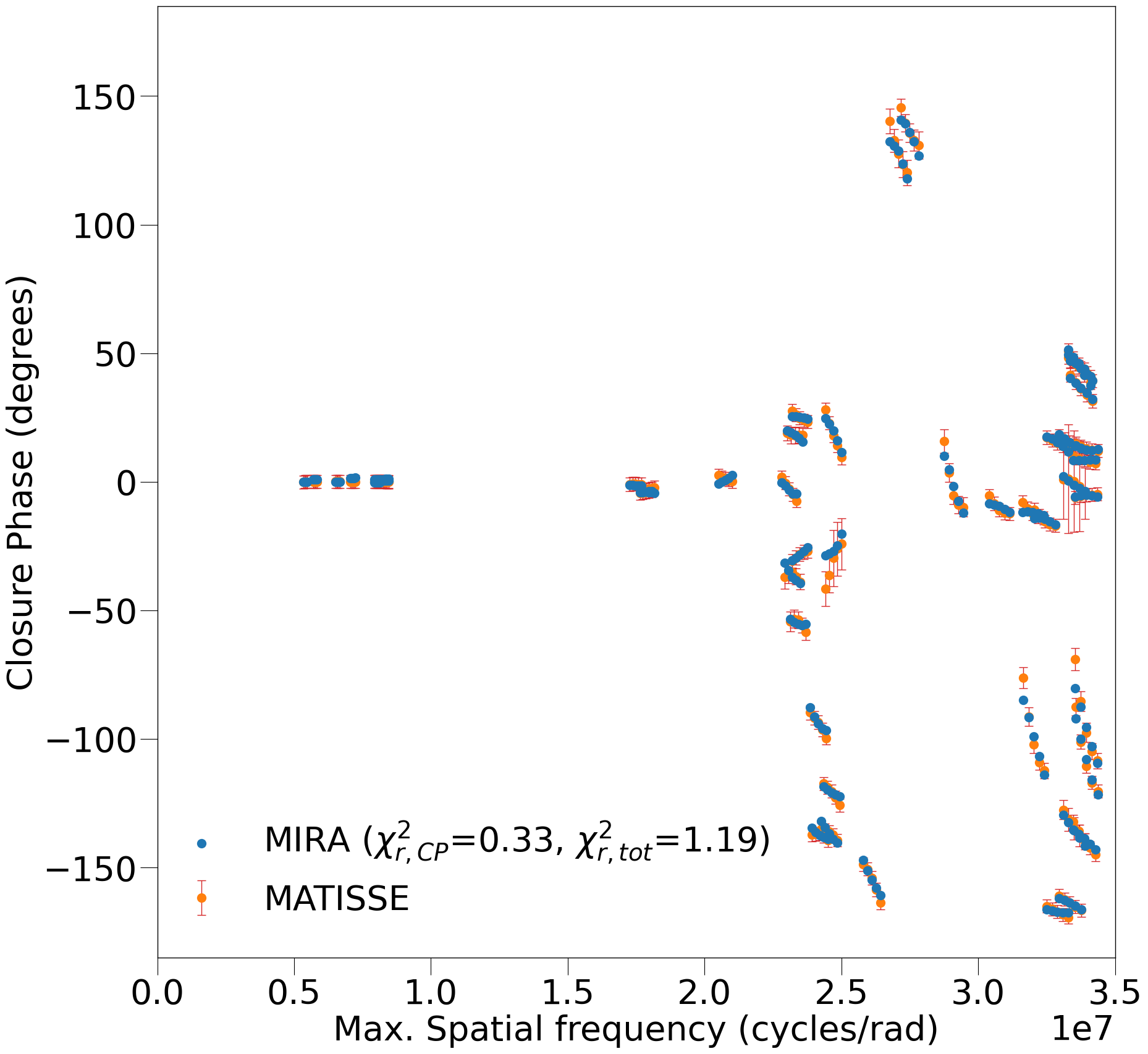}
            \put(17,23){\color{black}\bfseries 3.8 $\mu$m ($\mathrm{C_2H_2}$)}
            \end{overpic}%
        }
    \end{subfigure}
    \caption{Squared visibilities (left) and closure phases (right) of the reconstructed images (blue dots) compared to the input MATISSE data (orange dots).} \label{mira_inf_vs_input_lband}
\end{figure*} 

\begin{figure*}[htbp]
    \centering
    \begin{subfigure}{0.48\linewidth}
        \centering
        \makebox[\linewidth][c]{%
            \begin{overpic}[height=7.35cm]{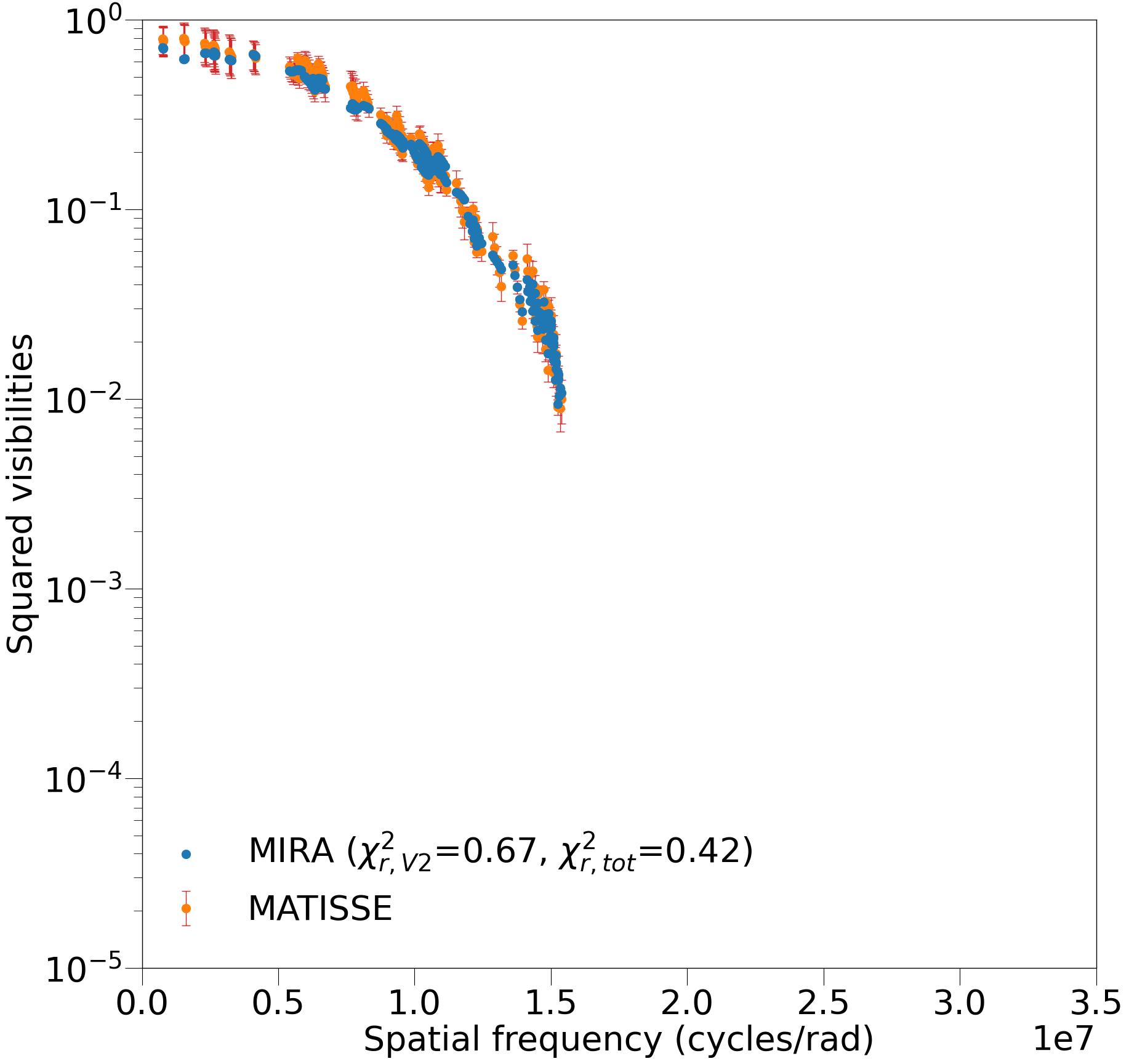}
            \put(17,23){\color{black}\bfseries 8.5 $\mu$m}
            \end{overpic}   \hfill
            \begin{overpic}[height=7.35cm]{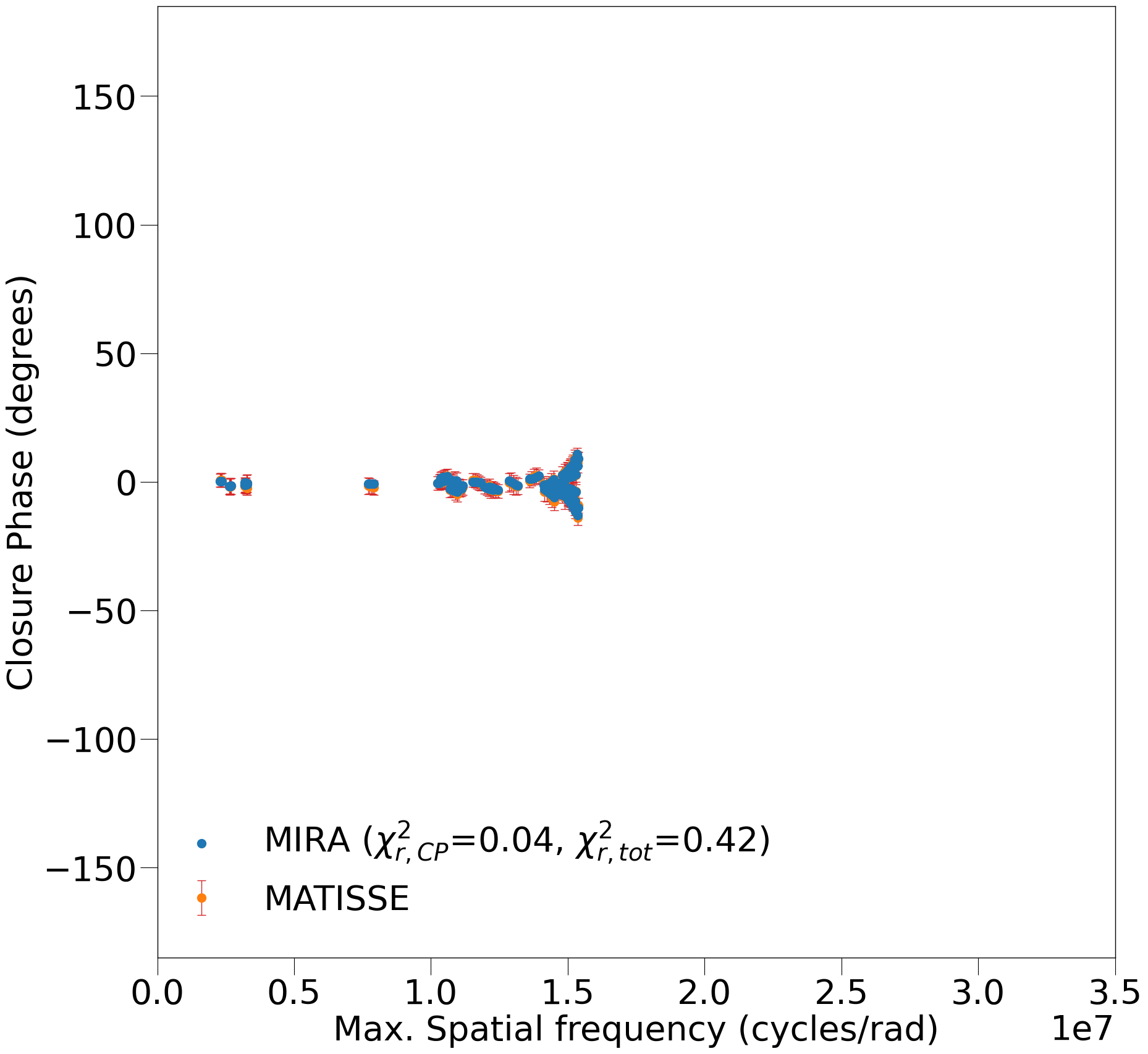}
            \put(17,23){\color{black}\bfseries 8.5 $\mu$m}
            \end{overpic}%
        } \label{}
    \end{subfigure}\\
    \begin{subfigure}{0.48\linewidth}
        \centering
        \makebox[\linewidth][c]{%
            \begin{overpic}[height=7.35cm]{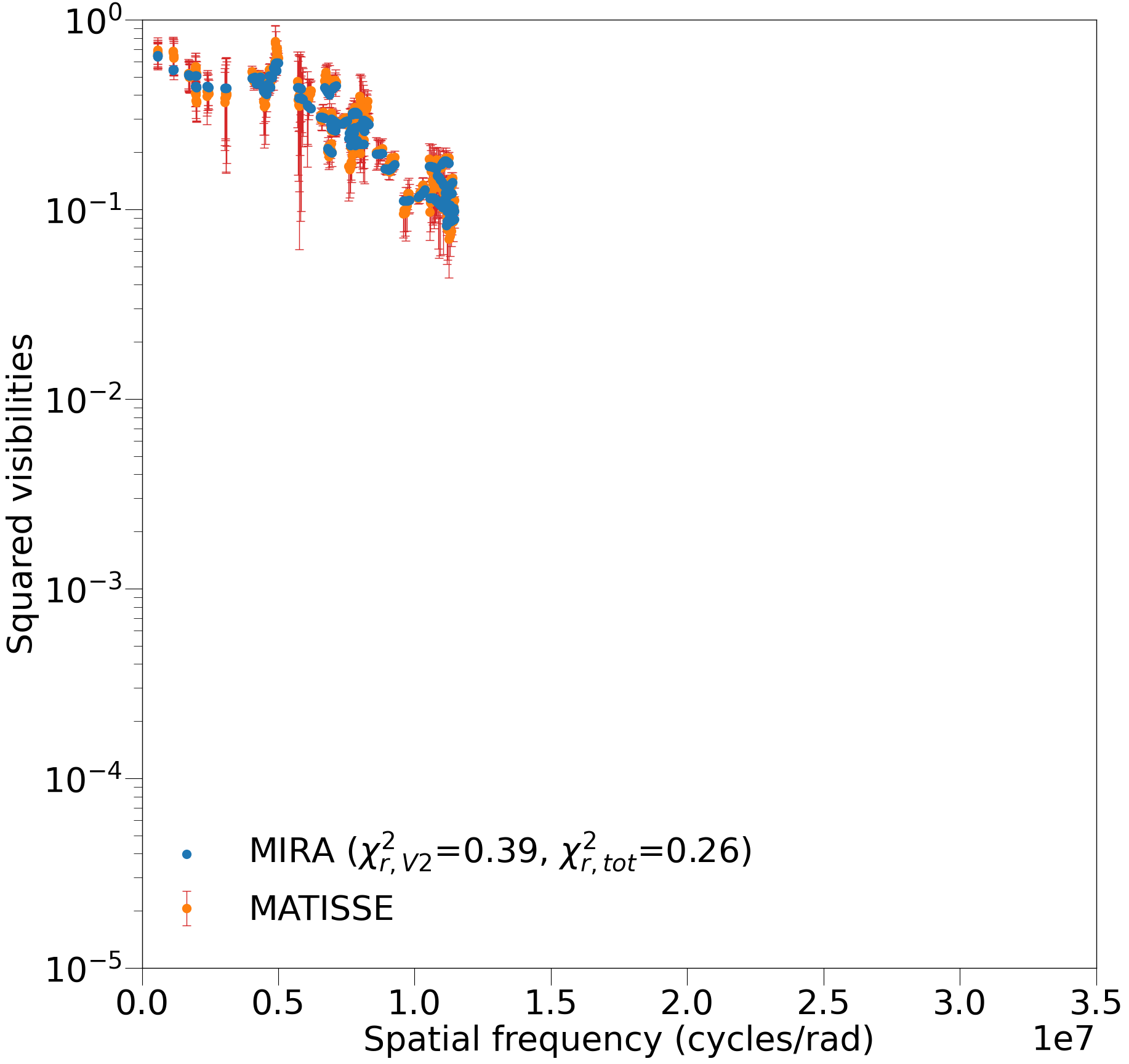}
            \put(17,23){\color{black}\bfseries 11.3 $\mu$m (SiC)}
            \end{overpic}   \hfill
            \begin{overpic}[height=7.35cm]{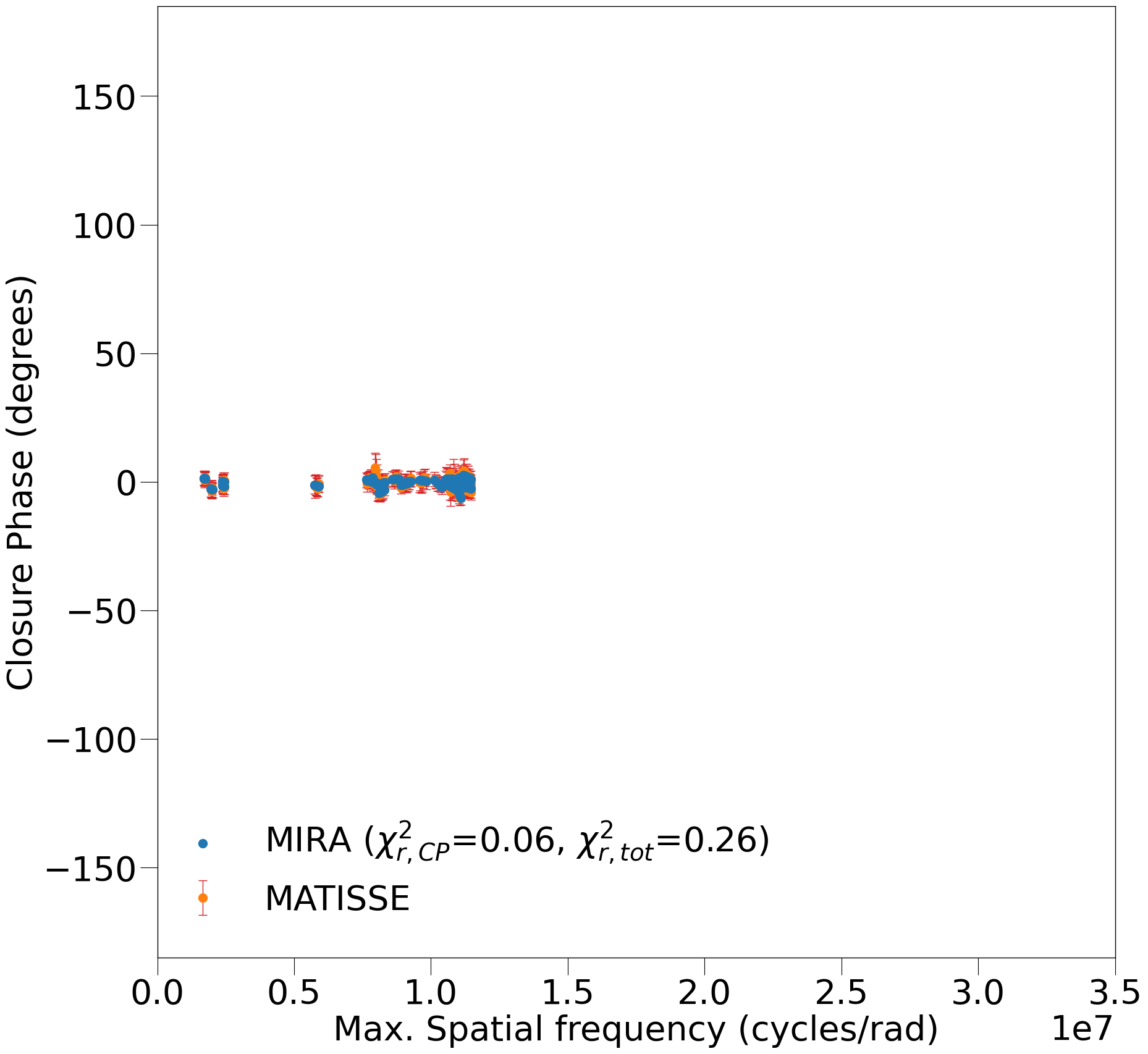}
            \put(17,23){\color{black}\bfseries 11.3 $\mu$m (SiC)}
            \end{overpic}%
        }
    \end{subfigure}
    \caption{Squared visibilities (left) and closure phases (right) of the reconstructed images (blue dots) compared to the input MATISSE data (orange dots).} \label{mira_inf_vs_input_nband}
\end{figure*} 

\FloatBarrier
\section{SQUEEZE results} \label{squeezults}

\indent\indent As opposed to MiRA, SQUEEZE fits the interferometric data via a Markov chain Monte Carlo algorithm. The total-variation regularisation function was used, this reconstruction method being suitable for cases where one is interested in keeping prominent features, as it handles well uniform areas that show steep (but localised) changes. Similar to the edge-preserving smoothness regularisation of MiRA, this method produces imprecise artefacts, while MiRA's quadratic compactness gives rise to artefacts in the shape of ripples around structures \citep{thiebaut2017}. The same parameters used for the MiRA reconstruction (pixel size and field-of-view) are used to reconstruct an image with Squeeze, with the final image representing the median of ten chains. This serves as an additional verification of the images. If the two algorithms converge on a similar solution, then the result can be considered reliable.

\par
\begin{figure*}[htbp]
    \centering
        \begin{minipage}[c]{0.95\textwidth}
           \centering
            \begin{subfigure}{.33\linewidth}
                \begin{overpic}[width=\linewidth]{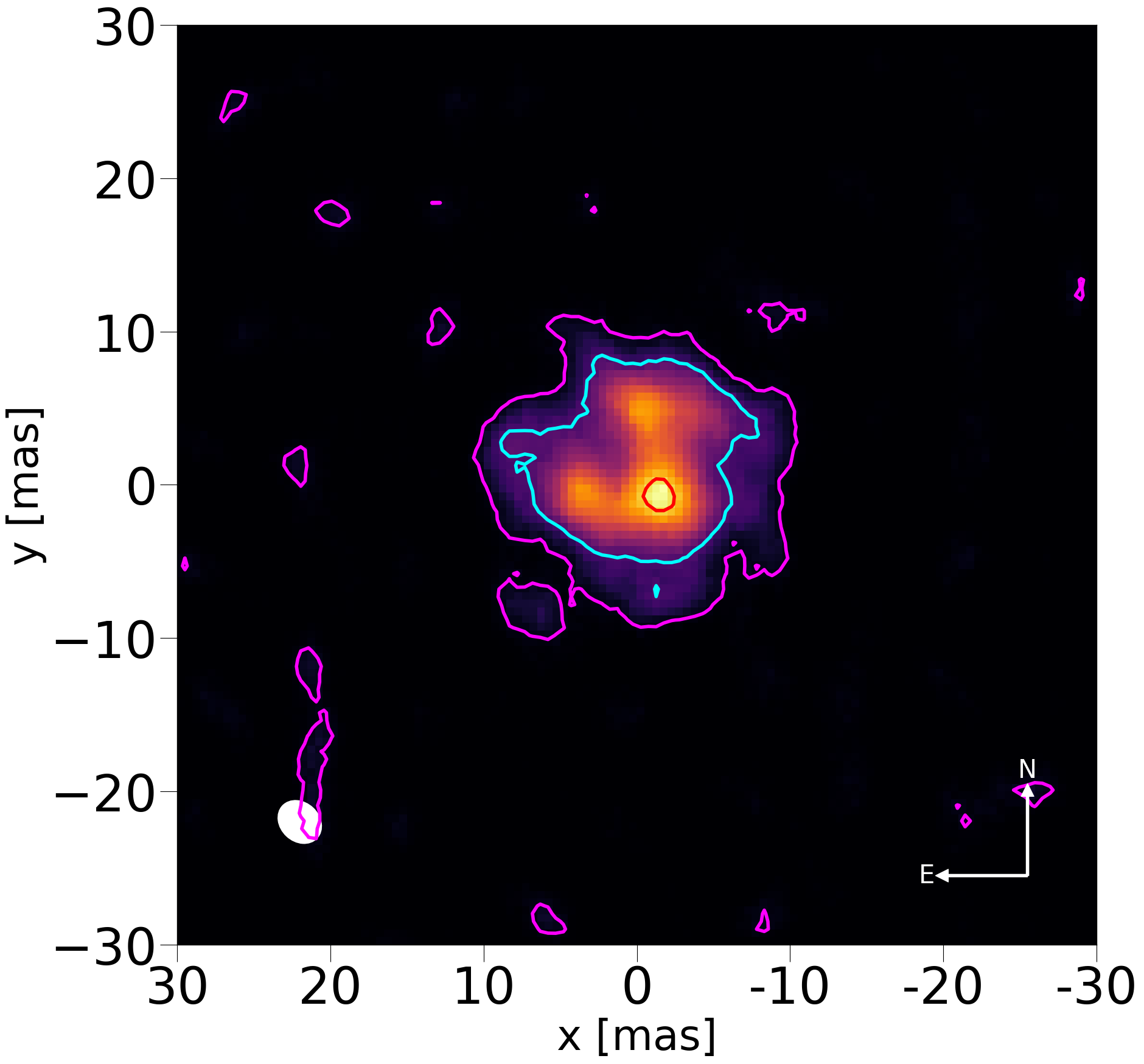}
                    \put(20,85){\color{white}\bfseries 3.1 $\mu$m}
                    \put(62.5,85){\color{white}\bfseries $\mathrm{C_2H_2+HCN}$}
                \end{overpic}
            \end{subfigure}
            \begin{subfigure}{.32\linewidth}
                \begin{overpic}[width=\linewidth]{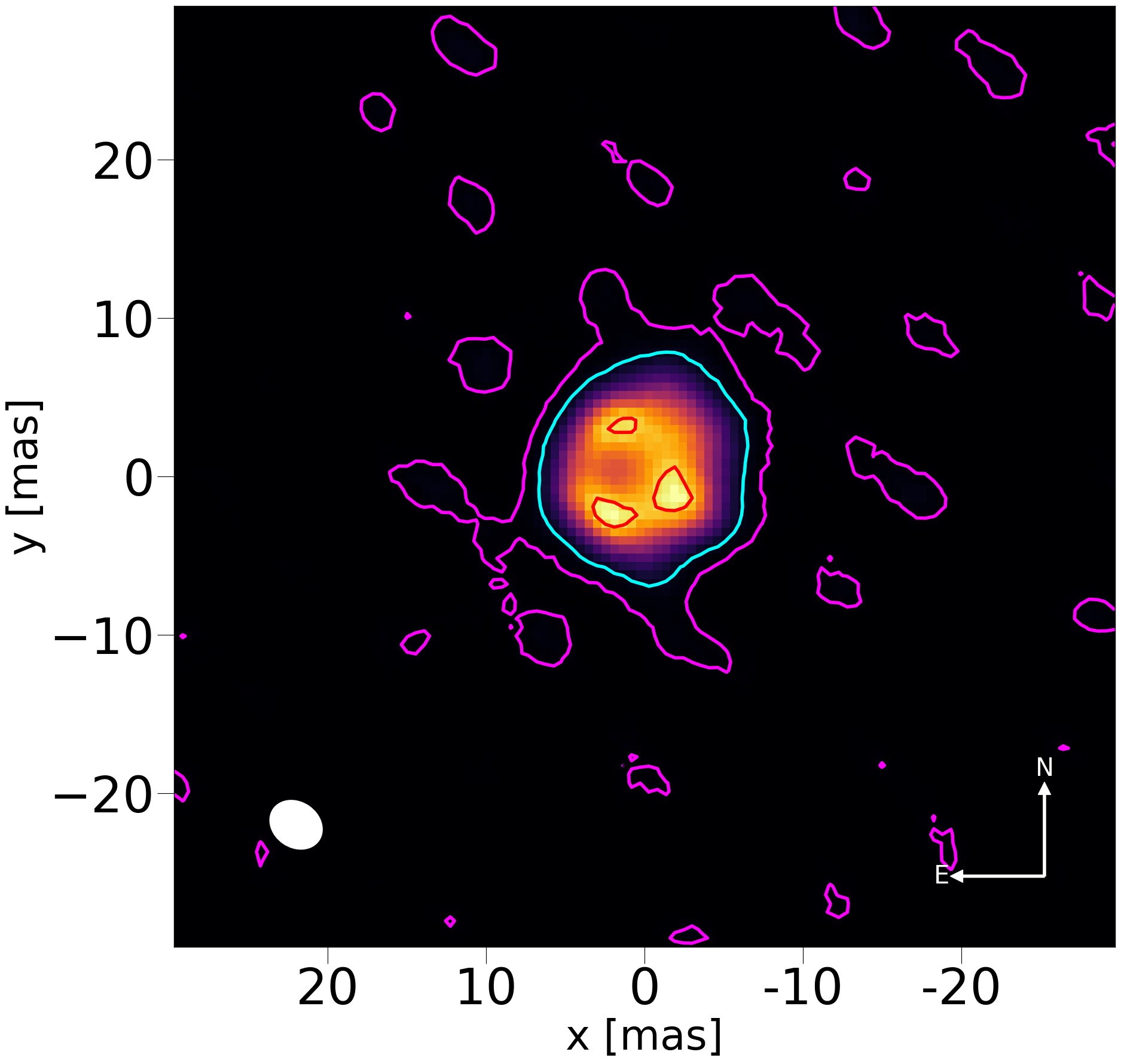}
                    \put(20,85){\color{white}\bfseries 3.5 $\mu$m}
                \end{overpic}
            \end{subfigure}
            \begin{subfigure}{.32\linewidth}
                \begin{overpic}[width=\linewidth]{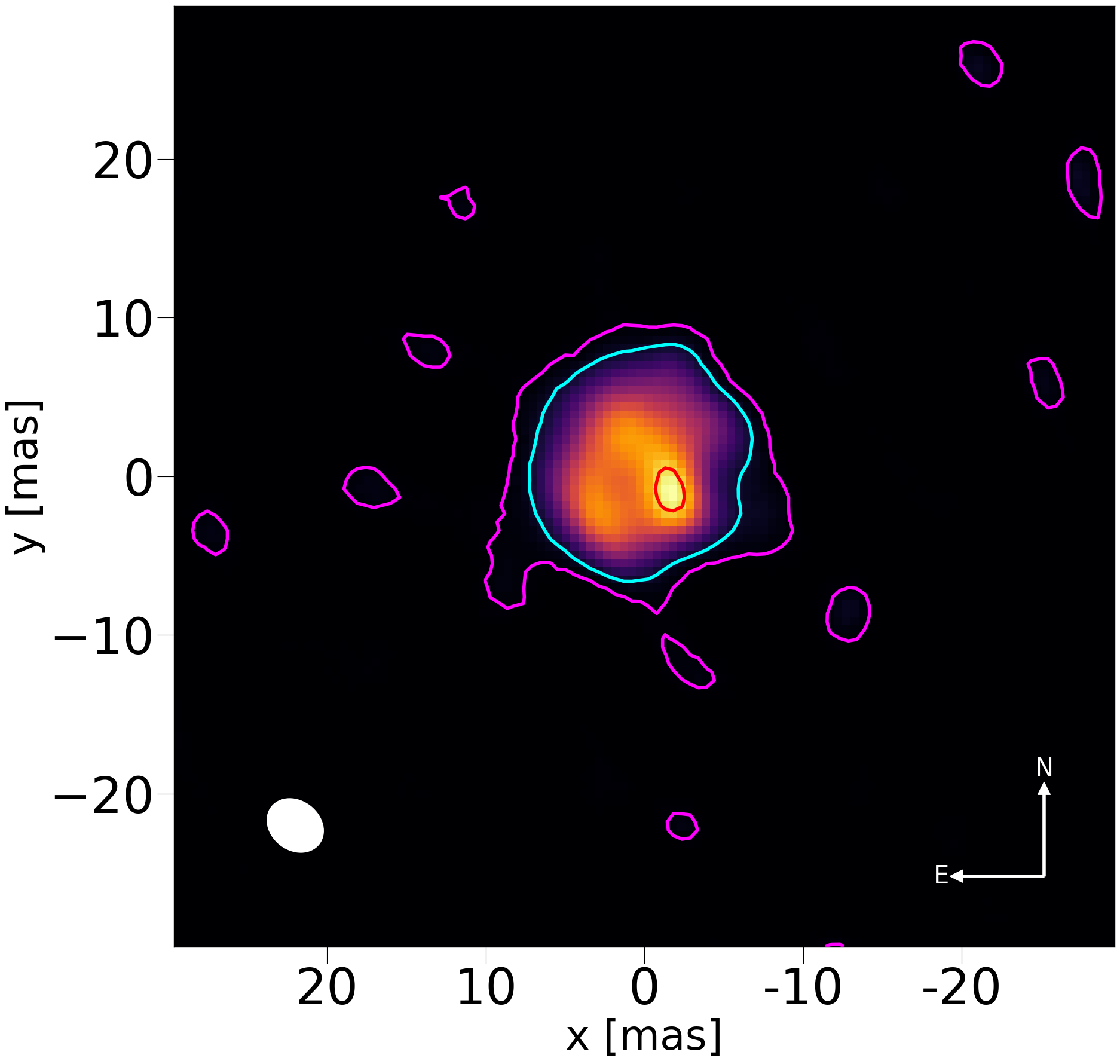}
                    \put(20,85){\color{white}\bfseries 3.8 $\mu$m}
                    \put(85,85){\color{white}\bfseries $\mathrm{C_2H_2}$}
                \end{overpic}
            \end{subfigure}           
            \par
            \begin{subfigure}{.33\linewidth}
                \begin{overpic}[width=\linewidth]{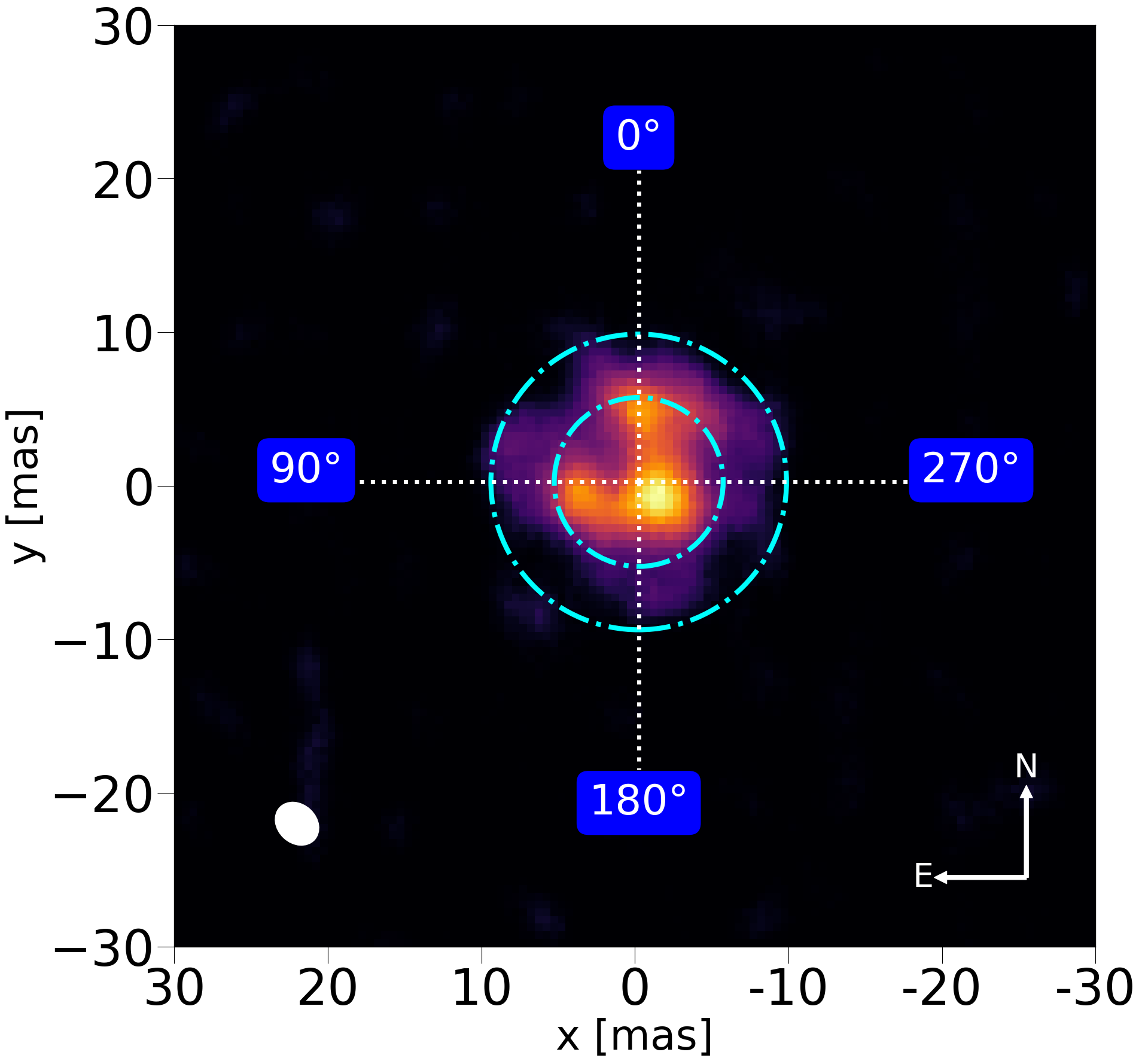}
                    \put(20,85){\color{white}\bfseries 3.1 $\mu$m}
                    \put(62.5,85){\color{white}\bfseries $\mathrm{C_2H_2+HCN}$}
                \end{overpic}
            \end{subfigure}
            \begin{subfigure}{.32\linewidth}
                \begin{overpic}[width=\linewidth]{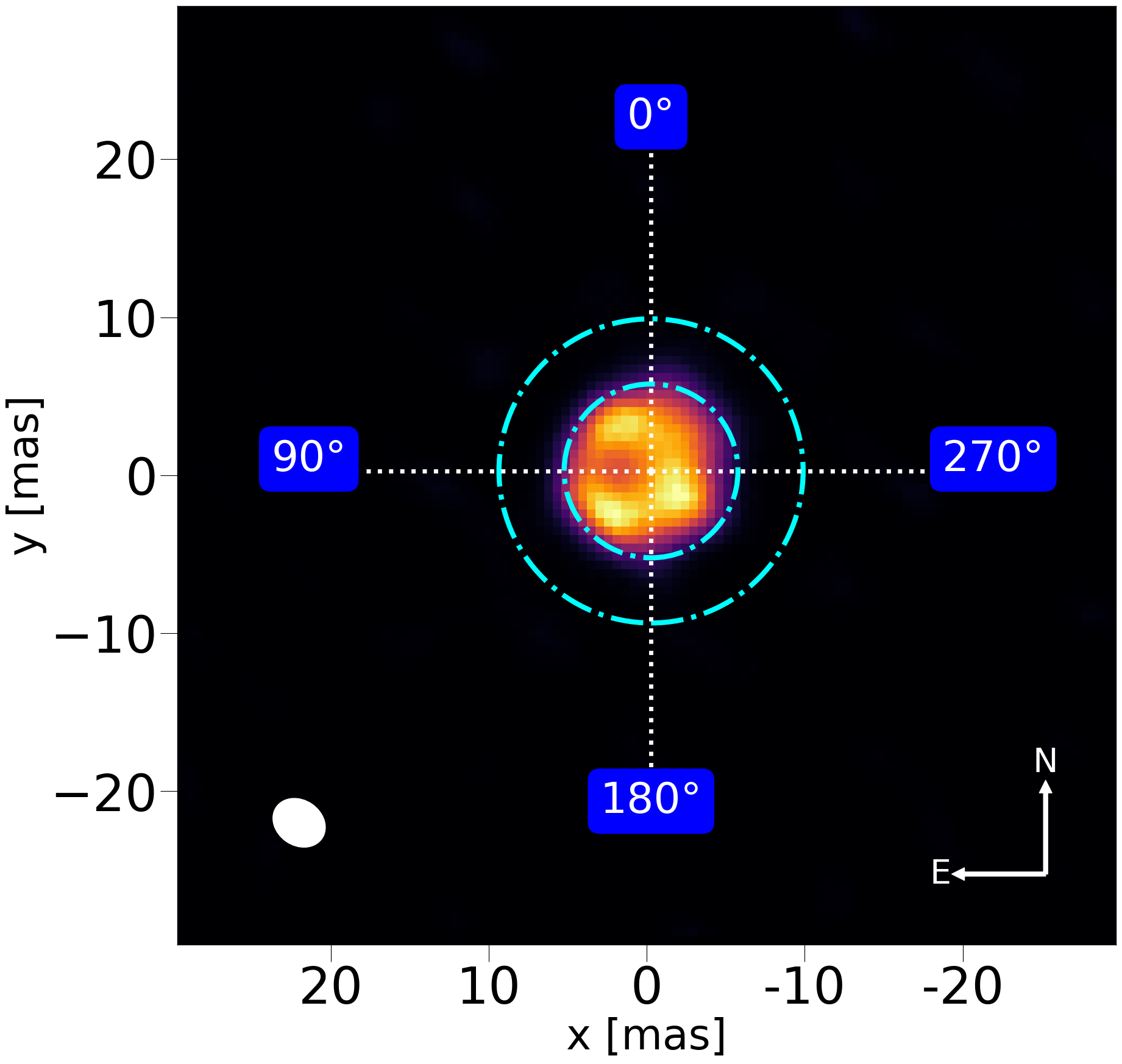}
                    \put(20,85){\color{white}\bfseries 3.5 $\mu$m}
                \end{overpic}
            \end{subfigure}        
            \begin{subfigure}{.32\linewidth}
                \begin{overpic}[width=\linewidth]{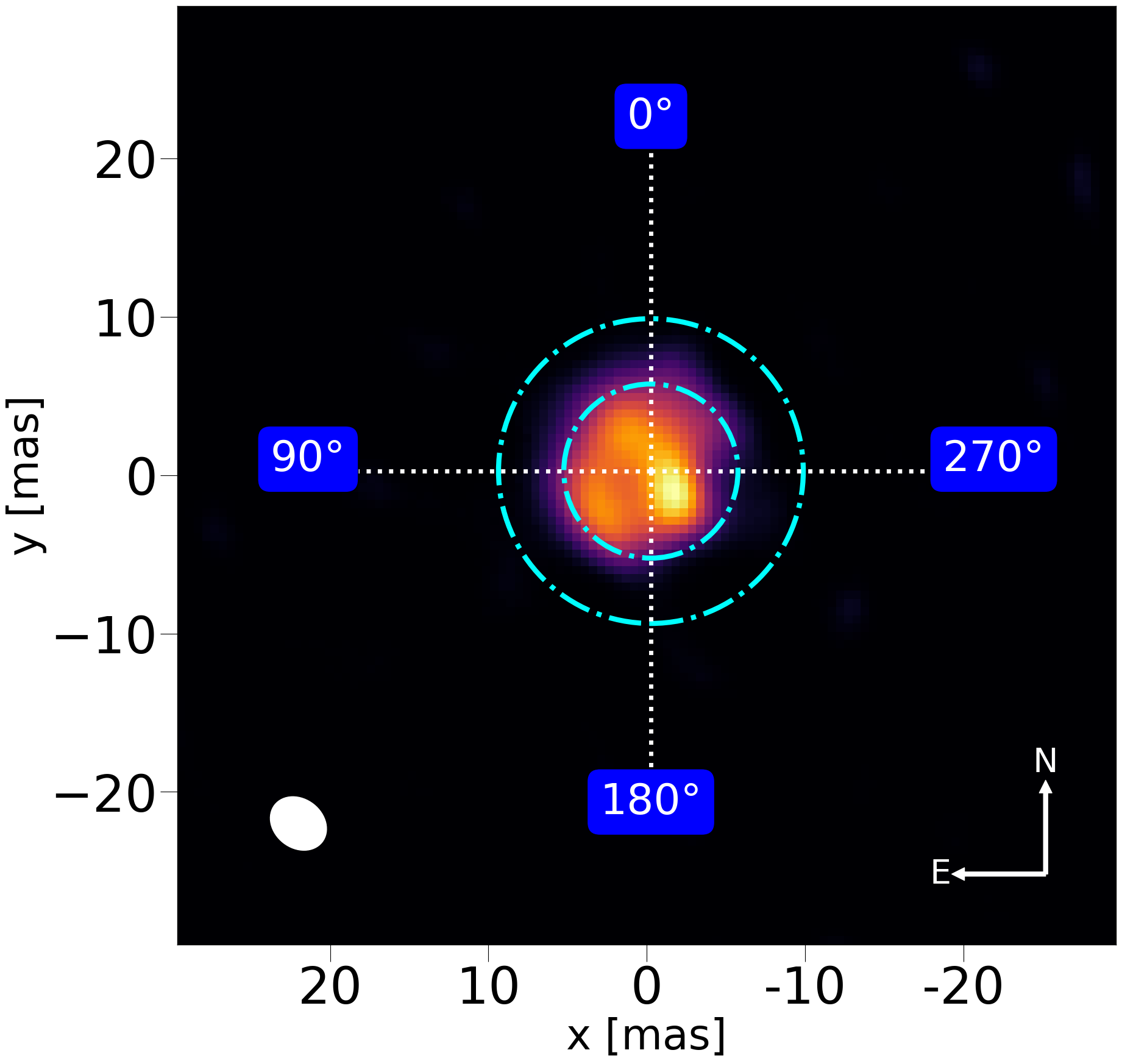}
                    \put(20,85){\color{white}\bfseries 3.8 $\mu$m}
                    \put(85,85){\color{white}\bfseries $\mathrm{C_2H_2}$}
                \end{overpic}
            \end{subfigure}        
            \par     
            \begin{subfigure}{.95\linewidth}
            \centering
            \hspace{0.85cm}
            \includegraphics[width=0.5\linewidth]{colorbar.png}
            \end{subfigure}
            \par  
            \hspace{0.55cm}
              \begin{subfigure}{.3\linewidth}
                      \includegraphics[width = \linewidth]{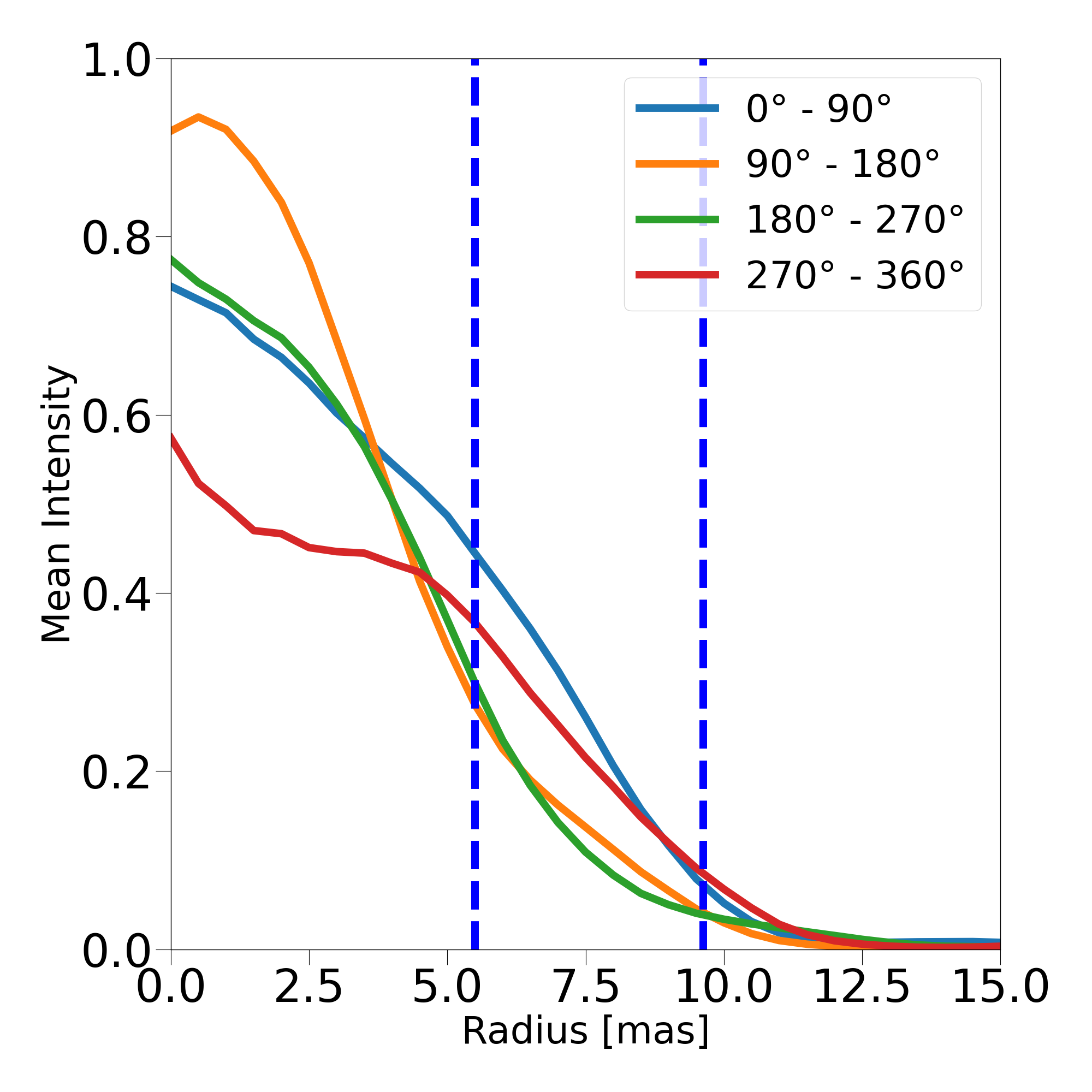}
                \caption{3.1 $\mu$m ($\mathrm{C_2H_2 + HCN}$)}
            \end{subfigure}     \hspace{0.25cm}       
        \begin{subfigure}{.3\linewidth}
                      \includegraphics[width = \linewidth]{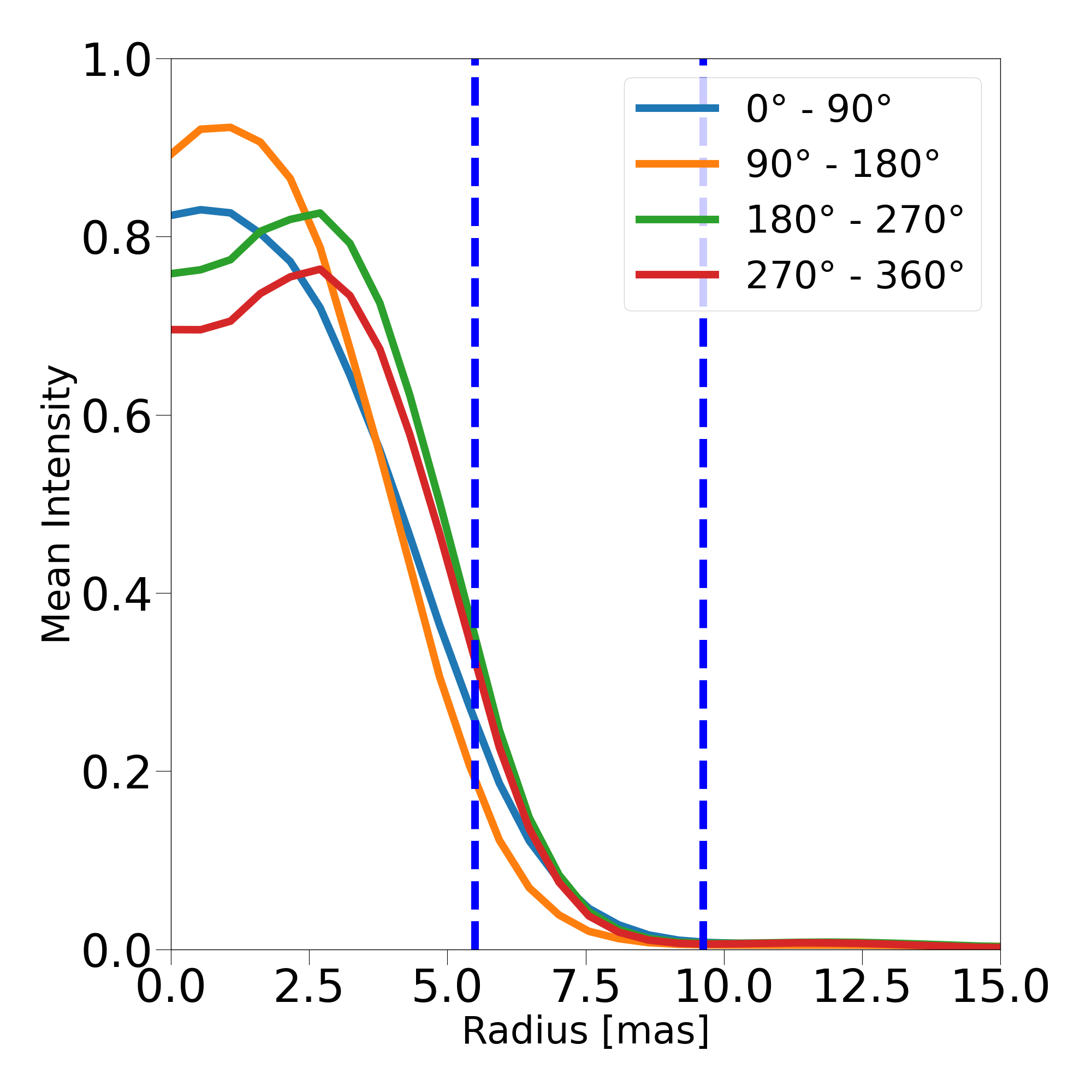}
                      \caption{3.5 $\mu$m (pseudo-continuum)}
        \end{subfigure} \hspace{0.15cm}
            \begin{subfigure}{.3\linewidth}
                \includegraphics[width = \linewidth]{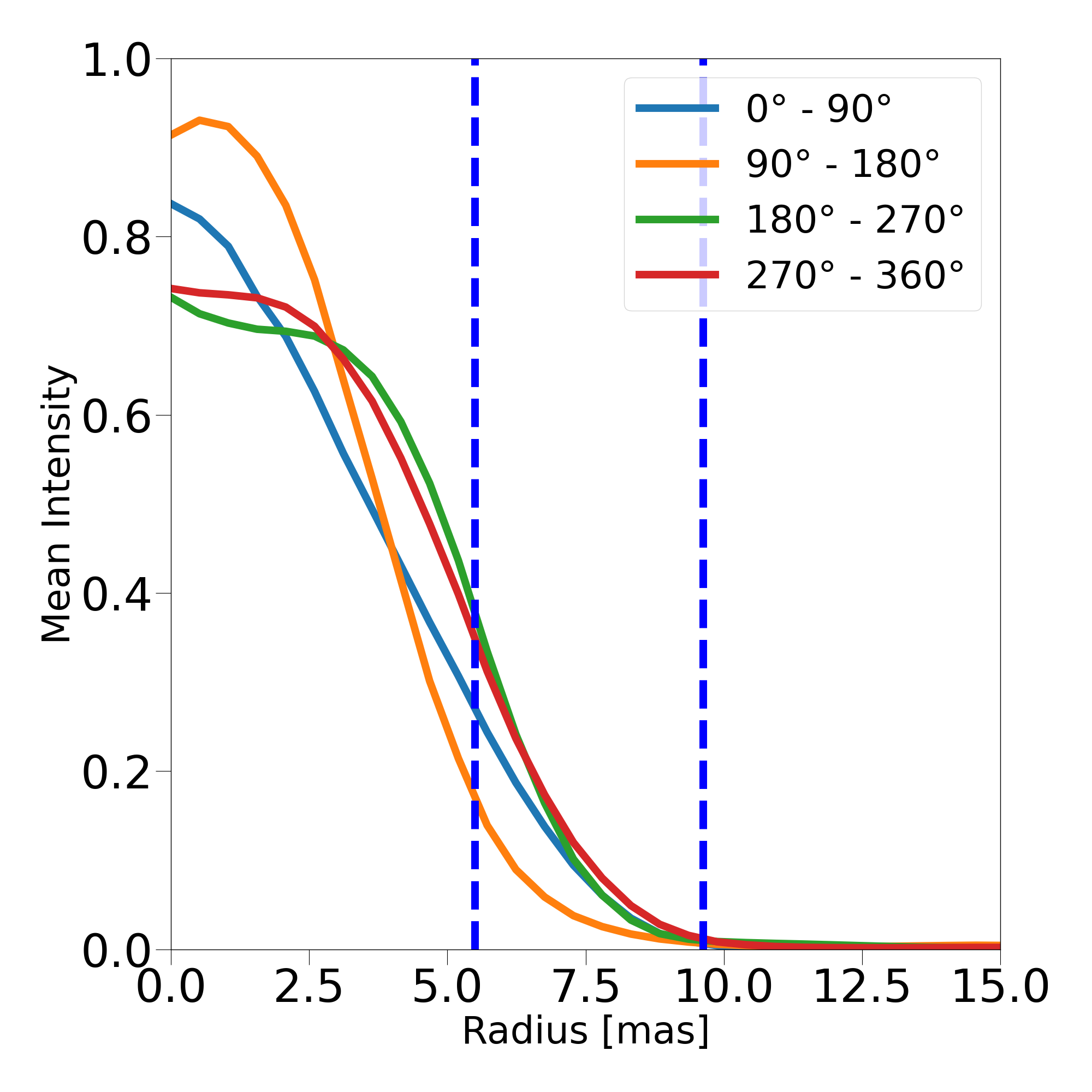}
                \caption{3.8 $\mu$m ($\mathrm{C_2H_2}$)}
            \end{subfigure}
        \end{minipage} 
    \caption{SQUEEZE reconstructed images for the L band data. See Fig.~\ref{mira_images_lband} for a full description of the contours, labels, and markers.} \label{squeezeults_lband}
\end{figure*}

 As seen in Fig.~\ref{squeezeults_lband}, the overall size of the object is the same as in the MiRA images. The more prominent features are also reproduced, such as the east-west elongation seen at 3.1 $\mu$m (although here it is part of a larger structure that also extends towards the north) and the south-westerly peak at 3.5 and 3.8 $\mu$m. In the case of SQUEEZE, the emission peak seen at 3.8 $\mu$m seems to also overlap with the one at 3.1 $\mu$m. The numerous small structures outside the stellar disk that reach the 5$\sigma$ level in the pseudo-continuum band are most likely artefacts resulting from the median averaging of the images. Combined with the background noise being much lower than in MiRA reconstructions, this could enhance the prominence of spurious background features.

\FloatBarrier
\section{Rotation tests} \label{rot_tests}

\indent\indent To assess the reliability of some features in the reconstructed images, rotation tests were used, following the approach of \citet{planquart}. This was done with the \texttt{OIFits modeler} software\footnote{Available at \href{https://amhra.oca.eu/AMHRA/oifits-modeler/input.htm}{https://amhra.oca.eu/AMHRA/oifits-modeler/input.htm}} \citep{amhra}, which creates synthetic data from the input uv-coverage and the assumed image. In this case, the uncertainties of the observational data are used, as error bars are not provided for the synthetic observables by \texttt{OIFits modeler}. The synthetic data are manipulated in a way that produces a rotated image upon a subsequent reconstruction attempt. When comparing the initial reconstruction with the rotated one, features that are present in both cases can be deemed as genuine, while the ones that disappear are most likely artefacts of the reconstruction process and can therefore not be trusted.

\par
\begin{figure*}[htbp]
    \centering
        \begin{minipage}[c]{0.95\textwidth}
           \centering
            \begin{subfigure}{.32\linewidth}
                \begin{overpic}[width=\linewidth]{snr5_contour_lm_band1_4p_compactness.png}
                    \put(20,85){\color{white}\bfseries 3.1 $\mu$m}
                    \put(65,85){\color{white}\bfseries $\mathrm{C_2H_2+HCN}$}
                \end{overpic}
            \end{subfigure}
            \begin{subfigure}{.32\linewidth}
                \begin{overpic}[width=\linewidth]{snr5_contour_lm_band2_compactness.png}
                    \put(20,85){\color{white}\bfseries 3.5 $\mu$m}
                \end{overpic}
            \end{subfigure}
            \begin{subfigure}{.32\linewidth}
                \begin{overpic}[width=\linewidth]{snr5_contour_lm_band4_compactness.png}
                    \put(20,85){\color{white}\bfseries 3.8 $\mu$m}
                    \put(85,85){\color{white}\bfseries $\mathrm{C_2H_2}$}
                \end{overpic}
            \end{subfigure}
            \par
            \begin{subfigure}{.32\linewidth}
                \begin{overpic}[width=\linewidth]{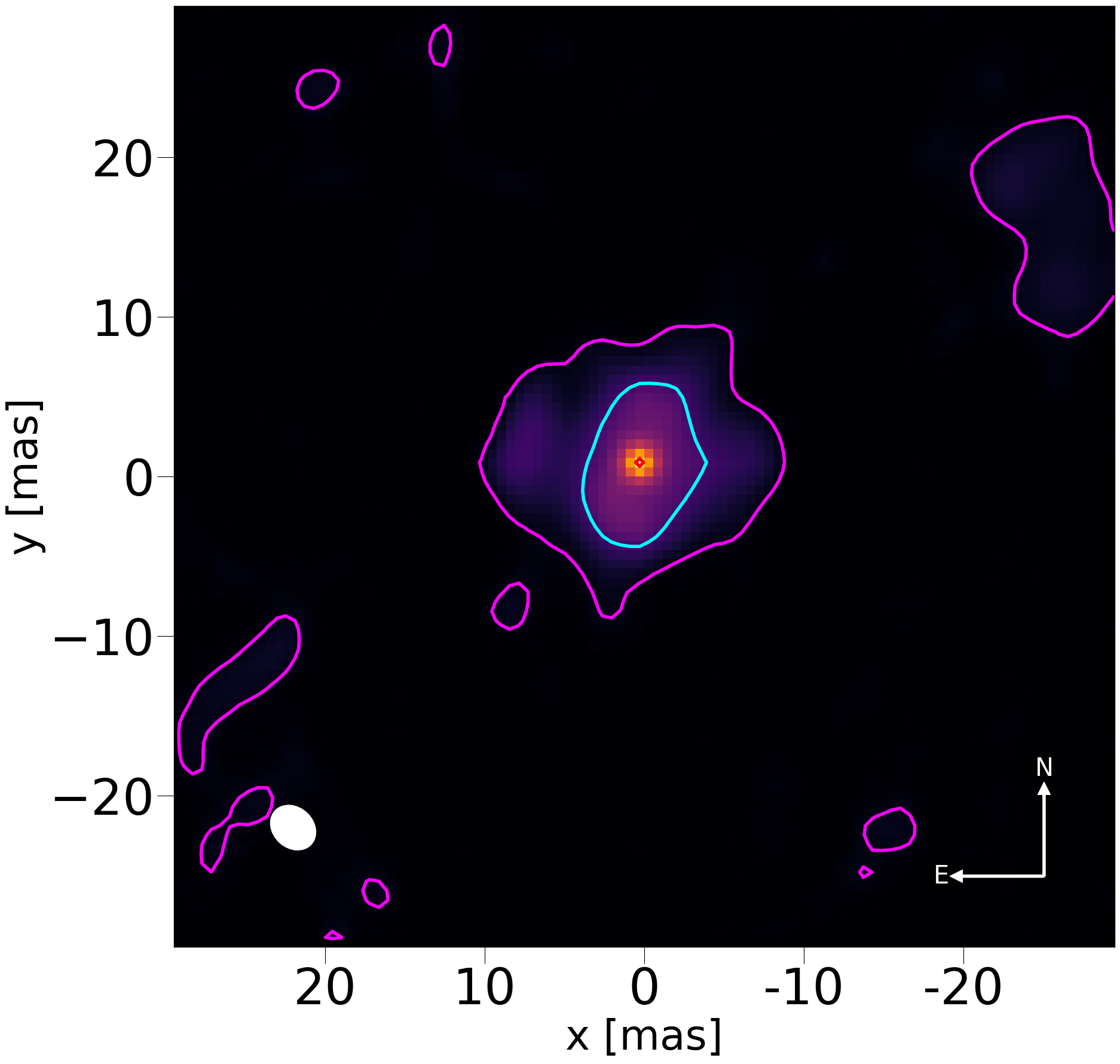}
                    \put(20,85){\color{white}\bfseries 3.1 $\mu$m}
                    \put(65,85){\color{white}\bfseries $\mathrm{C_2H_2+HCN}$}
                \end{overpic}
            \end{subfigure}
            \begin{subfigure}{.32\linewidth}
                \begin{overpic}[width=\linewidth]{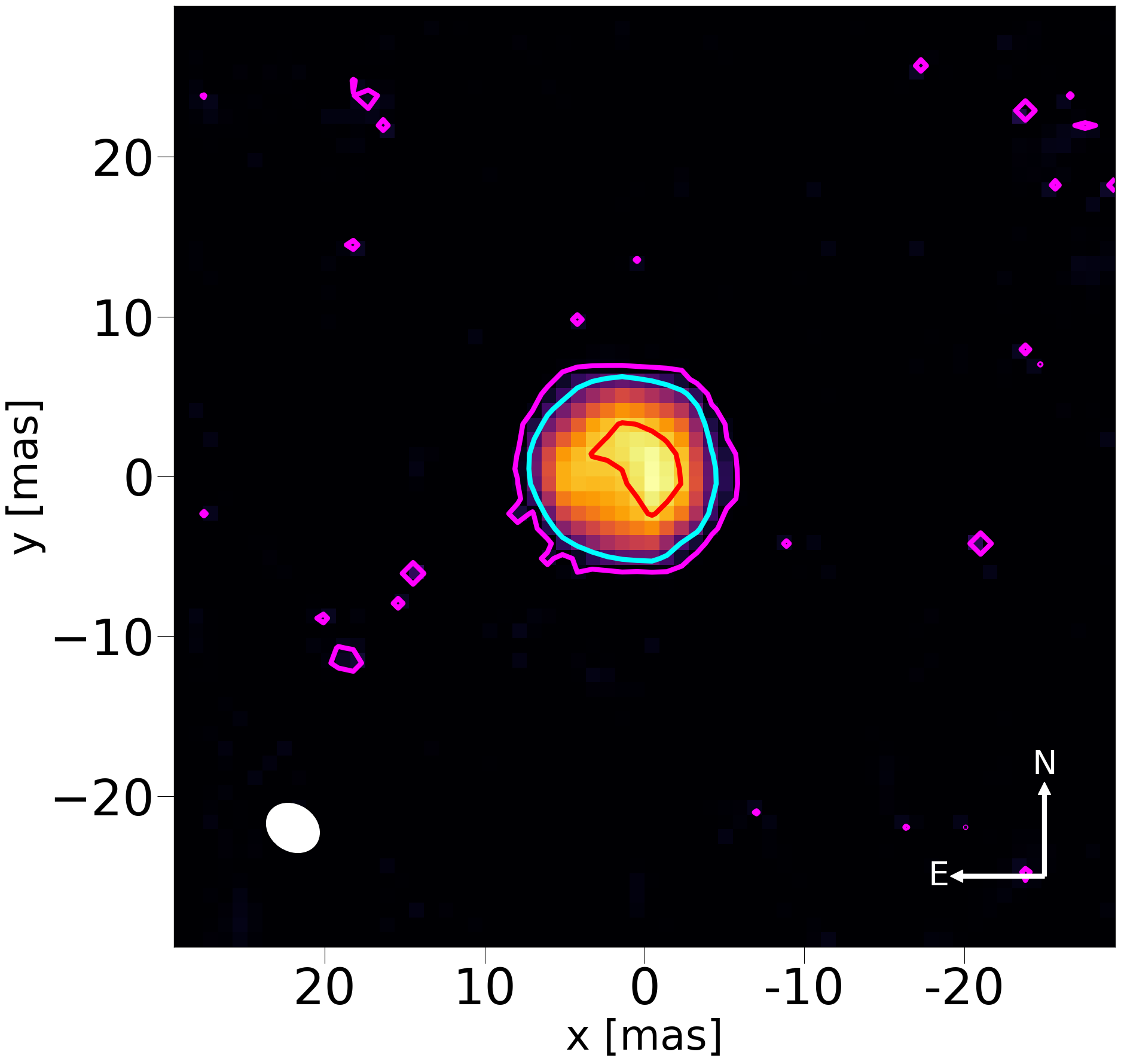}
                    \put(20,85){\color{white}\bfseries 3.5 $\mu$m}
                \end{overpic}
            \end{subfigure}
            \begin{subfigure}{.32\linewidth}
                \begin{overpic}[width=\linewidth]{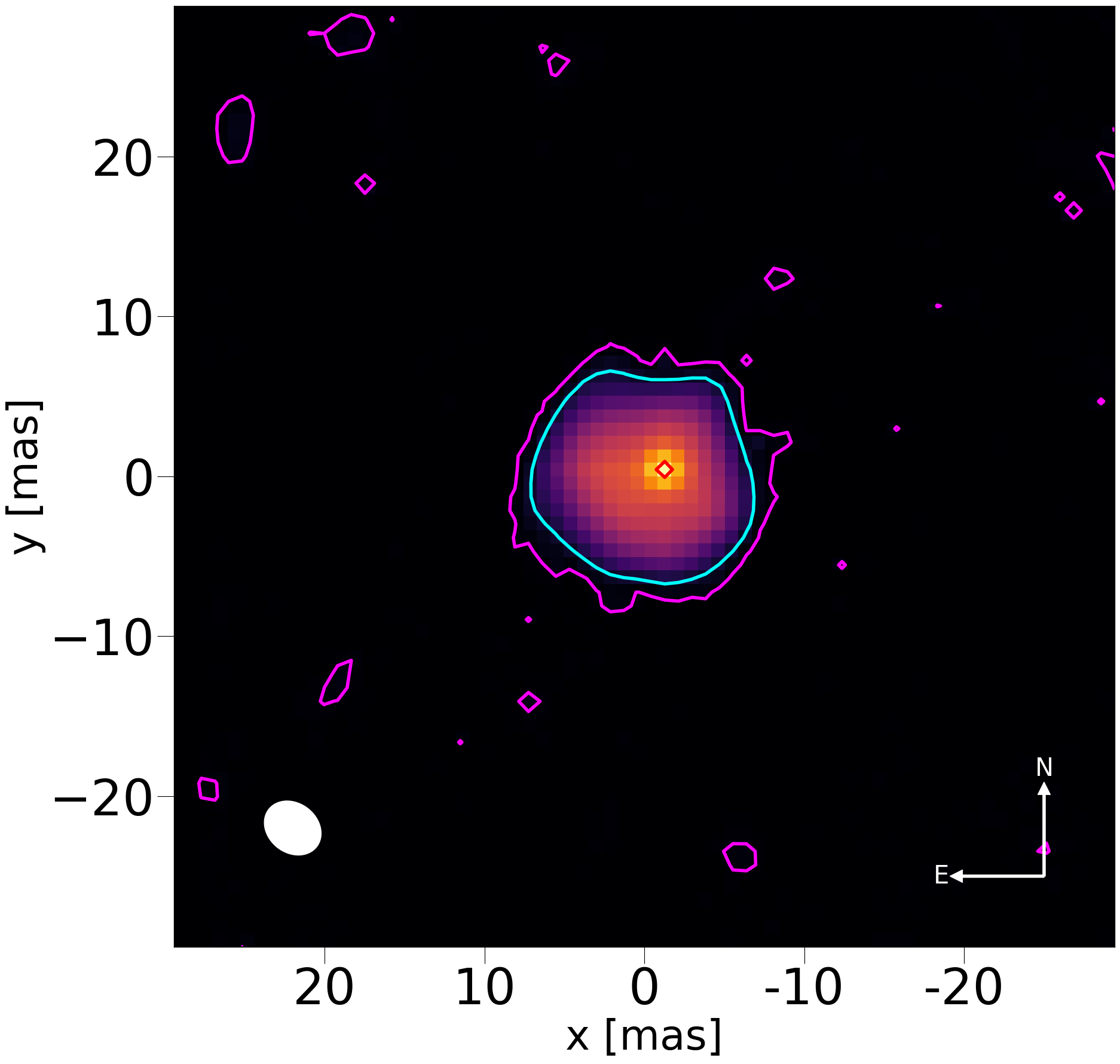}
                    \put(20,85){\color{white}\bfseries 3.8 $\mu$m}
                    \put(85,85){\color{white}\bfseries $\mathrm{C_2H_2}$}
                \end{overpic}
            \end{subfigure}           
            \par
            \begin{subfigure}{.32\linewidth}
                \begin{overpic}[width=\linewidth]{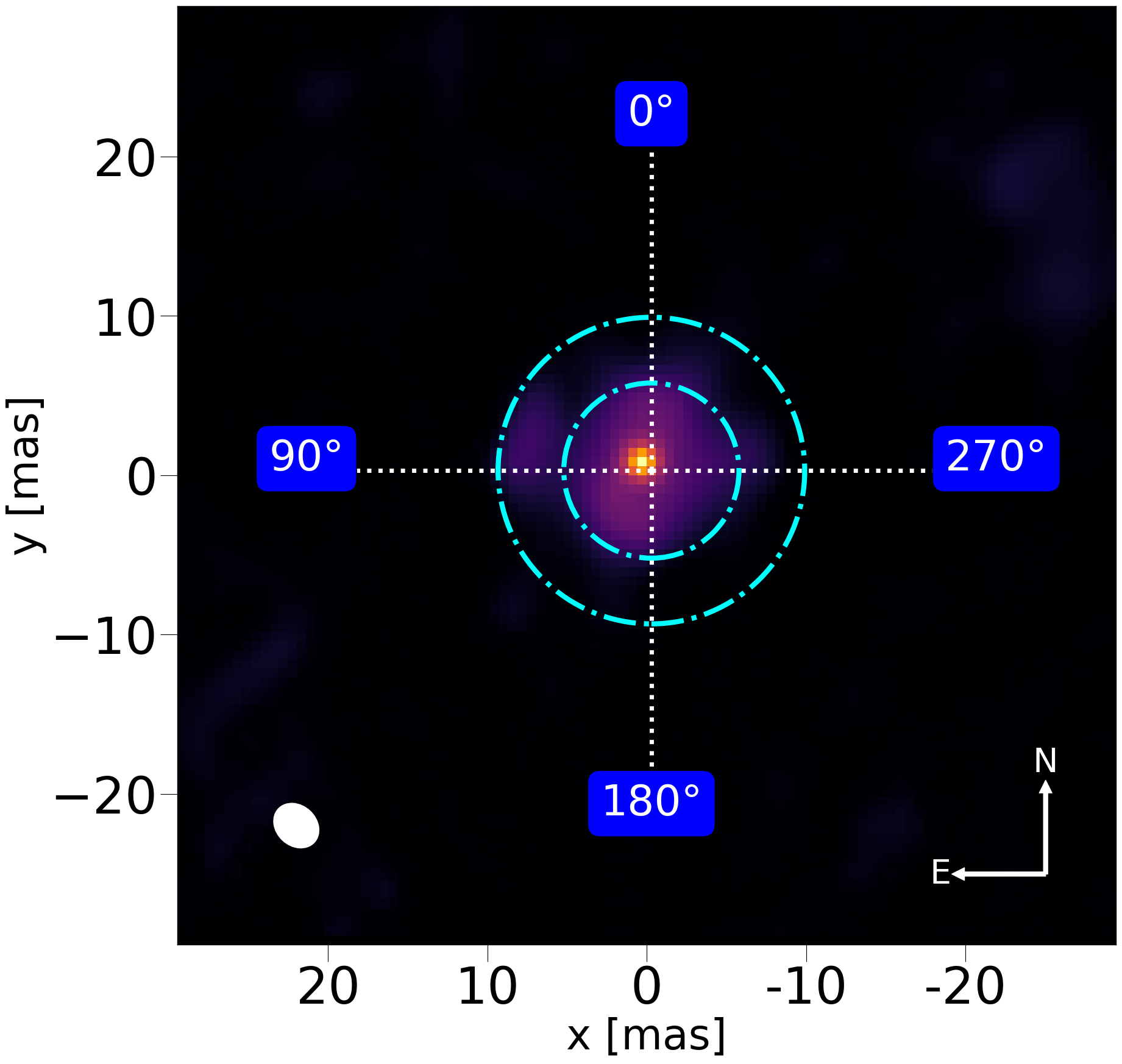}
                    \put(20,85){\color{white}\bfseries 3.1 $\mu$m}
                    \put(65,85){\color{white}\bfseries $\mathrm{C_2H_2+HCN}$}
                \end{overpic}
            \end{subfigure}
            \begin{subfigure}{.32\linewidth}
                \begin{overpic}[width=\linewidth]{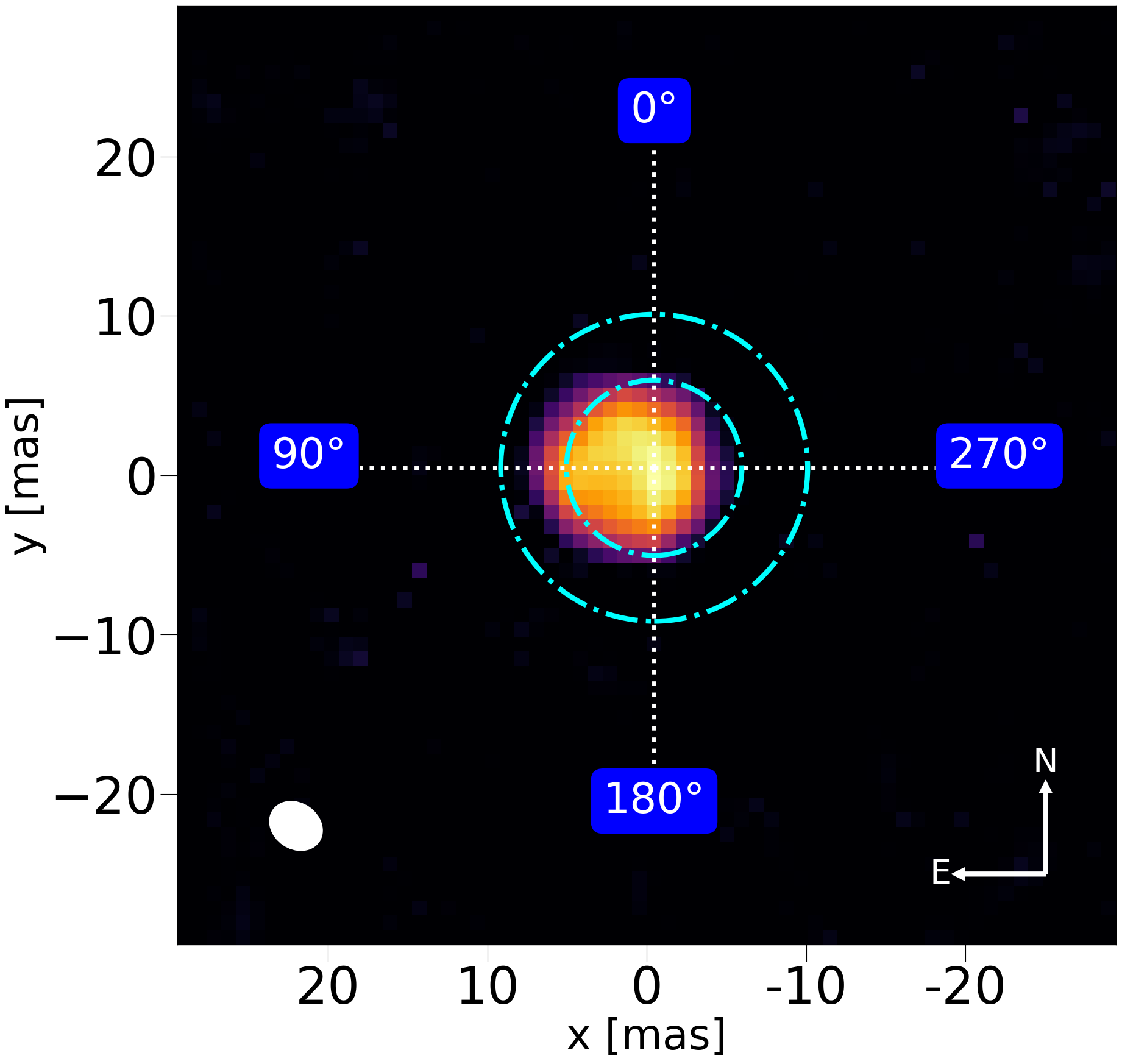}
                    \put(20,85){\color{white}\bfseries 3.5 $\mu$m}
                \end{overpic}
            \end{subfigure}        
            \begin{subfigure}{.32\linewidth}
                \begin{overpic}[width=\linewidth]{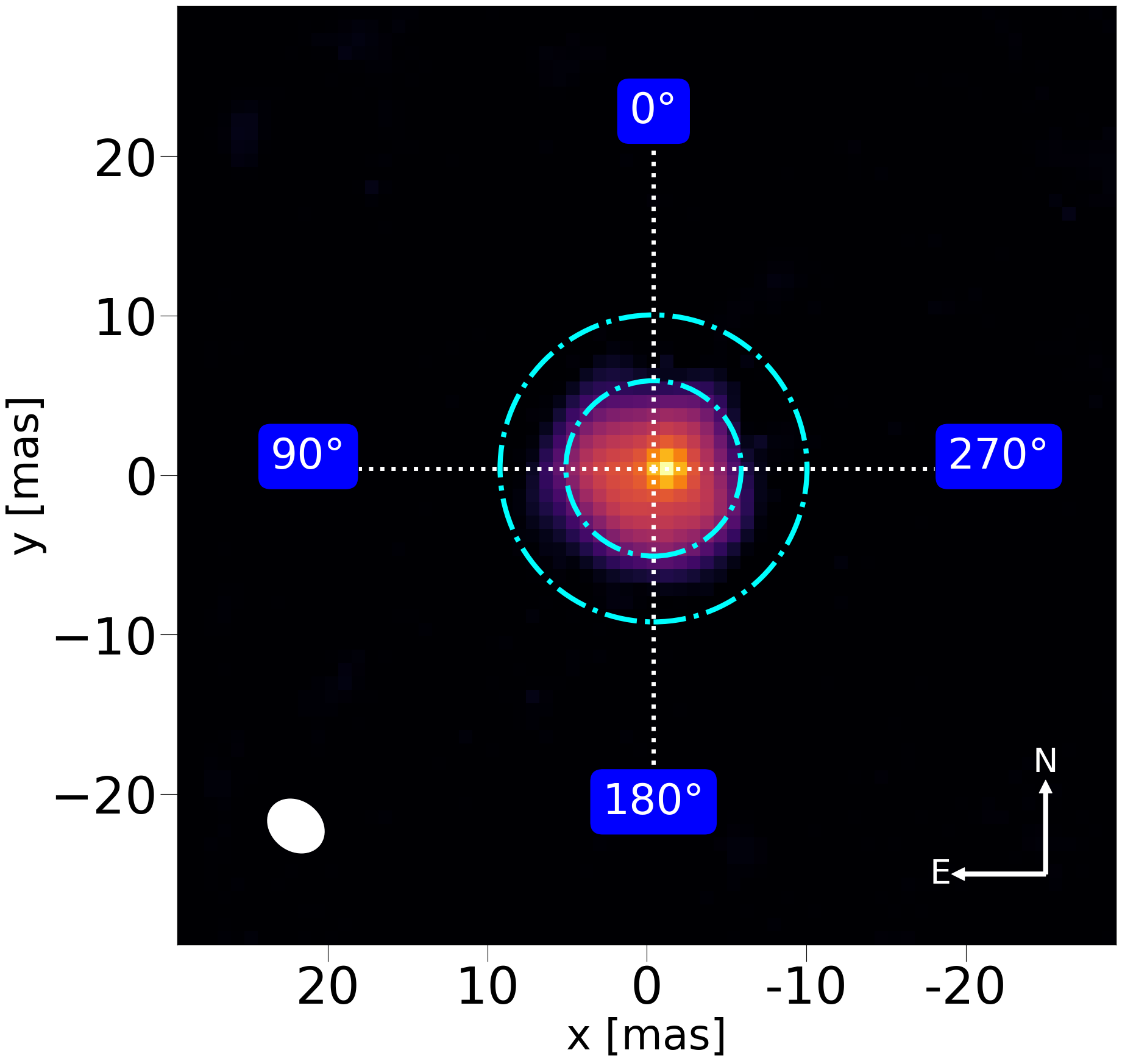}
                    \put(20,85){\color{white}\bfseries 3.8 $\mu$m}
                    \put(85,85){\color{white}\bfseries $\mathrm{C_2H_2}$}
                \end{overpic}
            \end{subfigure}        
            \par     
            \begin{subfigure}{.95\linewidth}
            \centering
            \hspace{0.85cm}
            \includegraphics[width=0.5\linewidth]{colorbar.png}
            \end{subfigure}
            \par  
            \hspace{0.55cm}
              \begin{subfigure}{.3\linewidth}
                      \includegraphics[width = \linewidth]{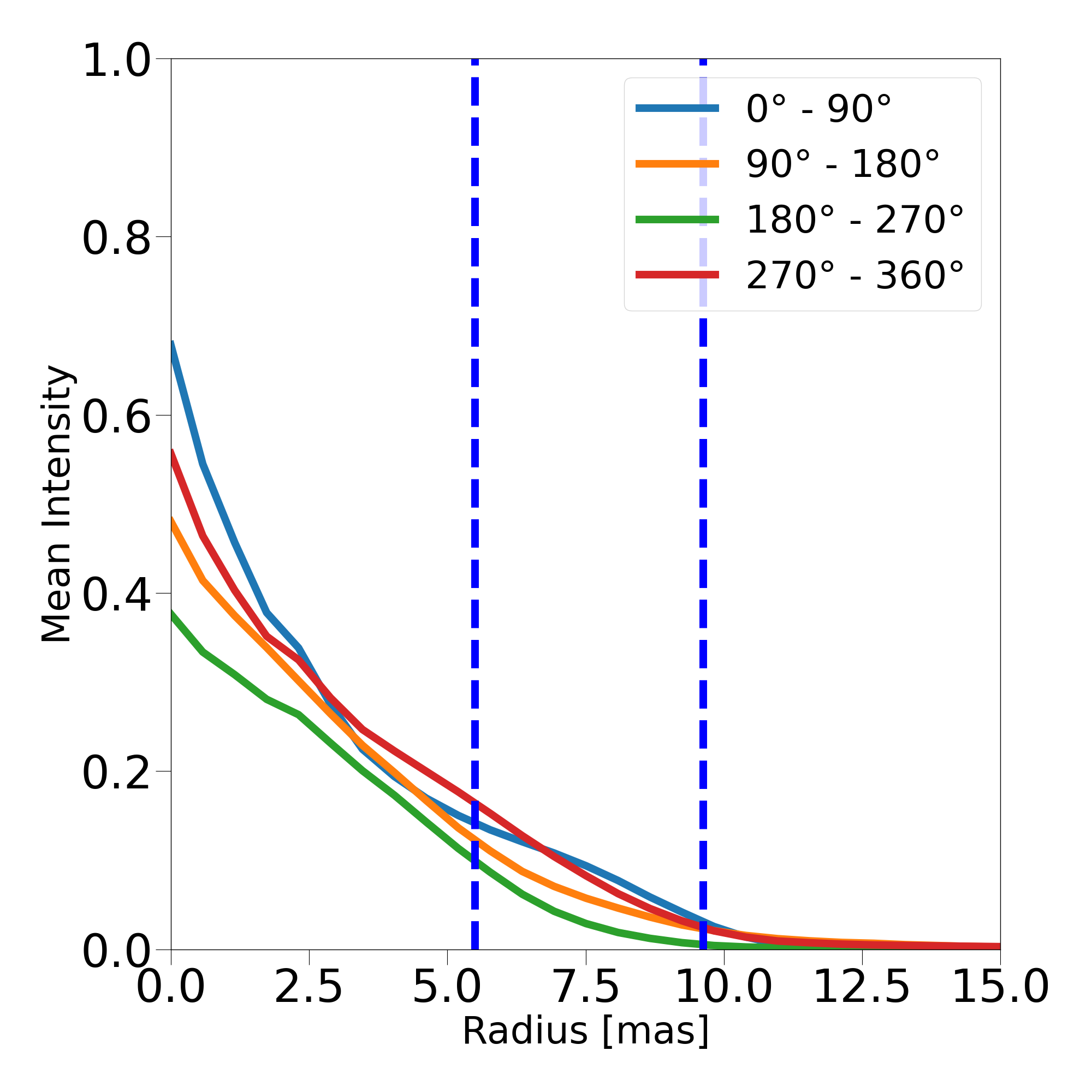}
                \caption{3.1 $\mu$m ($\mathrm{C_2H_2 + HCN}$)}
            \end{subfigure}     \hspace{0.25cm}       
        \begin{subfigure}{.3\linewidth}
                      \includegraphics[width = \linewidth]{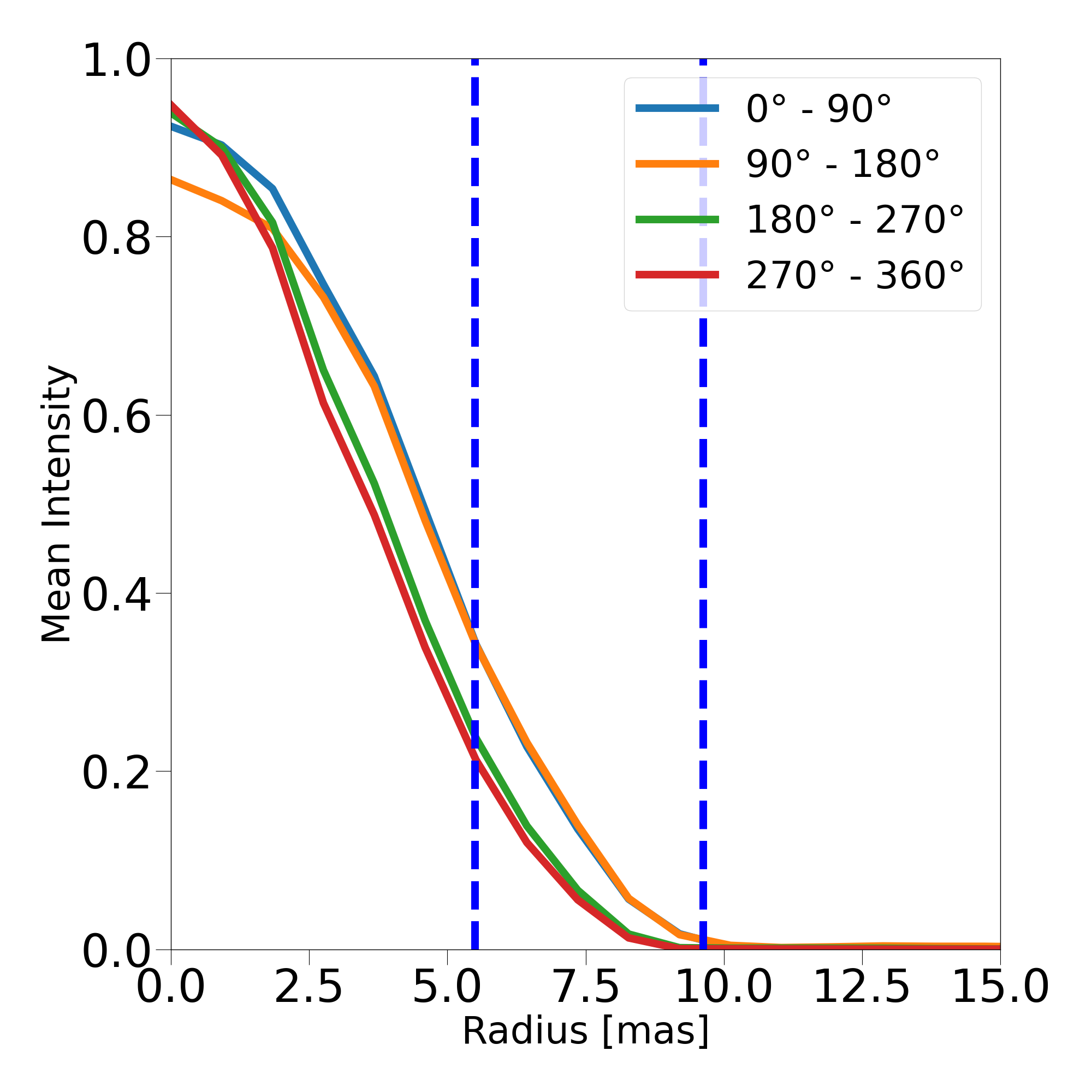}
                      \caption{3.5 $\mu$m (pseudo-continuum)}
        \end{subfigure} \hspace{0.15cm}
            \begin{subfigure}{.3\linewidth}
                \includegraphics[width = \linewidth]{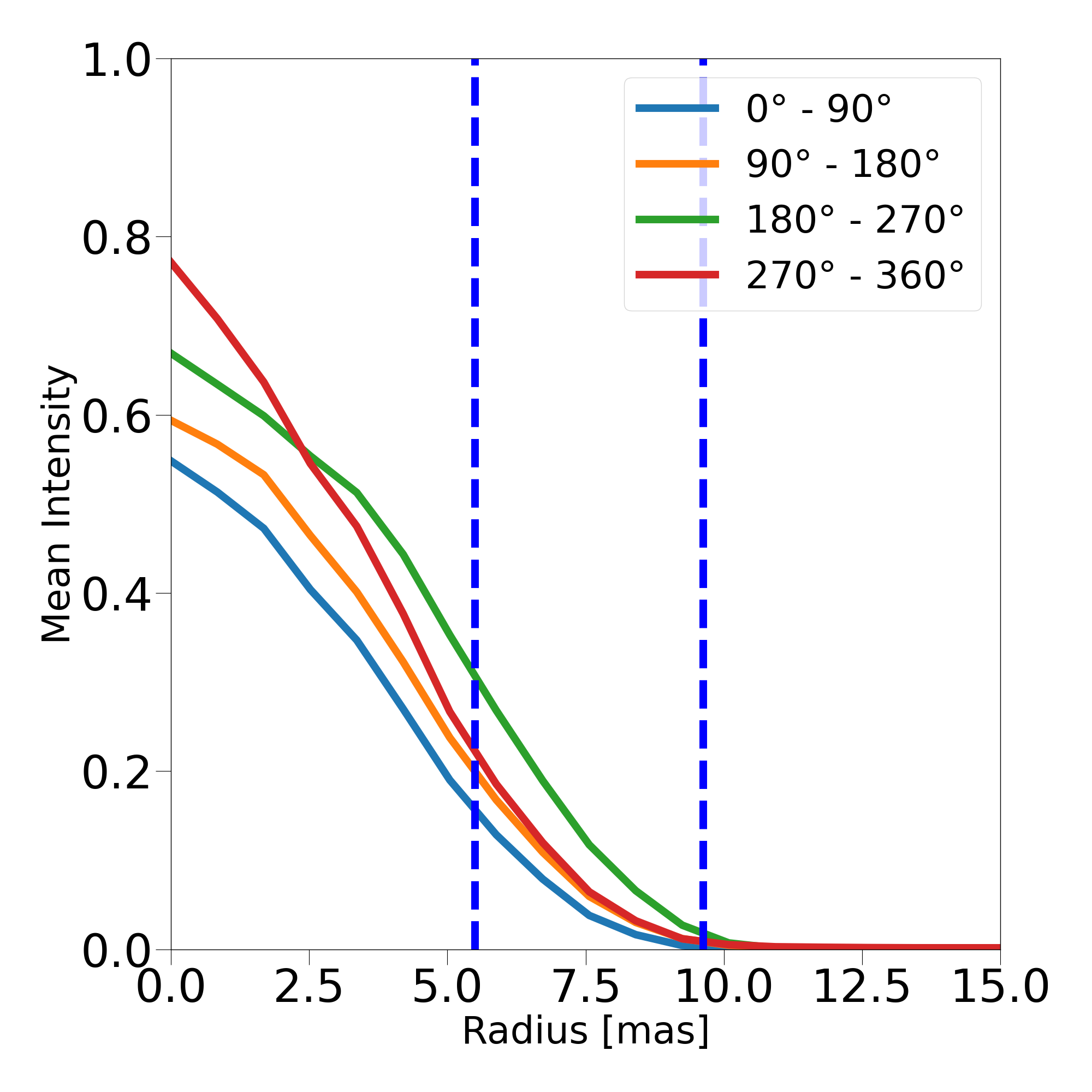}
                \caption{3.8 $\mu$m ($\mathrm{C_2H_2}$)}
            \end{subfigure}
        \end{minipage}
        \caption{MiRA original image reconstructions (top row) versus MiRA reconstruction of the \texttt{OIFits modeler} data, i.e. MiRA images rotated 90$\degr$ anti-clockwise (rows 2--4), for the MATISSE L band. See Fig.~\ref{mira_images_lband} for a full description of the contours, labels, and markers.} \label{mira_images_rot90_lband} 
\end{figure*}

\begin{figure*}[htbp]
    \centering
    \begin{minipage}[c]{0.95\textwidth}
        \centering
         \begin{subfigure}{.32\linewidth}
            \begin{overpic}[width=\linewidth]{snr5_contour_n_band1_extended_compactness.png}
                \put(20,85){\color{white}\bfseries 8.5 $\mu$m}
            \end{overpic}
        \end{subfigure}
        \begin{subfigure}{.32\linewidth}
            \begin{overpic}[width=\linewidth]{snr5_contour_n_band3_extended_compactness.png}
                \put(20,85){\color{white}\bfseries 11.3 $\mu$m}
                \put(85,85){\color{white}\bfseries SiC}
            \end{overpic}
        \end{subfigure}
        \par
        \begin{subfigure}{.32\linewidth}
            \begin{overpic}[width=\linewidth]{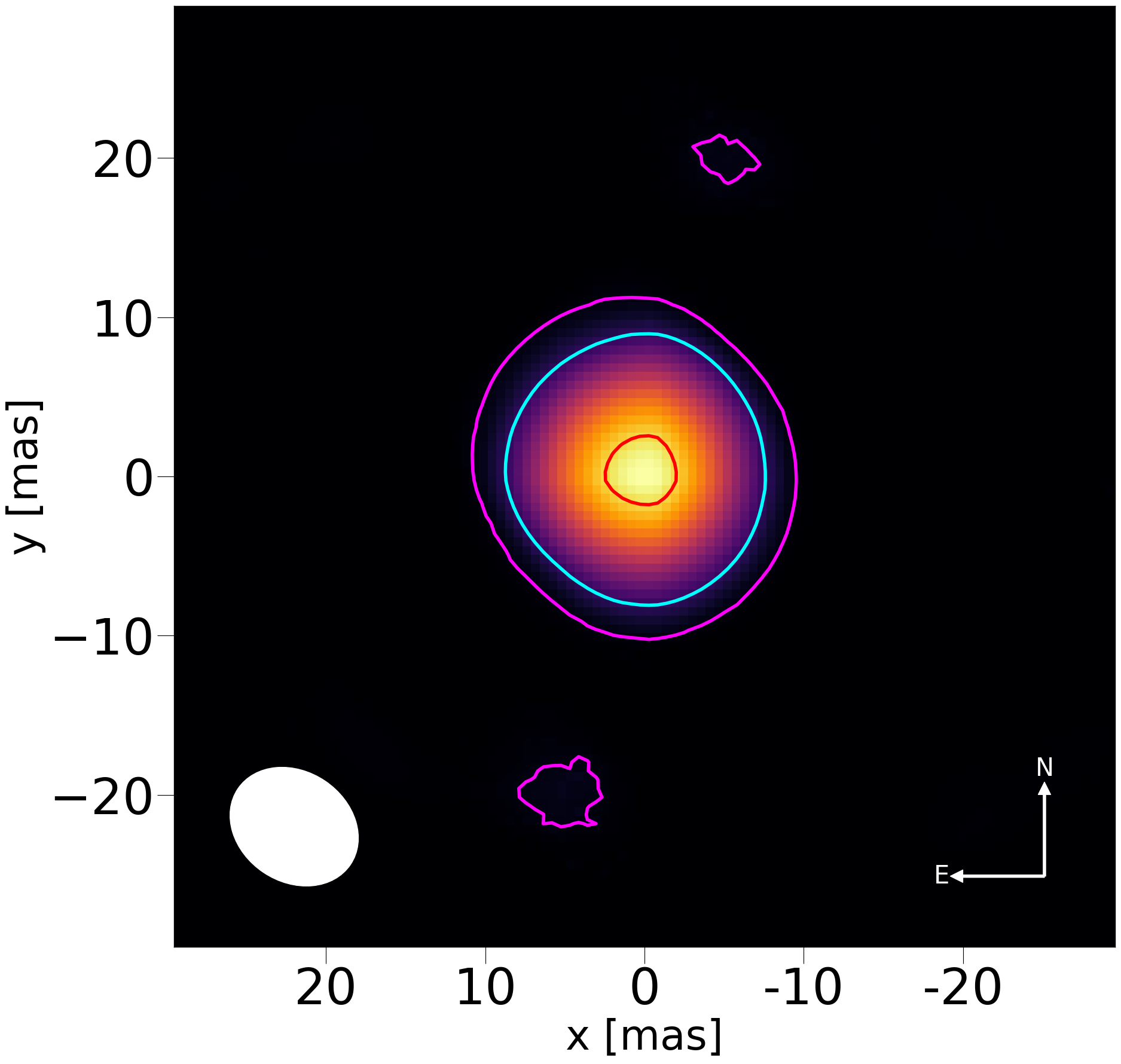}
                \put(20,85){\color{white}\bfseries 8.5 $\mu$m}
            \end{overpic}
        \end{subfigure}
        \begin{subfigure}{.32\linewidth}
            \begin{overpic}[width=\linewidth]{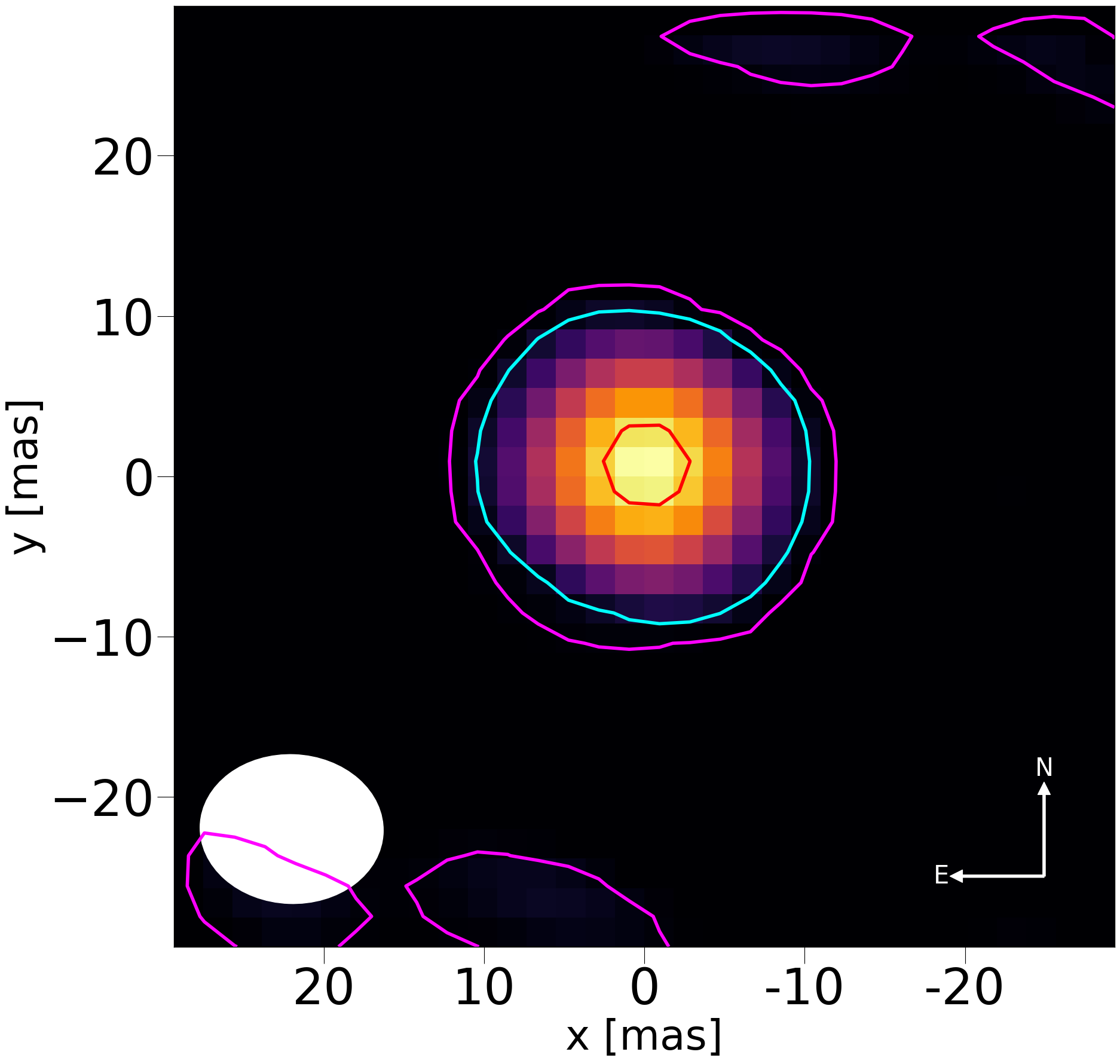}
                \put(20,85){\color{white}\bfseries 11.3 $\mu$m}
                \put(85,85){\color{white}\bfseries SiC}
            \end{overpic}
        \end{subfigure}
        \par
        \begin{subfigure}{.32\linewidth}
            \begin{overpic}[width=\linewidth]{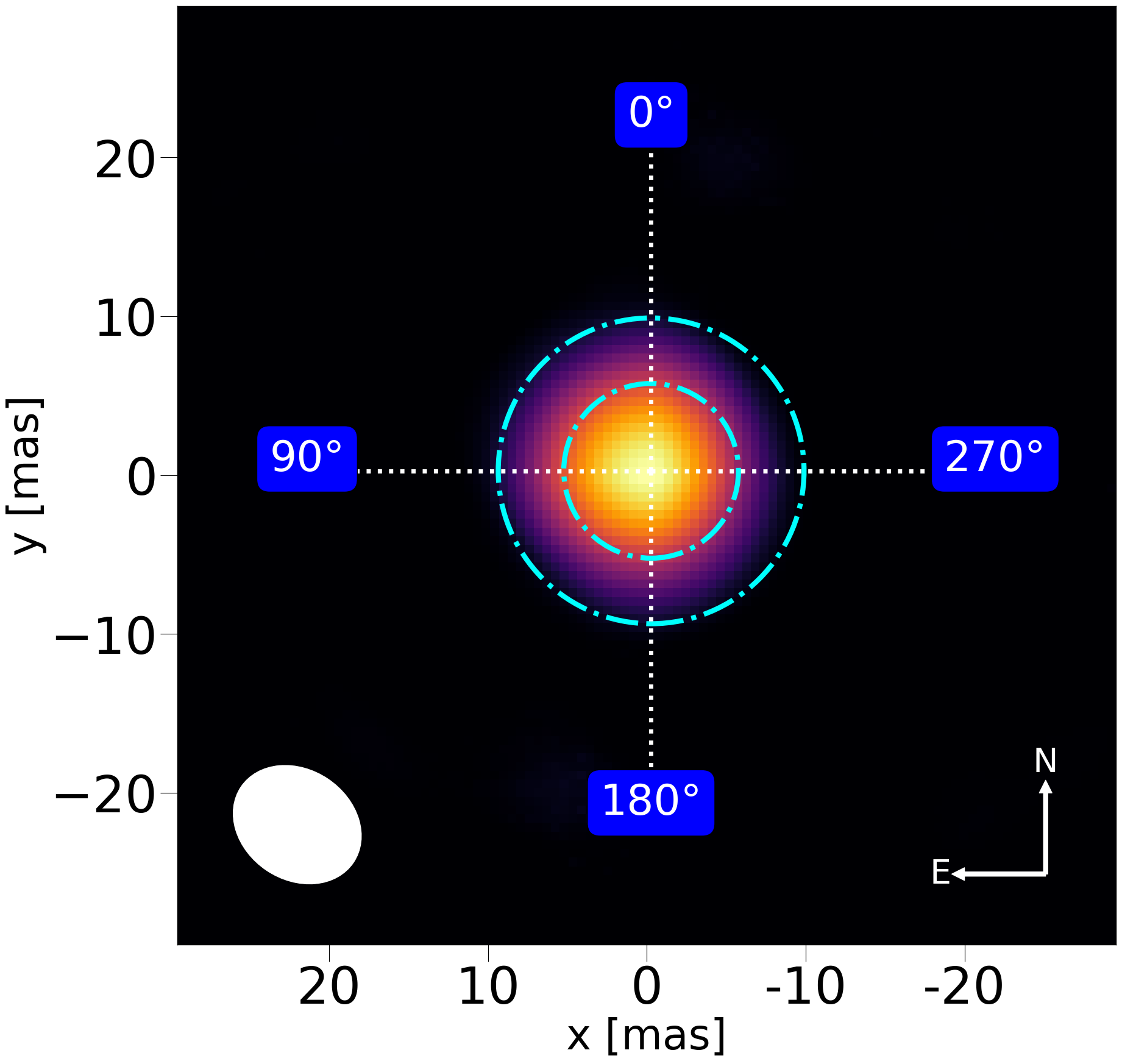}
                \put(20,85){\color{white}\bfseries 8.5 $\mu$m}
            \end{overpic}
        \end{subfigure}
        \begin{subfigure}{.32\linewidth}
            \begin{overpic}[width=\linewidth]{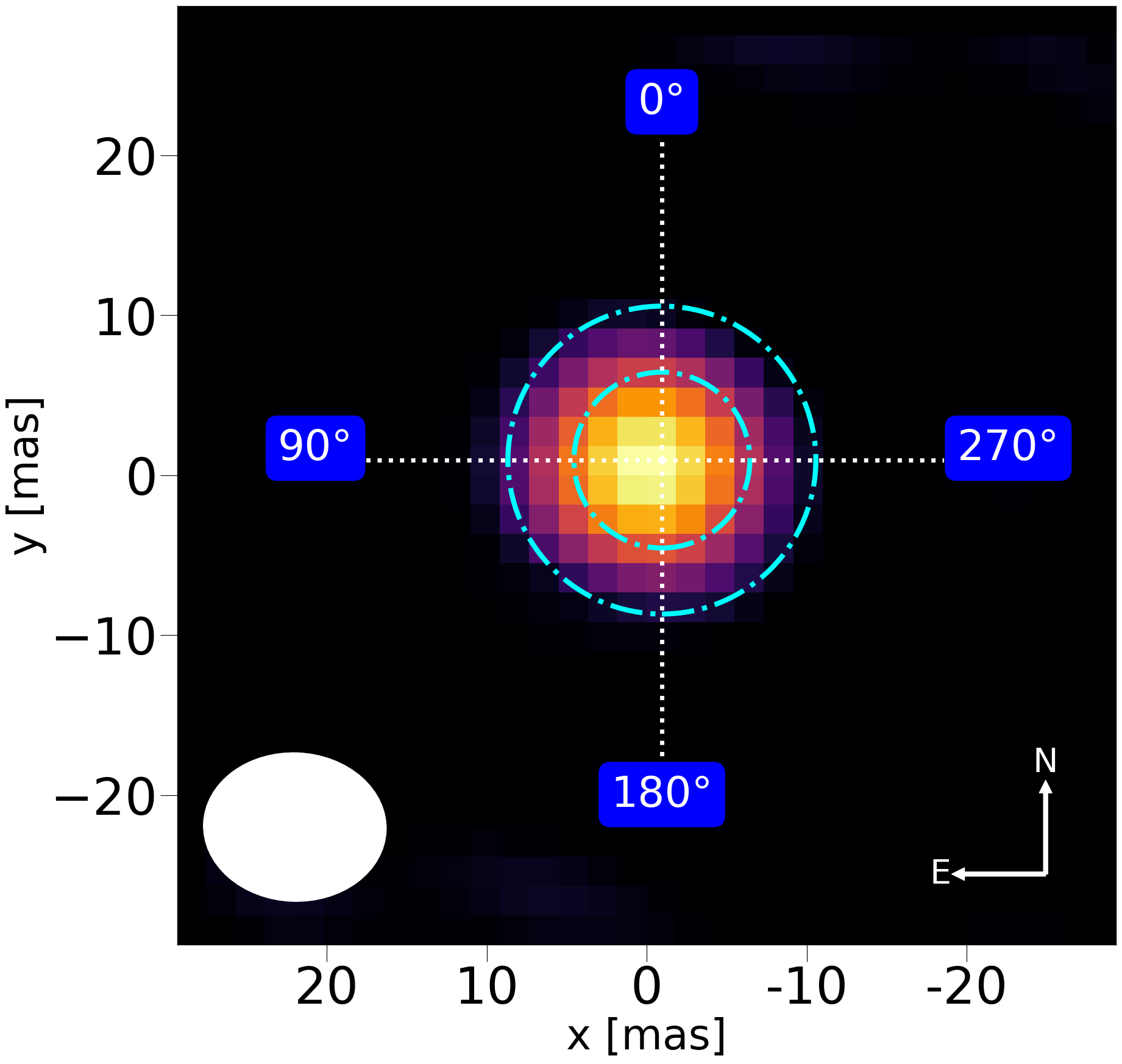}
                \put(20,85){\color{white}\bfseries 11.3 $\mu$m}
                \put(85,85){\color{white}\bfseries SiC}
            \end{overpic}
        \end{subfigure}
        \par
        \begin{subfigure}{.95\linewidth}
            \centering
            \hspace{0.85cm}
            \includegraphics[width=0.5\linewidth]{colorbar.png}
        \end{subfigure}
        \par
        \hspace{1cm}
        \begin{subfigure}{.32\linewidth}
            \includegraphics[width = \linewidth]{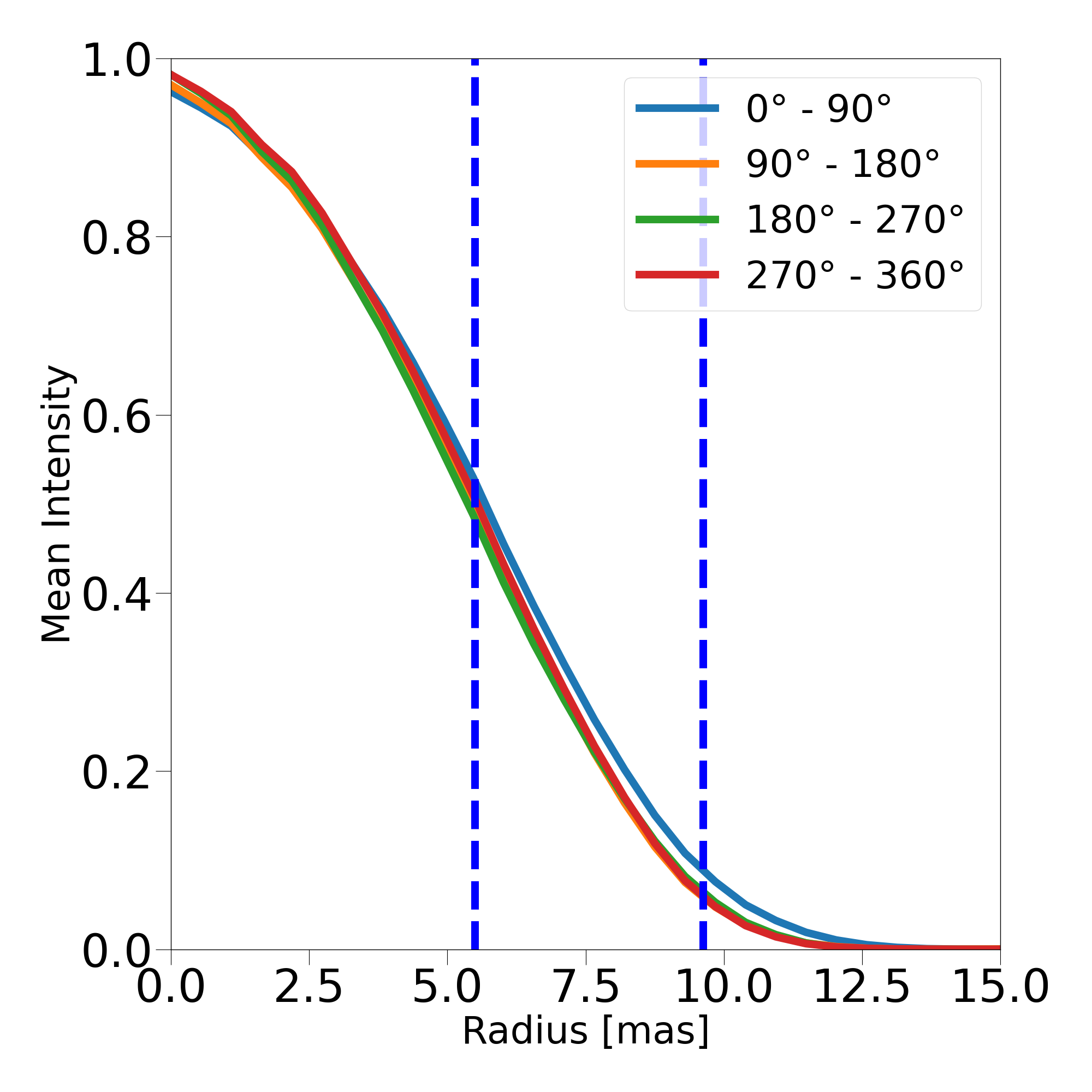}
            \caption{8.5 $\mu$m (pseudo-continuum)}
        \end{subfigure}
        \begin{subfigure}{.32\linewidth}
            \includegraphics[width = \linewidth]{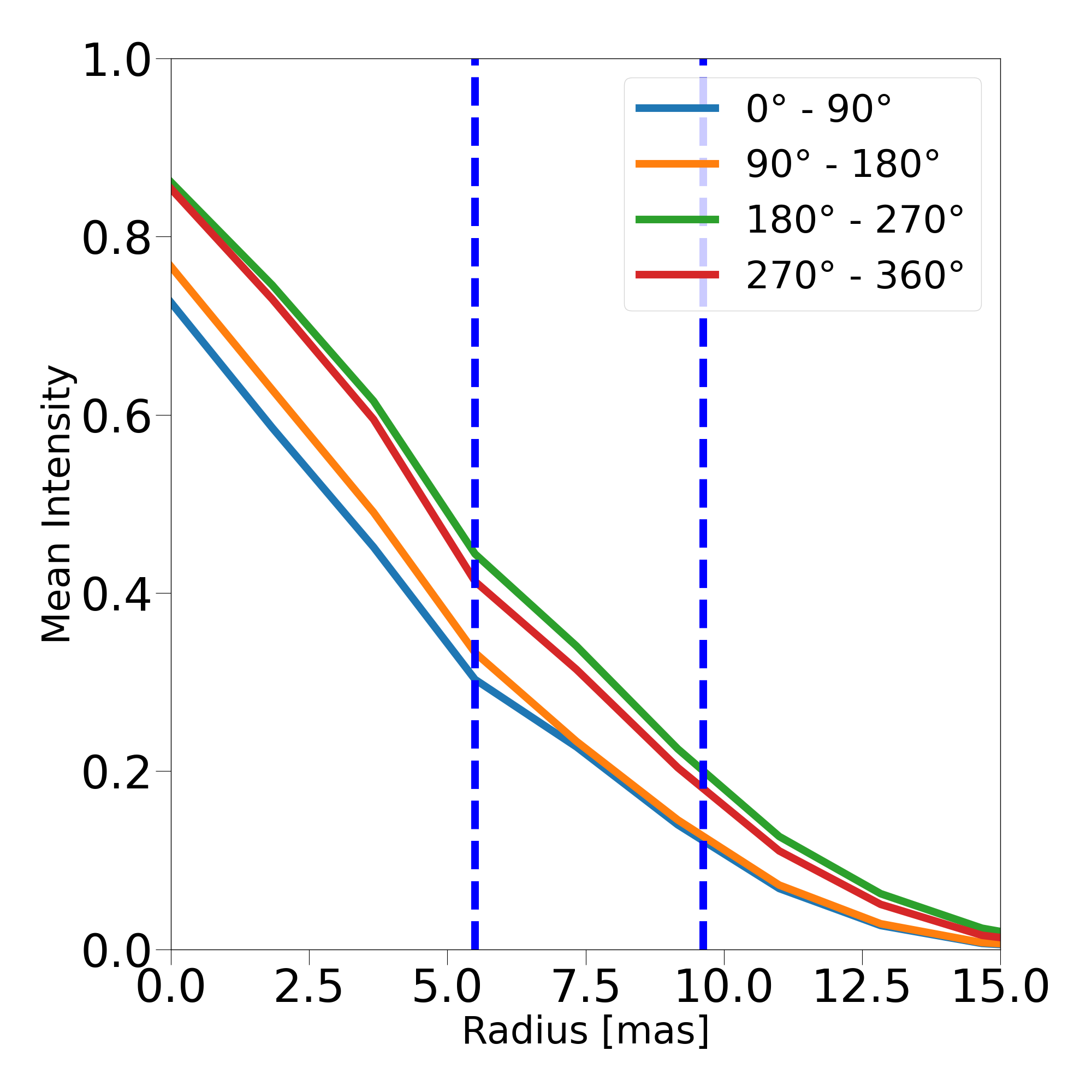}
            \caption{11.3 $\mu$m (SiC)}
        \end{subfigure} 
    \end{minipage}
    \caption{MiRA original image reconstructions (top row) versus MiRA reconstruction of the \texttt{OIFits modeler} data, i.e. MiRA images rotated 90$\degr$ anti-clockwise (rows 2--4), for the MATISSE N band. See Fig.~\ref{mira_images_nband} for a full description of the contours, labels, and markers.} \label{mira_images_rot90_nband} 
\end{figure*}

 As seen in Figs.~\ref{mira_images_rot90_lband} and \ref{mira_images_rot90_nband}, the data were rotated by 90$\degr$ anti-clockwise and most of the major features and peaks in intensity are recovered quite well. The 3.1 and 3.8 $\mu$m images shows some differences, specifically in intensity. The major features such as the protrusions and the elongation in the middle are still there, but the intensity is spread out more evenly across the entire surface of this layer. In the case of the 3.1 $\mu$m image, the intensity drop is quite significant, and some larger artefacts are visible towards the edges. This is most likely a product of the reconstruction algorithm, as the chosen regularisation function (quadratic compactness) can smooth out some of the finer details and sometimes can even lead to overly smoothed images.

\end{appendix}

\end{document}